\documentclass[fleqn, useAMS,usenatbib]{mnras_sjs}
\usepackage{amssymb, graphicx, verbatim}
\usepackage{upgreek, amsmath, ulem}

\newcommand{\e}[1]{\times 10^{#1}}

\newcommand{\wl}{$\lambda$}
\newcommand{\msun}{M$_\odot$}
\newcommand{\wll}{$\lambda \lambda$}

\setcitestyle{round,aysep={},yysep={,}}

\title{Nebular spectra of pair-instability supernovae}
\author[Anders Jerkstrand]{A. Jerkstrand$^{1}$\thanks{E-mail:a.jerkstrand@qub.ac.uk}, 
S. J. Smartt$^{1}$, A. Heger$^{2-4}$\\
$^1$Astrophysics Research Centre, School of Mathematics and Physics, Queen's University Belfast, Belfast BT7 1NN, UK
\\
$^2$Monash Centre for Astrophysics, School of Physics and Astronomy, Monash University, VIC 3800, Australia\\
$^3$University of Minnesota, School of Physics and Astronomy, Minneapolis, MN 55455, USA\\
$^4$Shanghai Jiao-Tong University, Department of Physics and Astronomy, Shanghai 200240, P.~R.~China\\
}
\begin{document}

\date{}

\pagerange{\pageref{firstpage}--\pageref{lastpage}} \pubyear{2002}

\maketitle

\label{firstpage}

\begin{abstract}
If very massive stars ($M \gtrsim 100~M_\odot$) can form and avoid too strong mass loss during their evolution, they are predicted to explode as pair-instability supernovae (PISNe). One critical test for candidate events is whether their nucleosynthesis yields and internal ejecta structure, being revealed  through nebular-phase spectra at $t \gtrsim 1$ yr, match those of model predictions. 
Here we compute theoretical spectra based on model PISN ejecta at 1-3 years post-explosion to allow quantitative comparison with observations.
The high column densities of PISNe lead to complete gamma-ray trapping for $t\gtrsim 2$ years which, combined with fulfilled conditions of steady state, leads to bolometric supernova luminosities matching the $^{56}$Co decay. Most of the gamma-rays are absorbed by the deep-lying iron and silicon/sulphur layers. The ionization balance shows a predominantly neutral gas state, which leads to emission lines of Fe~I, Si~I, and S~I. For low-mass PISNe the metal core expands slowly enough to produce a forest of distinct lines, whereas high-mass PISNe expand faster and produce more featureless spectra. Line blocking is complete below $\sim$5000 \AA~for several years, and the model spectra are red. The strongest line is typically [Ca~II]~\wll7291,~7323, one of few lines from ionized species. 
We compare our models with proposed PISN candidates SN 2007bi and PTF12dam, finding discrepancies for several key observables and thus no support for a PISN interpretation. We discuss distinct spectral features predicted by the models, and the possibility of detecting pair-instability explosions among non-superluminous supernovae.
\end{abstract}


\section{Introduction}

Recent observational results favour the existence of stars with $M \gtrsim 100$~\msun~in the local Universe. Dynamical weighing in spectroscopic binaries currently sets the highest directly inferred masses to about 100 \msun~\citep{Rauw2004, Bonanos2004, Schnurr2008}. The luminosity of Eta Carinae implies a current-day mass of the primary star of about 100 \msun \citep{Davidson1997, Hillier2001}, and spectroscopic modelling of very luminous stars in the Large Magellanic Cloud has determined luminosities requiring evolutionary masses of up to
300 \msun \citep{Crowther2010, Bestenlehner2011}. While spherically symmetric star formation models fail to produce stars more massive than about 20 \msun~\citep{Kahn1974,Wolfire1987} recent multidimensional simulations have succeeded in obtaining much higher masses \citep{Krumholz2009, Kuiper2010, Cunningham2011}. There is also a possibility that some very massive stars are formed via mergers \citep{Bonnell1998, PortegiesZwart1999, Pan2012}.

If very massive stars can form, their evolution and eventual fate depend sensitively on mass loss. 
If too strong mass loss can be avoided, massive helium cores, $M(\mbox{He}) \gtrsim 40$ \msun, may form which during core oxygen/neon burning reach a pair-formation instability that initiates collapse \citep{Barkat1967, Rakavy1967, Ober1983, Bond1984, ElEid1986, Heger2002, Waldman2008, Kozyreva2014a}. Because large amounts of thermonuclear fuel is still available, explosive burning sets in. For $M(\mbox{He}) \approx 40-65$ \msun, a violent eruption occurs, ejecting several solar masses of material. These eruptions, called pulsational pair-instability supernovae (SN), repeat until eventual core-collapse \citep{Heger2002, Woosley2007}.
For $M(\mbox{He}) \approx 65-130$ \msun, the first instability unbinds and explodes the entire star, producing a pair-instability SN \citep[PISN,][]{Heger2002}. For $M(\mbox{He}) \gtrsim 130$ \msun, photodisintegration losses causes the star to collapse directly to a black hole \citep{Fryer2001, Heger2002}.

Limitations in current understanding of mass loss makes it uncertain whether such massive helium cores are produced in Nature, and if so for which metallicities and rotation rates.  The highest-metallicity models presented in the literature that reach the PISN stage are models at $Z=0.43Z_\odot$ and $0.14 Z_\odot$  by \citet{Yusof2013}, and models at $Z=0.07 Z_\odot$ by \citet{Langer2007}. One should note, however, that these stars have large Eddington factors $\Gamma$ (in the \citet{Yusof2013} PISN progenitor models, $\Gamma \gtrsim 0.7$ on the main-sequence, increasing up to $\Gamma=0.96$ in the late burning stages) but the impact of boosted mass-loss rates \citep{Vink2011, Grafener2011}, sub-surface convection and envelope inflation \citep{Sanyal2015} and continuum-driven winds \citep{Owocki2004} are not included. Taking some of these effects into account, \citet{Kohler2015} excludes formation of PISNe at $Z \geq 0.4 Z_\odot$. On the other hand, uncertainties in the effects of clumping on main-sequence mass loss rates, and post-main sequence mass loss rates in general, could involve over-estimates of the mass-loss rate in some phases. One should also consider the possibility of late-time mergers for creating massive enough progenitors \citep{Justham2014}. 

Interest in PISNe has been rekindled by the possibility of detecting Pop III explosions with the James Webb Space Telescope and other high-z instruments \citep{Scannapieco2005, Whalen2013b}, the potential of their progenitors for re-ionizing the Universe \citep{Wyithe2003}, and from the discovery of long-duration superluminous SNe both at high \citep{Cooke2012} and low \citep{Smith2007, Gal-Yam2009,Young2010,Nicholl2013} redshift. These long-duration SNe, in particular the well-observed SN 2007bi and PTF12dam, have light curves with peak magnitudes and decline rates in broad agreement with theoretical PISN models. If they exist, PISNe would play an important role in galactic nucleosynthesis \citep{Umeda2002, Heger2002, Cooke2011, Morsony2014} as well as in feedback of momentum and energy into the ISM. 
%

Much work has been dedicated to modelling the appearance of PISNe during the shock breakout and diffusion phases \citep{Kasen2011, Dessart2012, Whalen2013, Dessart2013, Chatzopoulos2013, Whalen2014, Kozyreva2014b, Smidt2015, Chatzopoulos2015, Kozyreva2015}, and fitting of these models to observed candidates. The ability of central engine \citep[magnetar spin-down or accreting black hole,][]{Kasen2010, Woosley2010, Dexter2013} and circumstellar interaction \citep{Smith2007, Chevalier2011, Ginzburg2012}  models to produce similar light curves has, however, left interpretation of observed long-duration events ambiguous \citep{Chatzopoulos2013a, Nicholl2013, Moriya2013, McCrum2014}. In fact, either a central engine or circumstellar interaction scenario appears necessary to explain the majority of superluminous SNe, which have too short lightcurves to be fit with PISN or other $^{56}$Ni-powered models \citep{Quimby2011,Chomiuk2011,Inserra2013,Chornock2013,Benetti2014, Nicholl2014, Sorokina2015}. The pre-peak light curve of long-duration PTF12dam also appeared to rise too fast for PISN models \citep{Nicholl2013}.  

Given the ability of different model scenarios (PISN, magnetar, black hole, circumstellar interaction) to produce similar light curves, the unique signature of PISNe must be searched for in the spectra of candidate events, in particular at late times when the inner ejecta become visible and the nucleosynthesis can be analyzed. Recently, there has been considerable theoretical progress in computing late-time model spectra of SN explosion models taking non-thermal processes, Non-Local Thermodynamic Equilibrium (NLTE), and radiative transfer into account, and use of these to diagnose and analyze core-collapse and thermonuclear SNe \citep{Dessart2011, J11, Maurer2011, J12, Dessart2013, J14, J15a, J15b}. Our goal in this paper is to compute nebular-phase spectra of PISN explosion models, 
to compare these with observations of existing PISN candidates,
and to provide model predictions for future candidates to be tested against.

The paper is organized as follows. In Sect. \ref{sec:modelling} we describe our modelling setup. In Sect. \ref{sec:physical} we discuss physical conditions; energy deposition, temperature, ionization, and energy balance. In Sect. \ref{sec:lcands} we present and analyze the model light curves and spectra.
In Sect. \ref{sec:compdata} we compare the model spectra with proposed PISN candidates SN 2007bi and PTF12dam. In Sect. \ref{sec:discussion} we discuss uncertainties and extensions of the modelling, and in Sect. \ref{sec:conclusions} we summarize our findings.

\section{Modelling}
\label{sec:modelling}
We use the code described in \citet{J11} in its latest version \citep{J15a} to model the physical state of the ejecta and to compute model spectra. From this paper on, this code will be referred to as SUMO (SUpernova MOnte Carlo). SUMO computes the temperature and NLTE excitation/ionization solutions in each zone of the SN ejecta, taking a large number of physical processes into account. Specifically, the modelling consists of the following computational steps. (i) Emission, transport, and deposition of radioactive decay products (gamma-rays, X-rays, leptons). (ii) Determination of the distribution of non-thermal electrons created by the radioactivity \citep[by the method of][with atomic data updates]{Kozma1992}. (iii) Thermal equilibrium in each compositional zone. (iv) NLTE ionization balance for the 20 most common elements. (v) NLTE excitation structure for about 50 atoms and ions. (vi) Radiative transfer through the ejecta. The solutions are generally coupled to each other and global convergence is achieved by iteration.



\subsection{Ejecta models}
\label{sec:ejectamodels}
\begin{table*}
\centering
\caption{Properties of ejecta used in the modelling. The characteristic velocity is $V_{char}= \left(\frac{10}{3} \frac{E}{M_{ejecta}}\right)^{1/2}$. The masses of selected elements are listed.}
\begin{tabular}{|c|c|c|c|c|c|c|c|c|}
\hline
Model  & $M_{initial}(\mbox{He burn})$ & $M_{ejecta}$ & $E$ & $V_{char}$ & M(He) & M(O) & M(Si) & M($^{56}$Ni) \\
       & $(M_\odot)$               & $(M_\odot)$  &  $\left(10^{52}~\mbox{erg}\right)$   & (km s$^{-1}$)  & $(M_\odot)$  & $(M_\odot)$  & $(M_\odot)$ & $(M_\odot)$ \\
\hline
He80  & 80  & 79  & 1.6  & 5,800 & 0.78 & 47 & 14 & 0.13\\
He100 & 100 & 99  & 3.8  & 8,000 & 0.88 & 44 & 23 & 5.8\\
He130 & 130 & 129 & 8.1  & 10,200 & 2.1  & 33 & 24 & 40\\
\hline
\end{tabular}
\label{table:ejectaprop}
\end{table*}

As input we use three PISN ejecta from the \citet[][HW02 hereafter]{Heger2002} model grid; Models He80, He100, and He130. Beginning with almost pure helium composition ($Z=2\e{-4}$, mostly carbon which is produced during H burning of a $Z=0$ gas), these stars are evolved and exploded in 1D with KEPLER \citep{Weaver1978, Heger2000}. As SUMO requires the ejecta in the coasting phase, the original models (which were run to 100s) were evolved further up to 12d after explosion, at which point they have reached homology. Some basic properties of these ejecta are listed in Table \ref{table:ejectaprop}. 

As described above, helium cores of 65-130~\msun\ cover the range of possible progenitors for PISN explosions. PISN from $M(He)<$80 \msun~cores produce almost no $^{56}$Ni and are thus dim and unlikely to be observable in the nebular phase. As our lowest mass model we therefore take He80, which makes 0.13 \msun~of $^{56}$Ni, similar to normal core-collapse SNe. This model will not produce a particularly luminous supernova (see \citet{Kasen2011} and Sect. \ref{sec:normal-lum}). 
%

Other PISN models have been presented in the literature \citep[e.g.,][]{Umeda2002, Kasen2011, Chatzopoulos2013, Dessart2013, Yusof2013, Whalen2014, Kozyreva2014a}. 
If PISNe occur in the nearby Universe (redshift $z \ll 1$), they are unlikely to have progenitors with $Z = 0$ as modelled by HW02. In order to make a meaningful comparison of metal-free models with local candidate events, it is important that the metal-free models (with artifical envelope stripping) are not too structurally different from metal-containing models evolved through the main sequence (with envelope stripping by line-driven winds). 
We investige the sensitivity of explosion models to metallicity, rotation, and explosion codes in Sect. \ref{sec:discussion}, finding only minor differences for the same final He core masses. The HW02 models are therefore fully applicable to comparison with local SN events.

\begin{figure}
\includegraphics[width=1\linewidth]{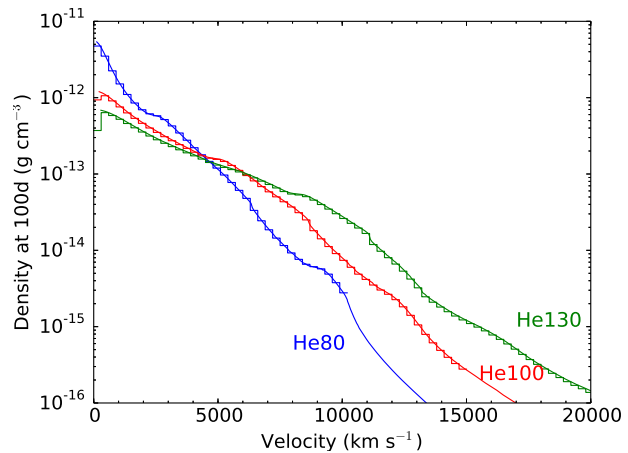} 
\caption{Density profiles of ejecta. The smooth lines are the input models and the discretized lines are the zoned models.}
\label{fig:density}
\end{figure}

\begin{figure}
\includegraphics[width=1\linewidth]{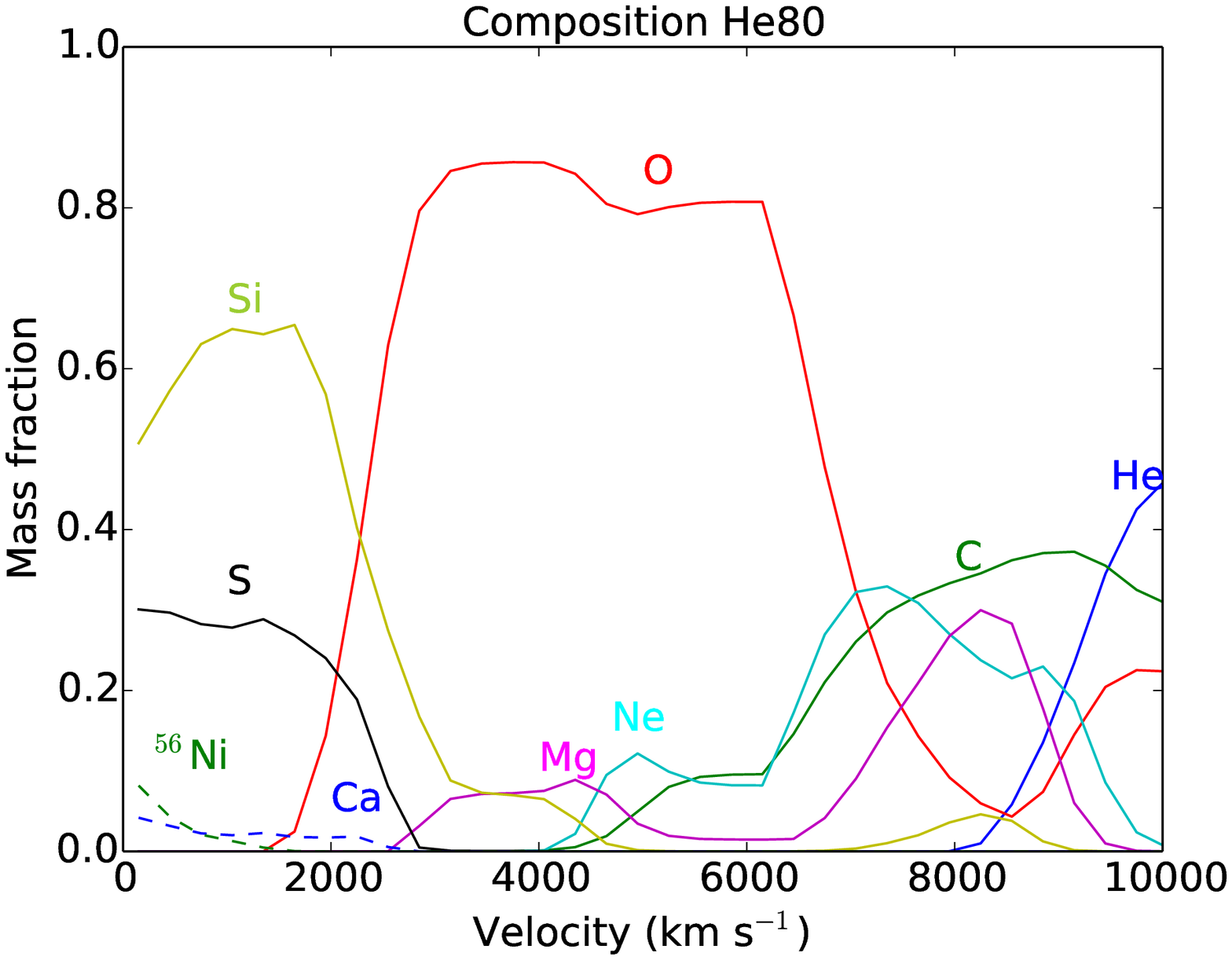} 
\includegraphics[width=1\linewidth]{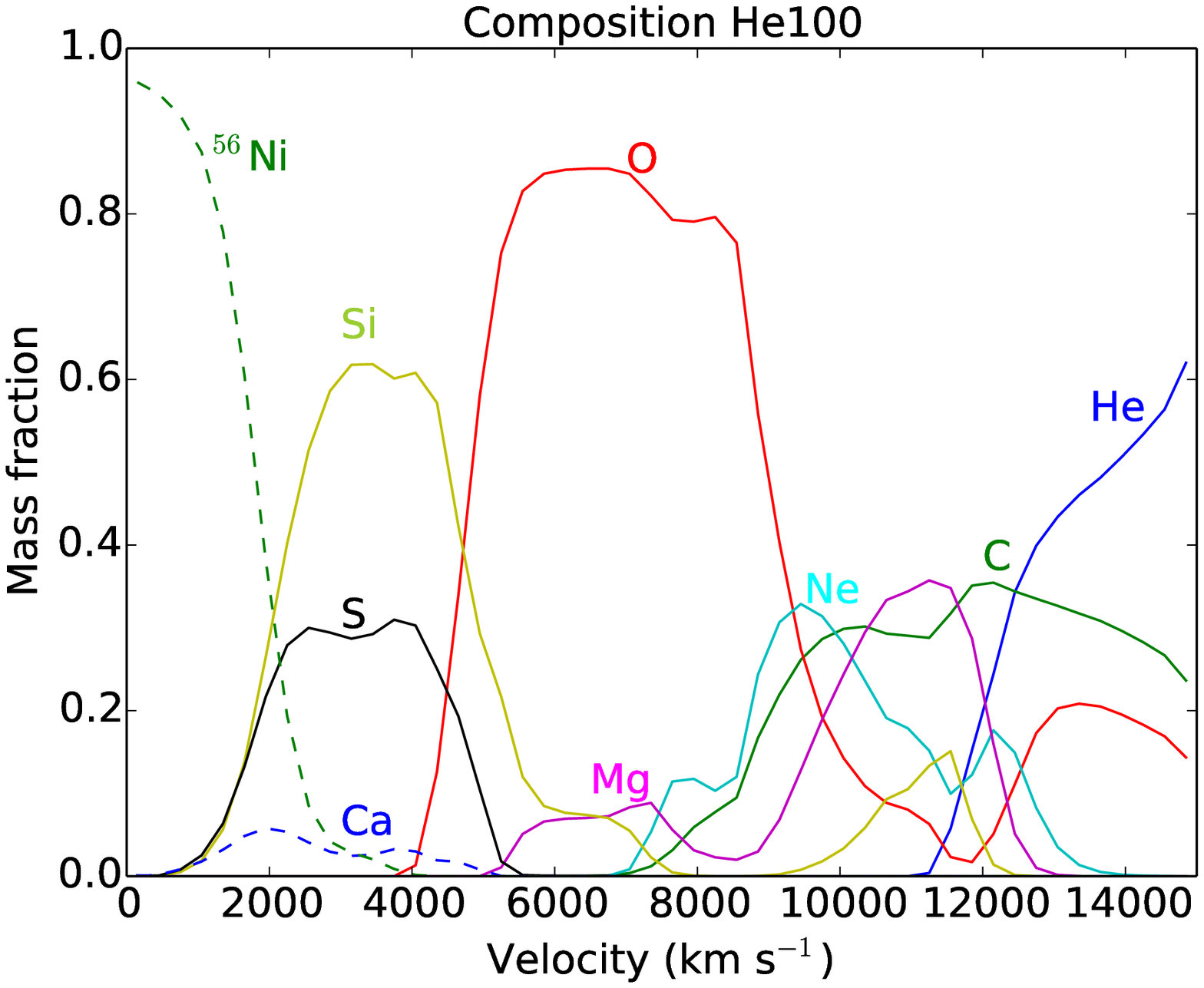} 
\includegraphics[width=1\linewidth]{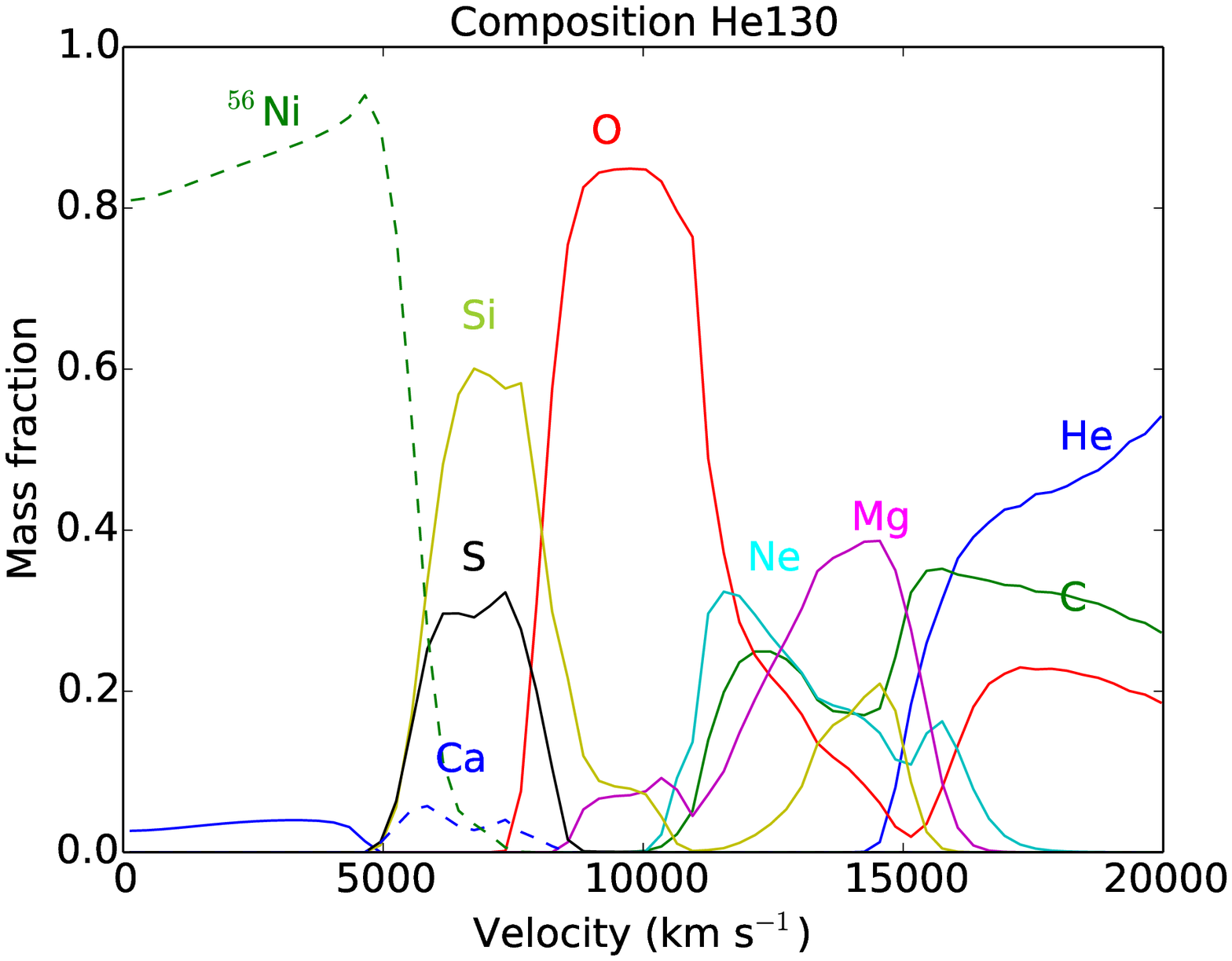} 
\caption{Composition (selected elements) of the three PISN models used for input into SUMO. These show the velocity structure after 12d when the models have reached homologous expansion.}
\label{fig:comp}
\end{figure}

We zone the homologously expanding ejecta into shells of 300 km s$^{-1}$ thickness. We include material out to 10,000 km s$^{-1}$ (Model He80), 15,000 km s$^{-1}$ (Model He100), and 20,000 km s$^{-1}$ (Model He130). These cut-offs include over 99\% of the mass in each input model, and we show in Sect. \ref{sec:lastlineblocking} that effects of higher-velocity material is minor in the nebular phase. Figure \ref{fig:density} shows the density profiles of the models, and Fig. \ref{fig:comp} shows the composition profiles for the most common elements.


\subsubsection{Mixing}
\label{sec:mixing}

Hydrodynamic mixing in PISNe has been investigated by \citet{Joggerst2011}, who simulated the evolution of blue supergiant (BSG) and red supergiant (RSG) explosions with a 2D hydrodynamic code. The BSG progenitors showed neglegible mixing. The RSG progenitors showed neglegible or weak mixing in some cases, but significant mixing in other cases where pulsational phases in the pre-SN evolution had given regions of steeply increasing $\rho r^3$ in the outer parts of the H/He envelope. The mixing was still, however, limited to the O/He/H layers (in one case also involving the Si layer). In no case was any significant mixing of $^{56}$Ni obtained. \citet{Chatzopoulos2013} obtain similar results of mild mixing for both non-rotating and rotating cores. Further work in this area was done by \citet{Chen2014}, who extended the simulations to include the early oxygen-burning phase. While this energy generation did trigger some further instabilites, it was not strong enough to lead to any significant enhancements of the mixing. Finally, \citet{Whalen2014} confirmed that PISNe from compact progenitors experience neglegible mixing by simulating the explosion of He cores in 2D. 

The mixing is not likely to alter significantly in 3D, as an already moderately higher Rayleigh-Taylor growth rate is counteracted by a rate reduction due to finger interaction \citep{Joggerst2010,Joggerst2011}. It therefore appears to be a robust result that 1D explosion models provide an accurate approximation for the structures of PISN ejecta, at least for the explosions of Wolf-Rayet and BSG progenitors. We therefore limit ourselves here to modelling 1D explosion models without any artificial mixing applied.

\section{Physical conditions}
\label{sec:physical}
For each ejecta we model, we compute spectra at 400d, 700d, and 1000d post-explosion.
Before we examine spectra and light curves, we consider here some of the physics of the spectral formation process.

\subsection{Gamma-ray deposition}
\label{sec:gammadep}
\begin{figure}
\includegraphics[width=1\linewidth]{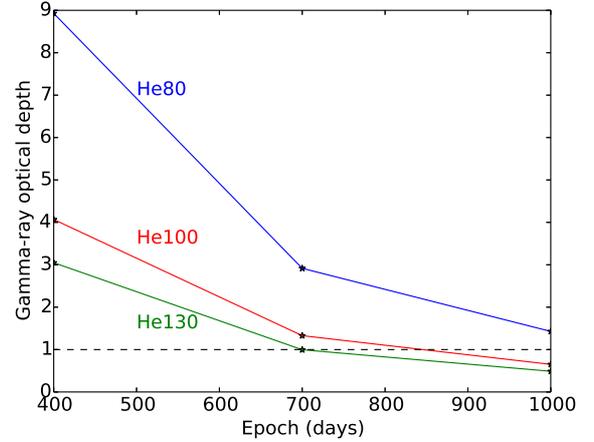} 
\caption{Gamma-ray optical depth as function of time.}
\label{fig:gammatau}
\end{figure}

The gamma-ray optical depth can be estimated by the uniform sphere formula
\begin{equation}
\tau_\gamma = 8.6\e{-3} \left(\frac{M}{1~M_\odot}\right)^2 \left(\frac{E}{10^{51}~\mbox{erg}}\right)^{-1} \left(\frac{t}{1~\mbox{year}}\right)^{-2}
\end{equation}
where we have used an effective gray opacity of $\kappa_\gamma = 0.03$ cm$^2$~g$^{-1}$ \citep{Colgate1980}. Type Ia SNe ($M^2/E \sim 1$) become optically thin at $\sim$1 month, Type II SNe ($M^2/E \sim 100$) at $\sim$1 year, and PISNe ($M^2/E \sim 300$) at $\sim$2 years. Figure \ref{fig:gammatau} shows the optical depth evolution in the three models illustrating this behaviour. Note that more massive PISNe have lower $M^2/E$ values and therefore lower gamma-ray optical depth at any given epoch.

\begin{figure}
\includegraphics[width=1\linewidth]{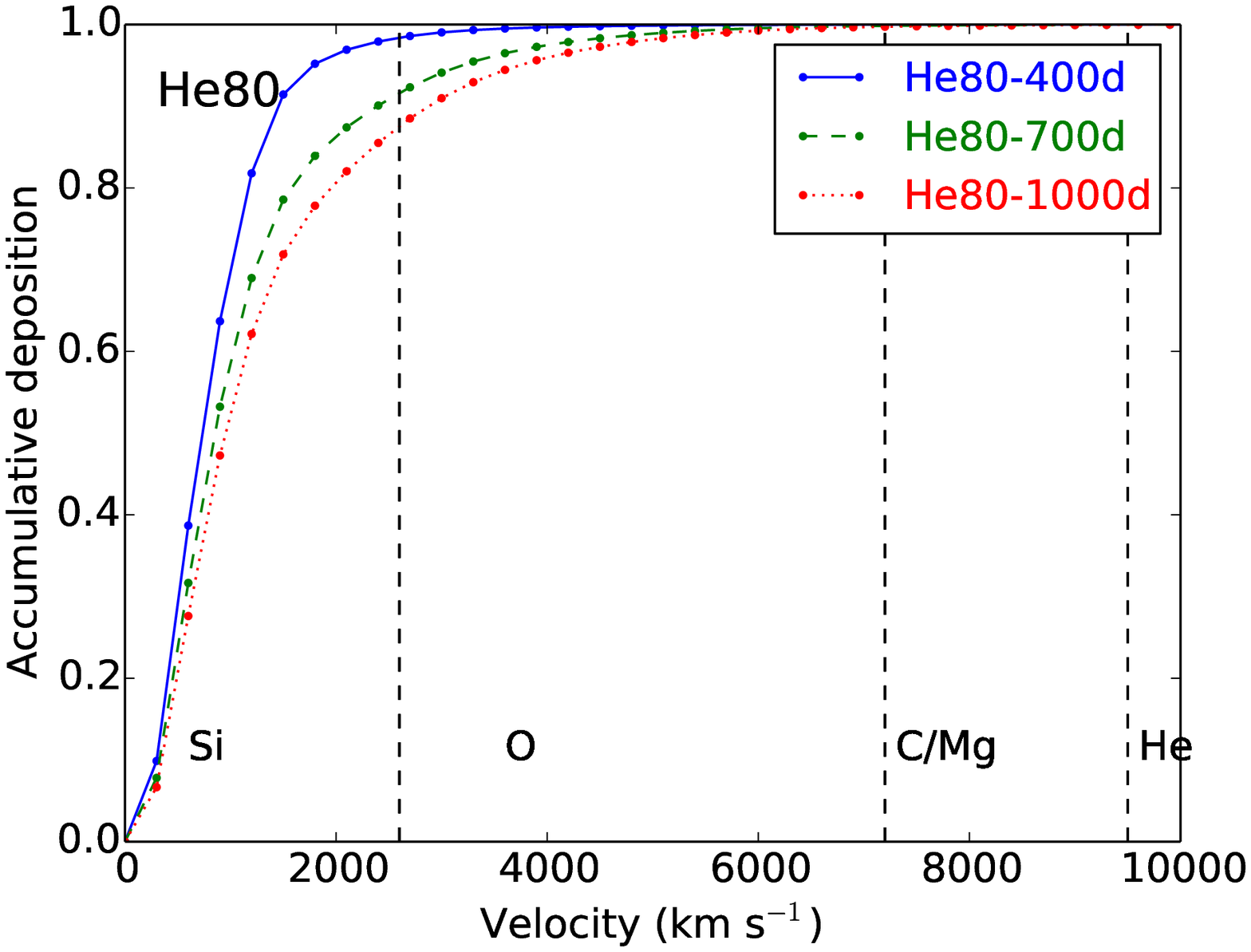} 
\includegraphics[width=1\linewidth]{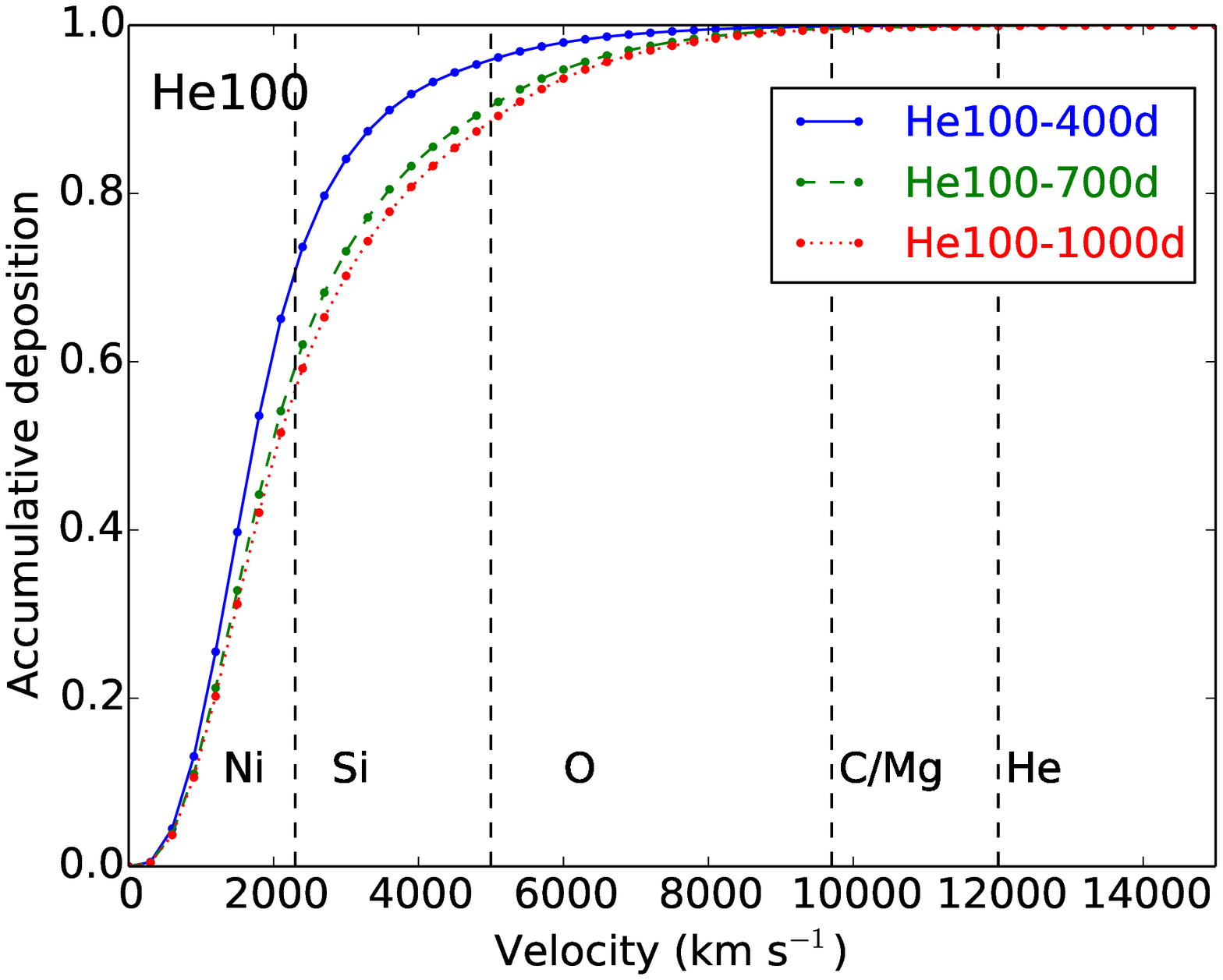} 
\includegraphics[width=1\linewidth]{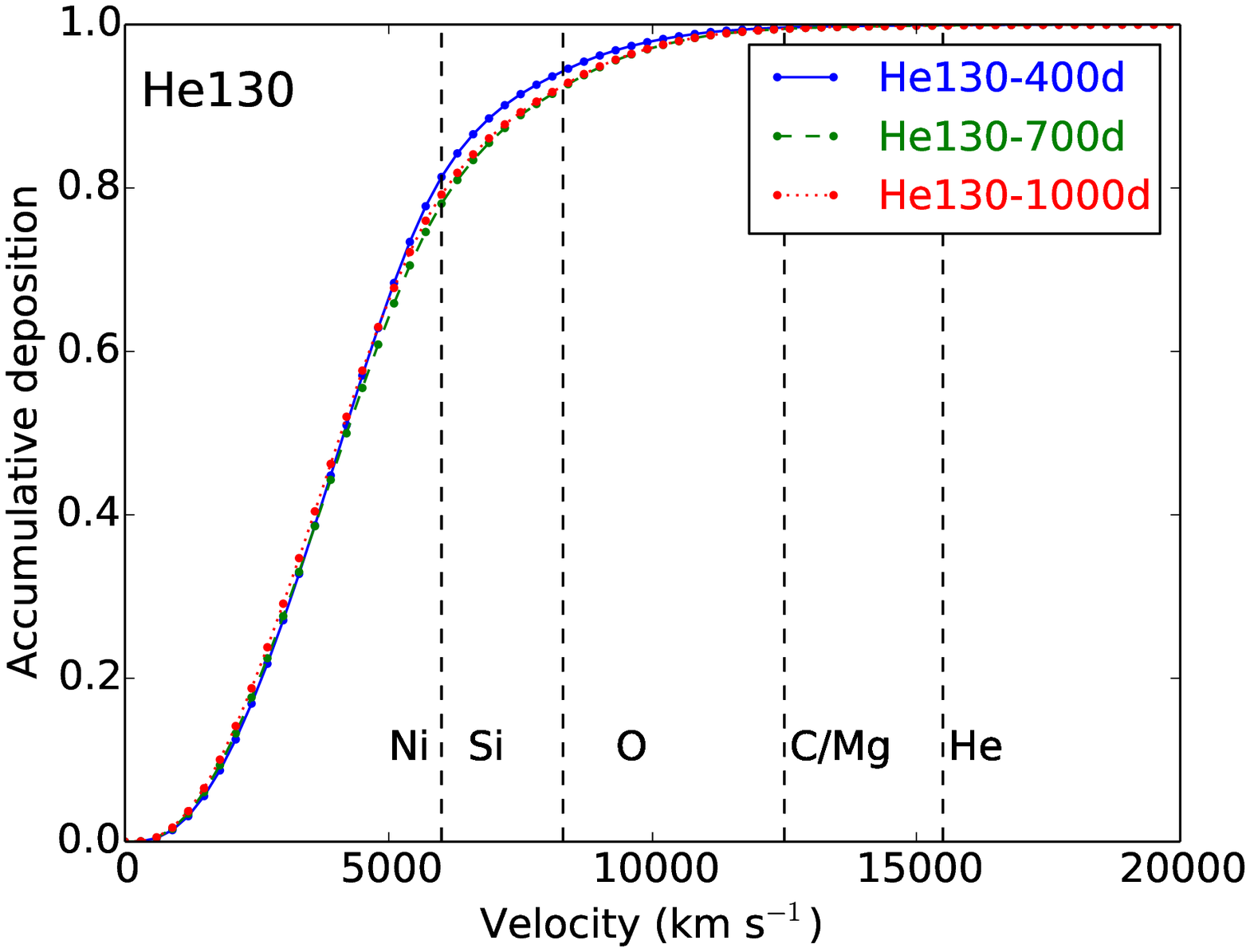} 
\caption{Accumulative energy deposition from radioactive decay products. The zones are labelled with their most abundant element.}
\label{fig:gammaaccum}
\end{figure}

The high column densities and weak mixing of PISNe mean that the gamma-rays are quite efficiently trapped in the $^{56}$Ni regions of the ejecta. Some transport will occur up to the overlying silicon/sulphur layers, but the oxygen/magnesium/carbon layers will not be exposed to much gamma-ray irradiation. Figure \ref{fig:gammaaccum} demonstrates this by showing the accumulated deposition as function of velocity coordinate in the three models. Considering He100 as an example, at 400 days only 5\% of the energy is deposited outside 5000 km s$^{-1}$ (where oxygen becomes abundant), rising to 15\% at 1000 days. The fractions deposited outside the $^{56}$Ni and Si/S regions are similar for He80 and He130. From this consideration only, nebular-phase PISN spectra can be expected to display emission lines from deep-lying heavy elements (calcium, silicon, sulphur, iron, nickel), but with weaker contribution from lighter elements at higher velocity (helium, carbon, oxygen, magnesium).

\subsection{Temperature and ionization}
\begin{figure}
\includegraphics[width=1\linewidth]{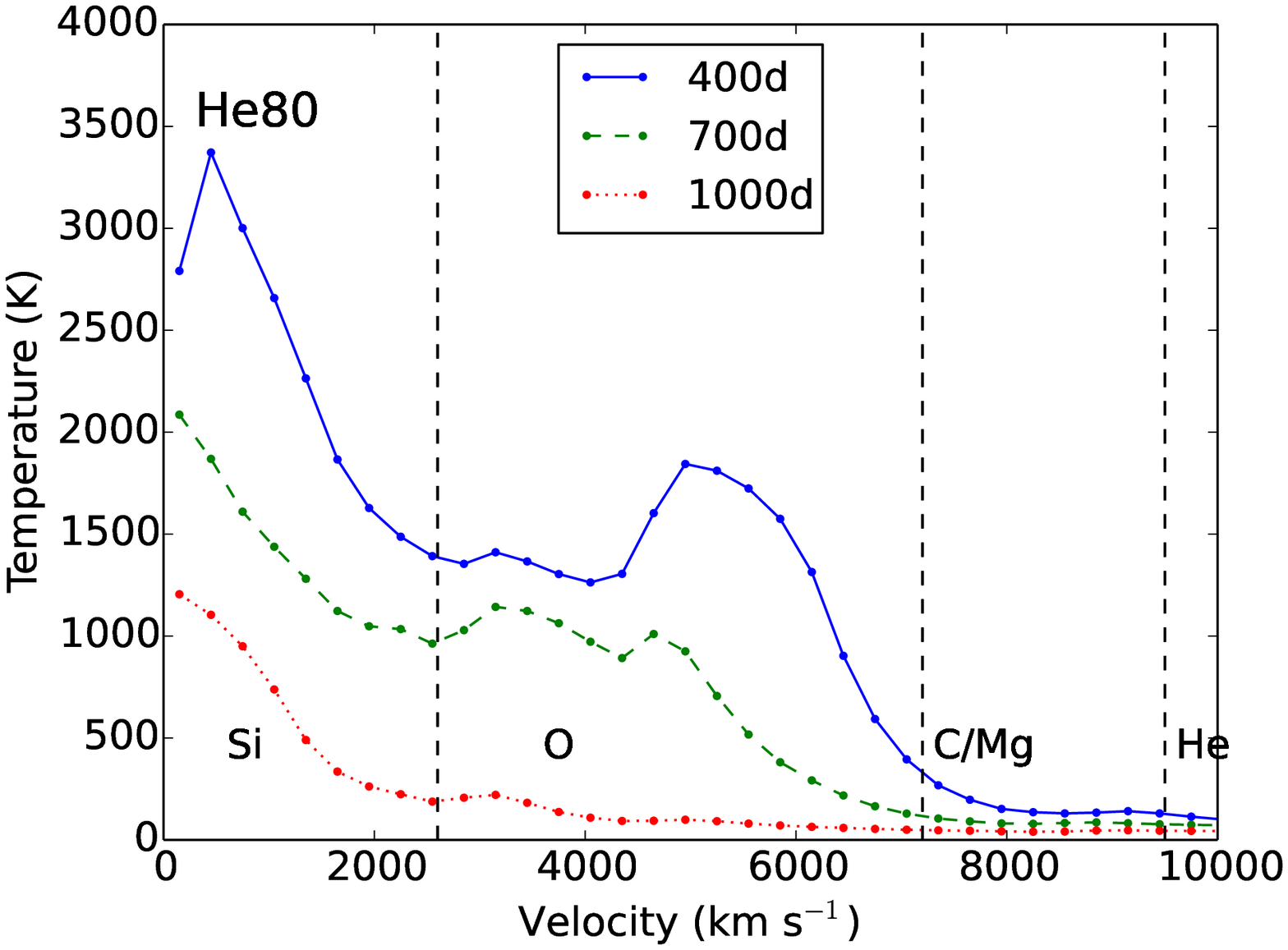} 
\includegraphics[width=1\linewidth]{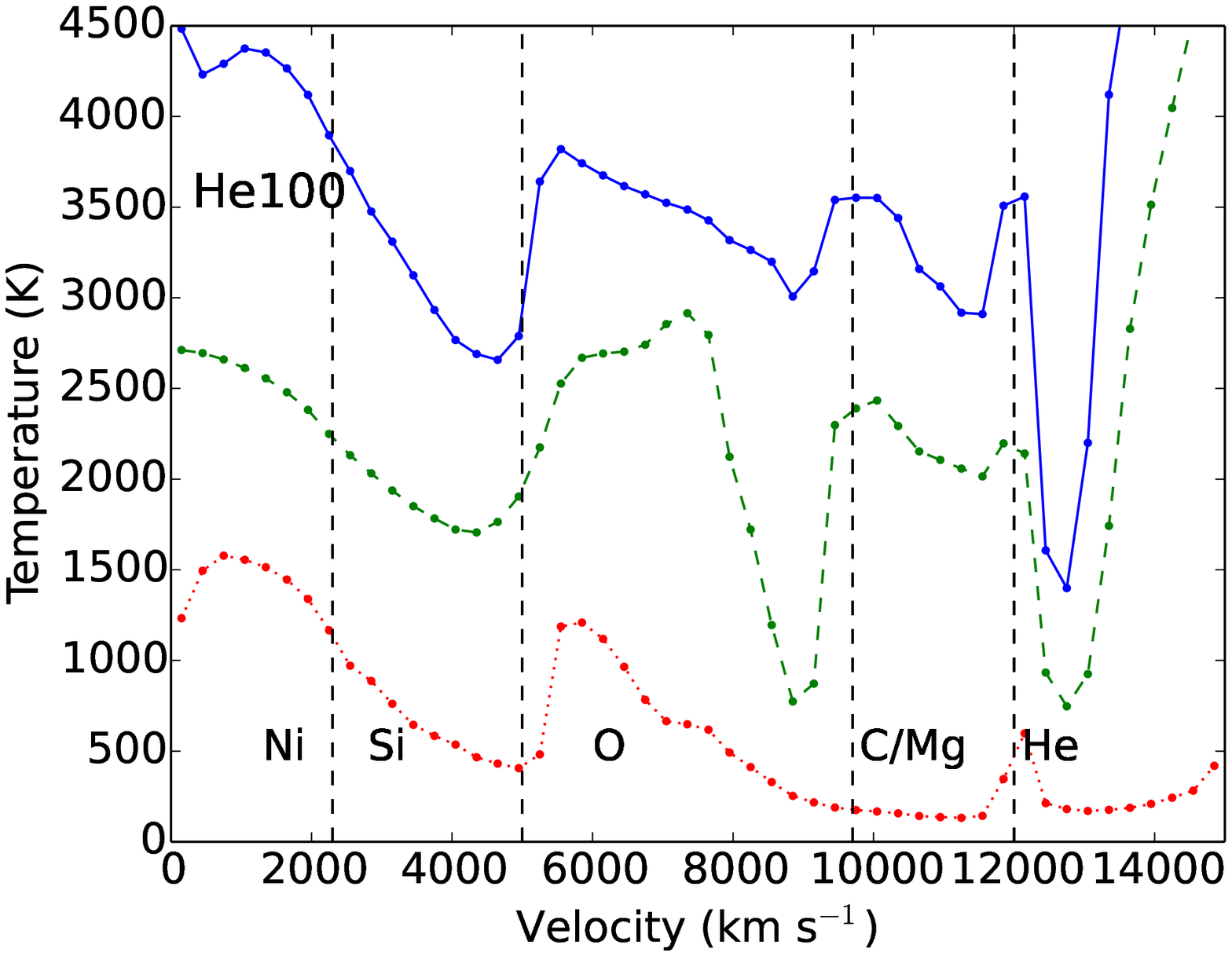} 
\includegraphics[width=1\linewidth]{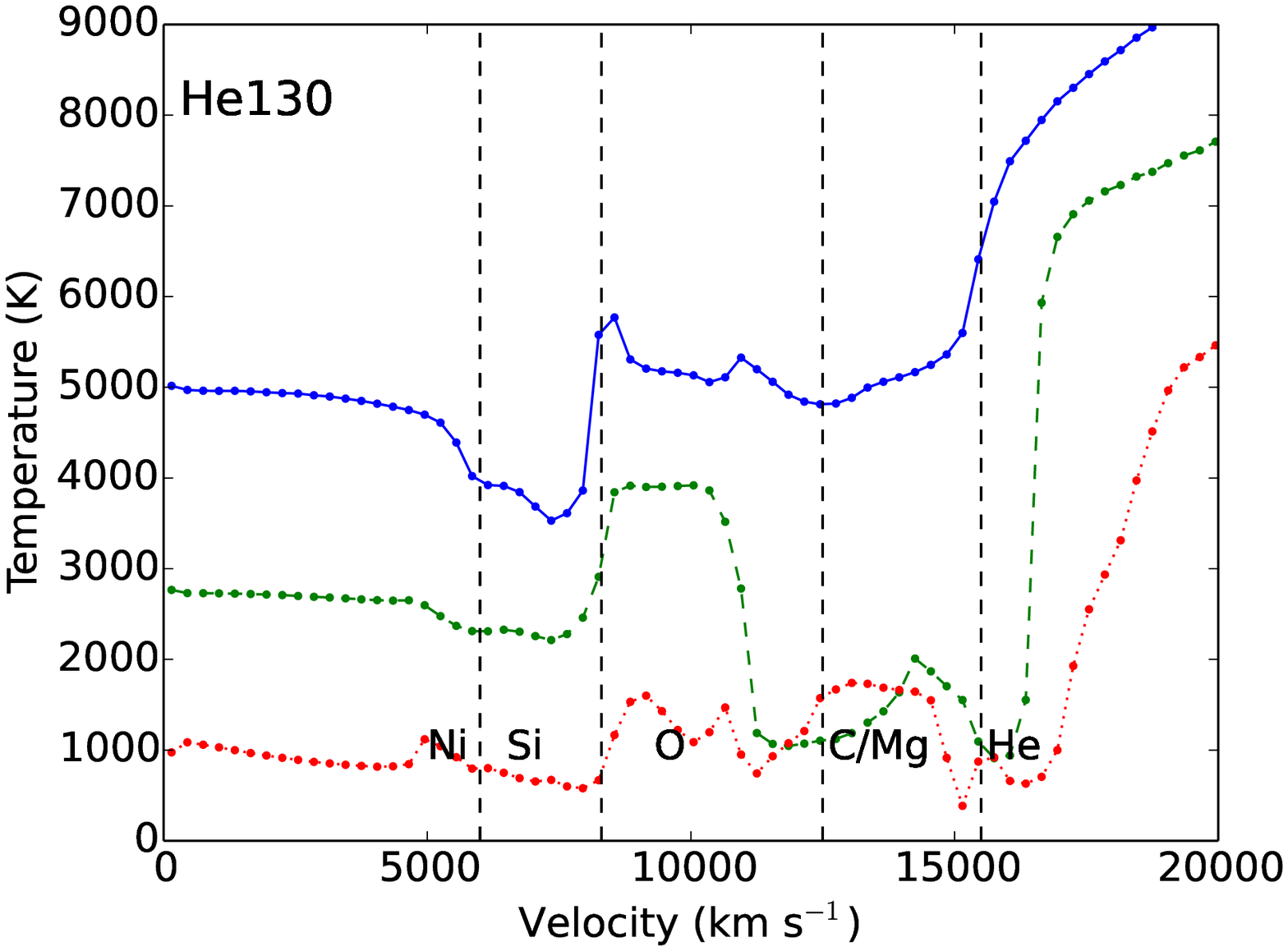} 
\caption{Temperature in the models.}
\label{fig:temps}
\end{figure}

\begin{figure}
\includegraphics[width=1\linewidth]{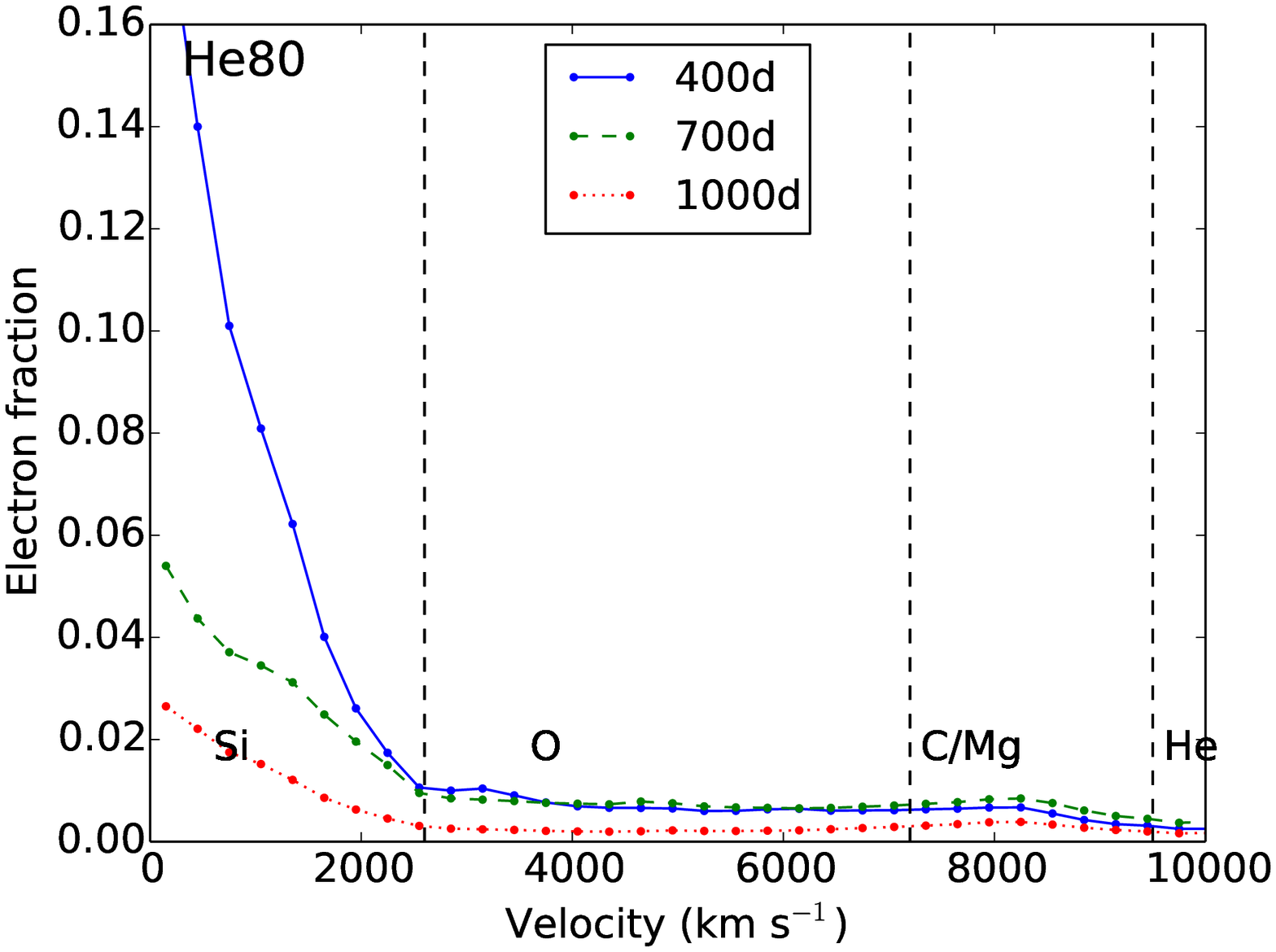} 
\includegraphics[width=1\linewidth]{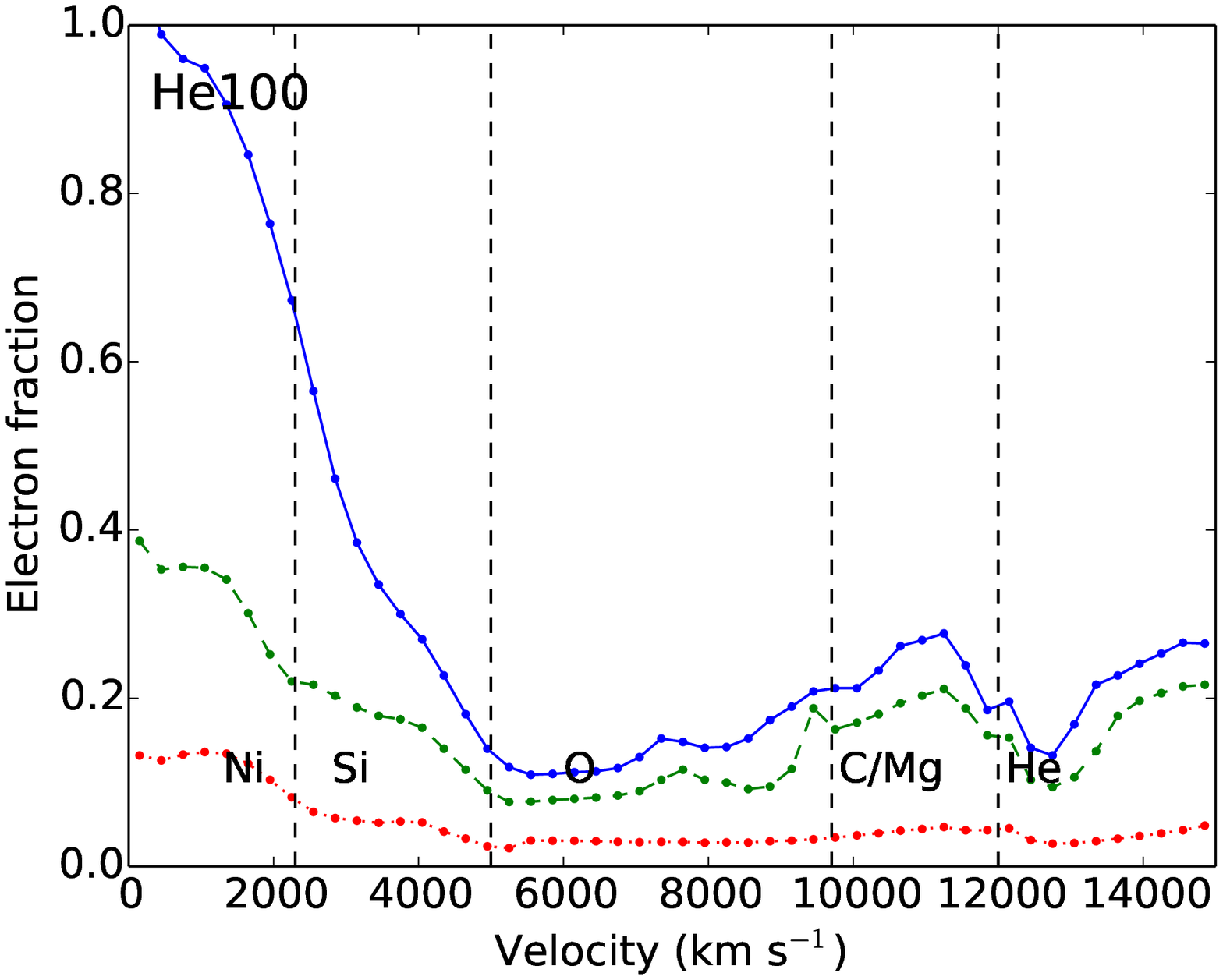} 
\includegraphics[width=1\linewidth]{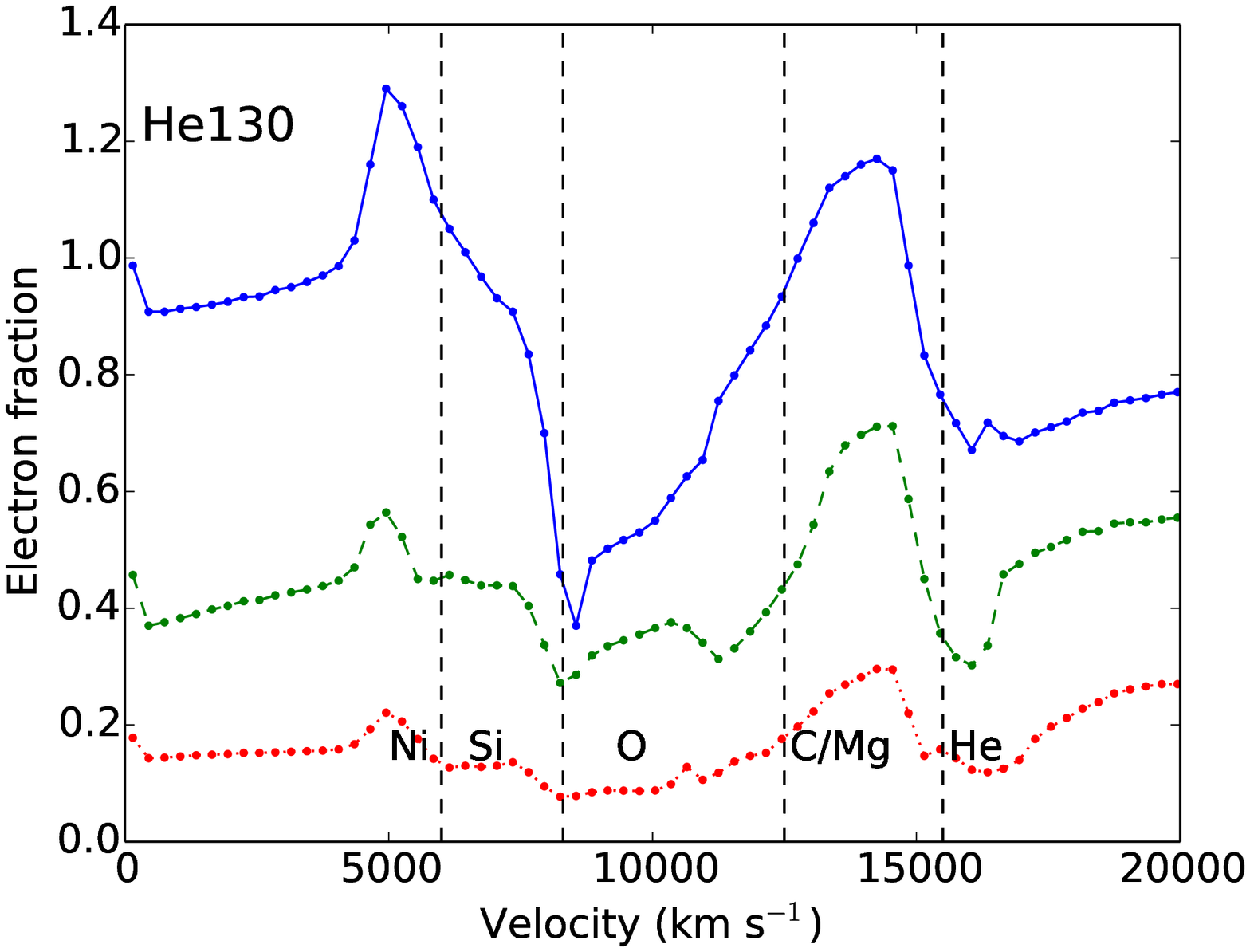} 
\caption{Electron fraction in the models.}
\label{fig:xe}
\end{figure}

Figures \ref{fig:temps} and \ref{fig:xe} show the temperature and ionization in the models. The temperature is fairly uniform throughout the ejecta, with higher gamma-ray deposition in the deeper layers being compensated by more efficient cooling by iron-group lines. At 400 days model He100 has an ejecta temperature of 3000-4000 K, falling to 200-1500 K at 1000 days. Some of the cooling at these quite low temperatures occur in the near-infrared rather than in the optical, especially approaching 1000d.
The ionization is also quite low, with the electron fraction (number of electrons divided by the number of atoms and ions) mostly below unity. The ejecta are thus composed of mainly neutral species, with 1-50\% singly ionized. Combined with the deposition occuring mainly in iron and silicon/sulphur gas, lines of Fe I, Si I, and S I can be predicted to be prominent.



\subsection{Line opacity}
\label{sec:opacity}

\begin{figure}
\includegraphics[width=1\linewidth]{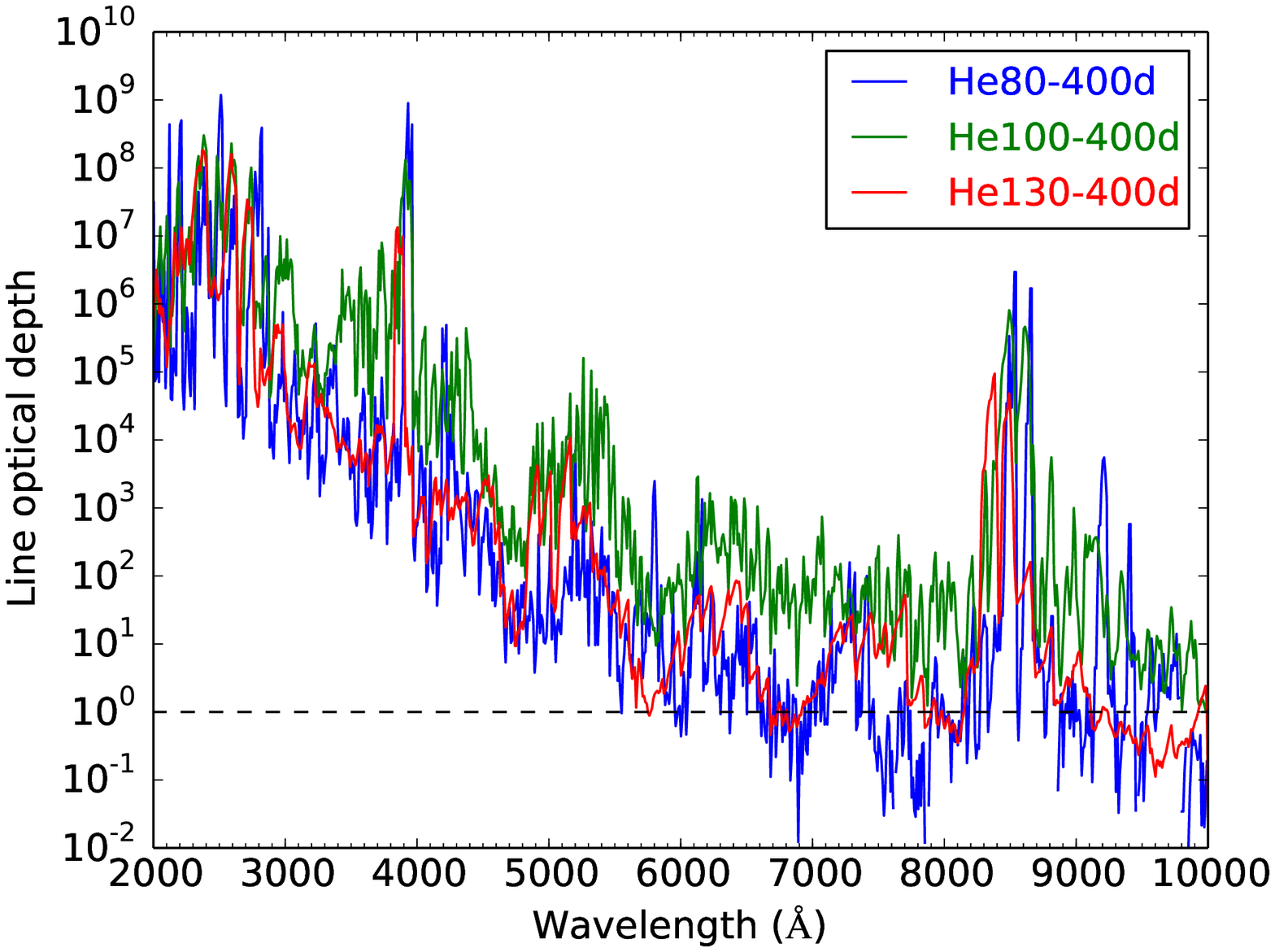} 
\includegraphics[width=1\linewidth]{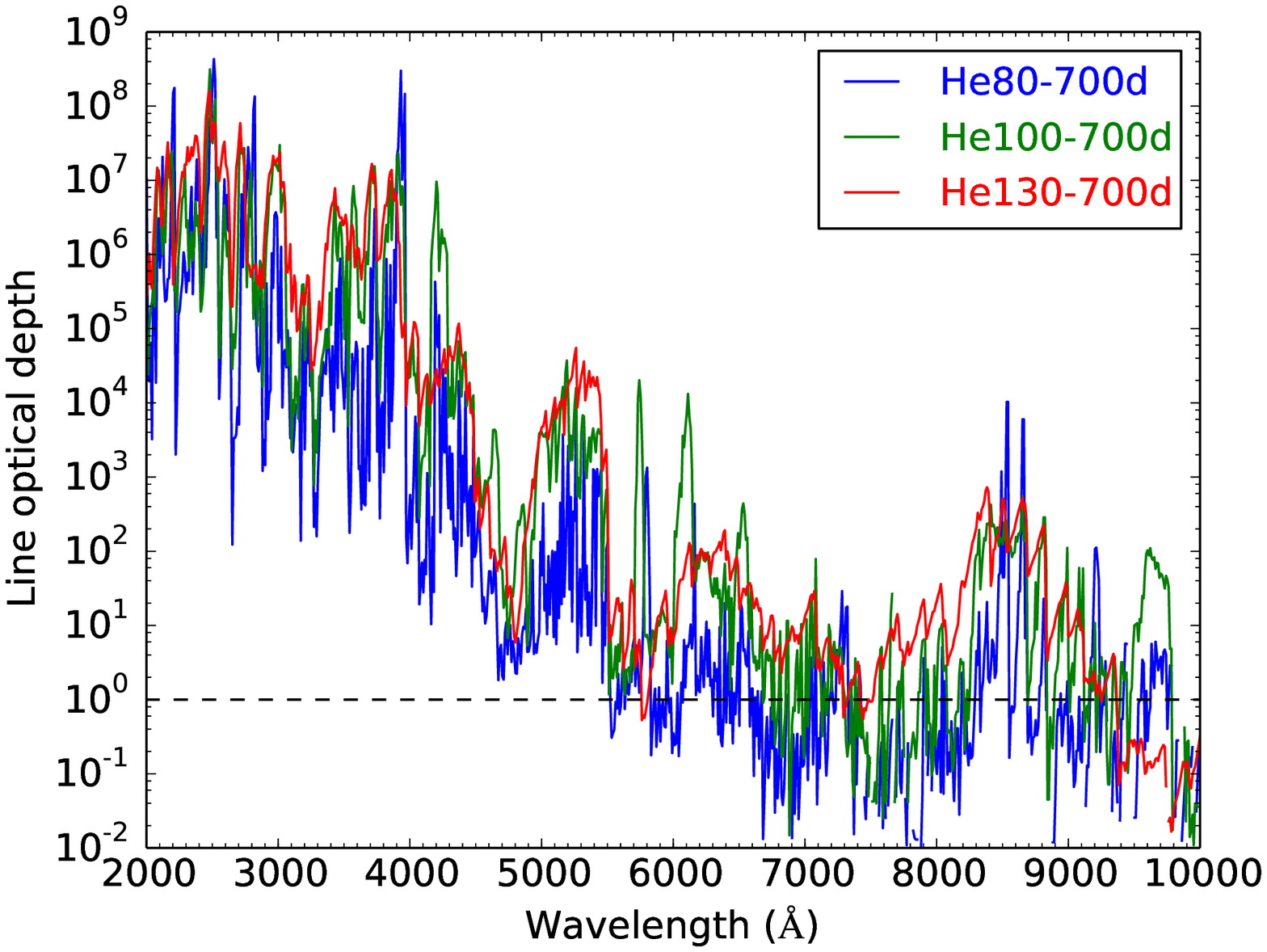} 
\includegraphics[width=1\linewidth]{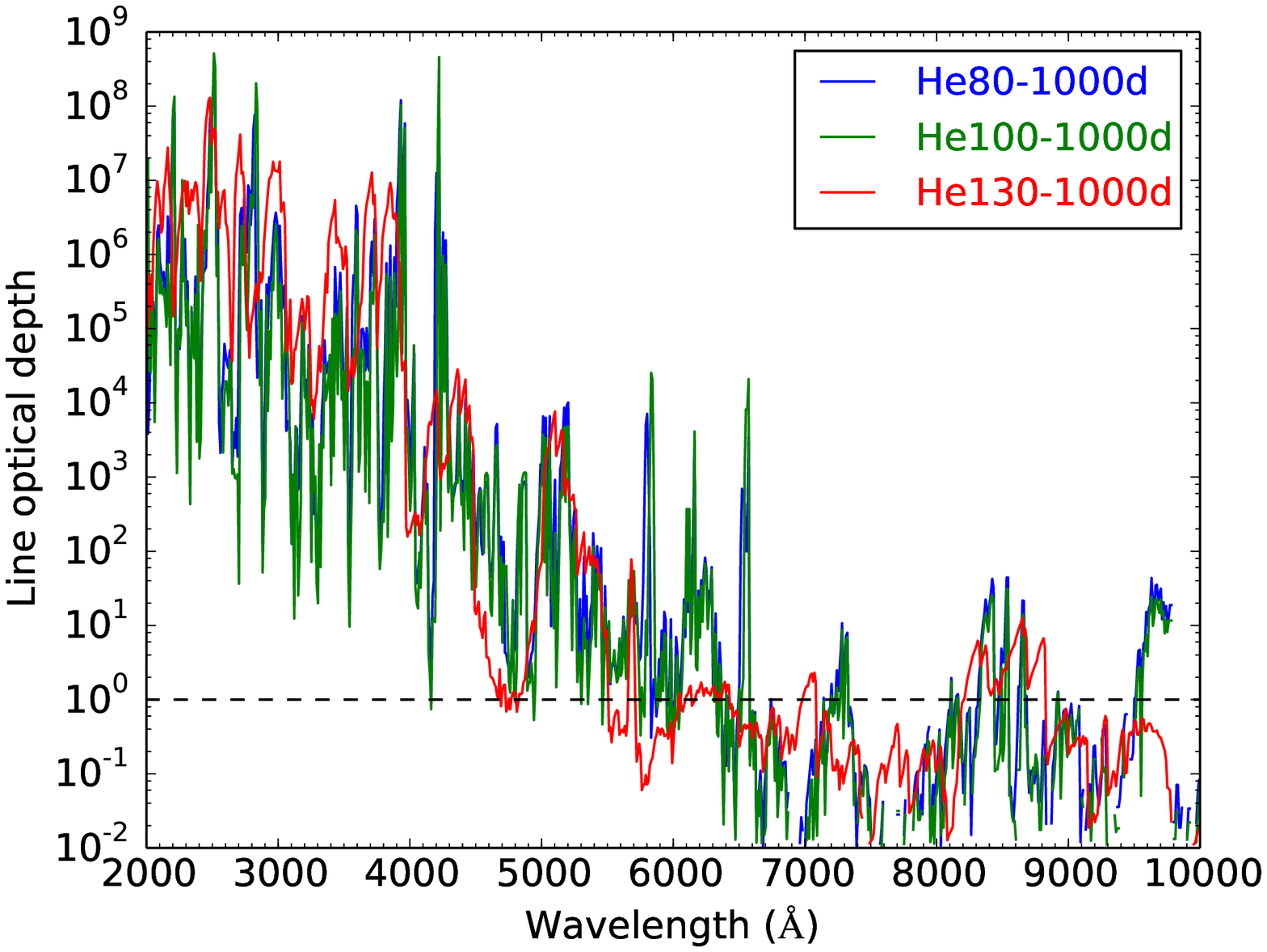} 
\caption{Line optical depth. The values are the sum of (comoving) line optical depths a photon sees as it travels from the centre to the edge of the ejecta.}
\label{fig:pesc}
\end{figure}

Figure \ref{fig:pesc} shows the line optical depth as function of wavelength. At 400 days the ejecta are still optically thick throughout the optical, and essentially no photons can traverse the ejecta without having at least one line interaction. The optical depths decrease over time, roughly as $t^{-2}$; at 1000 days the wavelength region up to 6000 \AA~is still fully blocked, whereas redward of this there is partial blocking. The implication is that radiative transport maintains a crucial influence for the spectral formation even several years post-explosion, as previously has been demonstrated in other contexts \citep{J11,J15a}, and that PISNe will be dim in the UV/blue also at late times.

\subsubsection{Line blocking outside last zone}
\label{sec:lastlineblocking}
\begin{figure}
\includegraphics[width=1\linewidth]{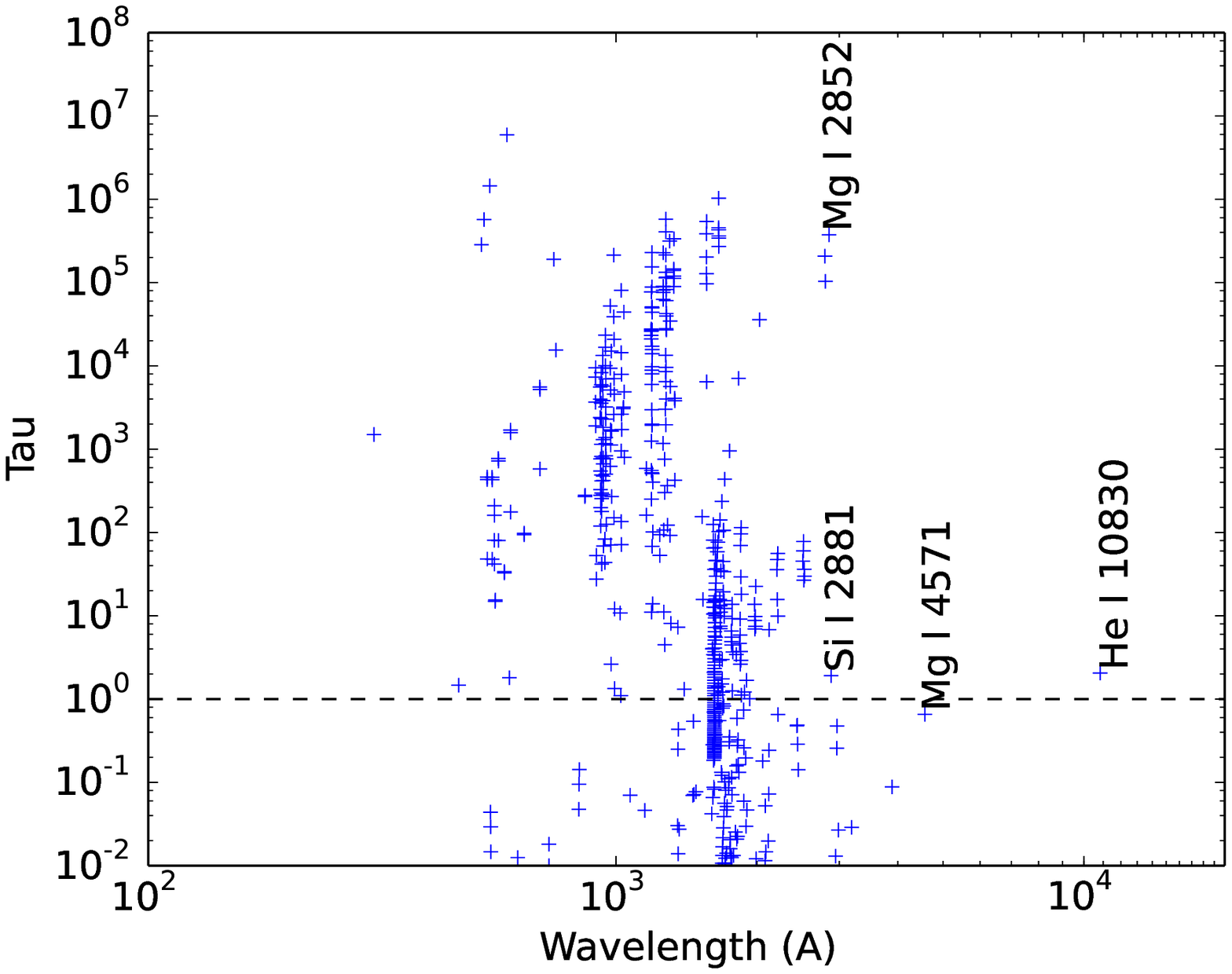} 
\caption{Line optical depths in the last zone in Model He80, at 400d.}
\label{fig:last}
\end{figure}

Because we truncate the ejecta at maximum velocities ($1.0\times 10^4, 1.5\times 10^4, \mbox{and}~2.0\times 10^4$ km s$^{-1}$ in He80, He100, and He130, respectively), it is important to consider what the omission of the highest-velocity material entails. Figure \ref{fig:last} shows the line optical depths in the outermost zone in model He80 at 400d. It is safe to assume that the vast majority of lines will show decreasing (Sobolev) line depths with velocity coordinate (because the density is rapidly decreasing), so these can be considered to be upper limits to the line optical depths encountered outside the model edge. The figure shows that there is still a large number of optically thick lines below 3000 \AA. This means that our predicted flux below 3000 \AA~may be somewhat overestimated; some fraction of this flux will scatter and fluoresce in material outside our model edge. Importantly, however, this does not introduce any significant uncertainty for the $>3000$ \AA~flux, since this is already much stronger than the $<3000$ \AA~flux. Radiative transfer of the $<3000$ \AA~flux could also only lead to addition of very broad scattering/fluorescence lines from the high-velocity region, unable to change the characteristics of the spectrum. Thus, the effect of ignoring the highest-velocity material is to introduce a possible overestimate of the already weak UV flux. We made some tests of this conclusion by comparing models with cut-offs at $1.0\times 10^4$ and $1.5\times 10^4$ km s$^{-1}$ for He100, finding very minor differences throughout the spectral range.

It is of interest to note that throughout the optical/near-infrared, only one line, He I $\lambda$10830, remains optically thick in the outermost ejecta layers. Detailed modelling of this line would therefore require inclusion of higher velocity material.

\subsection{Steady-state verification}
The line formation model assumes steady state for level populations, temperature, and radiation field, i.e. that the time-derivative term in the corresponding equations can be ignored. Before proceeding we pause to assess the validity of these assumptions for the PISN ejecta modelled here. The first criterion requires $\tau_{\rm rec}/\tau_{\rm 56Co} \ll 1$, where $\tau_{\rm rec}$ is the recombination time scale and $\tau_{\rm 56Co}$ is the $^{56}$Co decay time-scale of 111 days. The second requires $\tau_{\rm cool}/\tau_{\rm 56Co} \ll 1$ where $\tau_{\rm cool}$ is the cooling time-scale. The third requires $\tau_{\rm rad}/\tau_{\rm 56Co} \ll 1$, where $\tau_{\rm rad}$ is the radiative transfer time-scale. See \citet{Jerkstrand2011PhD} for more in-depth discussion about these requirement.

\begin{figure}
\includegraphics[width=1\linewidth]{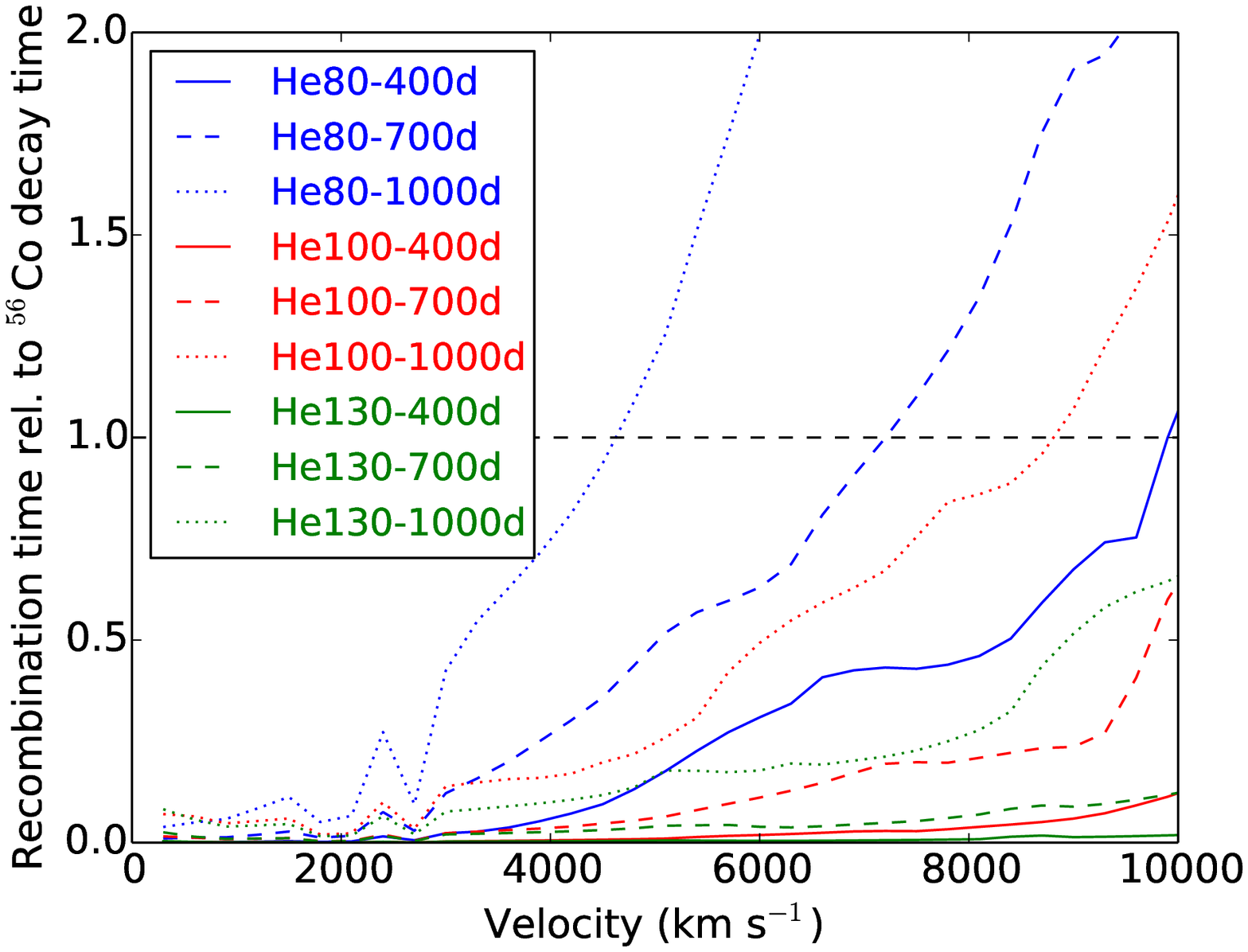} 
\includegraphics[width=1\linewidth]{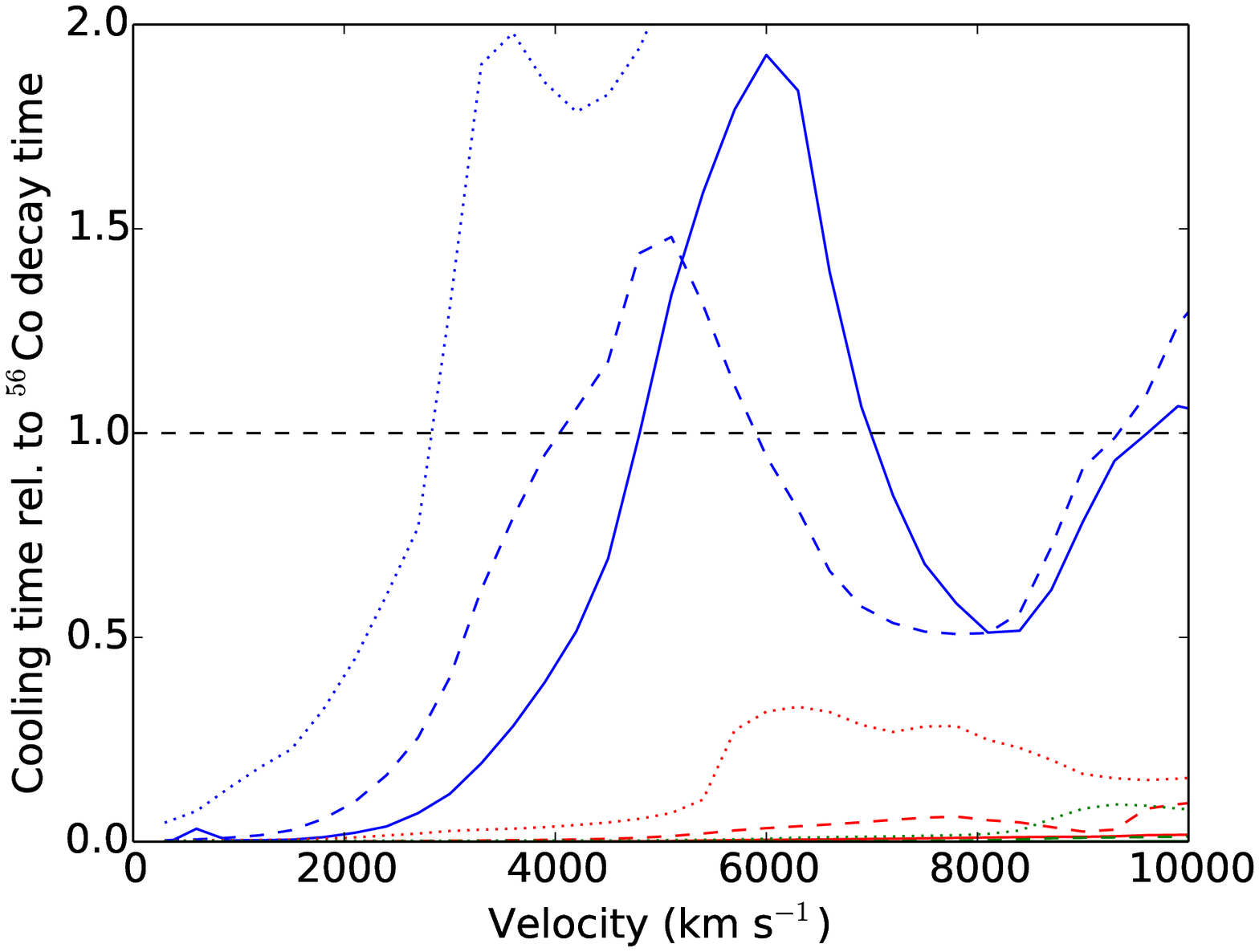} 
\caption{Recombination timescale (top) and cooling time-scale (bottom) relative to the radioactive decay time scale.}
\label{fig:rectime}
\end{figure}

Figure \ref{fig:rectime} shows the relative recombination time scale $\tau_{\rm rec}/\tau_{\rm 56Co}$. The important velocity range to examine is that over which the bulk of the gamma-ray/positron deposition occurs; as Fig. \ref{fig:gammaaccum} shows, for all models and epochs over $95\%$ of the deposition occurs in layers below $10^4$ km s$^{-1}$. For models He100 and He130, steady-state conditions are fulfilled at all times, although marginally for He100 at 1000d. For He80, steady-state prevails at 400d, but the outer layers freeze out at 700 and 1000d. As energy deposition and spectral formation occurs mainly below the freeze-out velocities of $>5000$ km s$^{-1}$ (Fig. \ref{fig:gammaaccum}), this probably has limited impact, but some caution is still needed for the He80-700d and He80-1000d models.

Figure \ref{fig:rectime} (bottom) shows the relative cooling time scale $\tau_{\rm cool}/\tau_{\rm 56Co}$. Steady-state is fulfilled for He100 and He130 at all times. For He80, the conditions are marginally satisfied at 400 and 700d, and there is significant breakdown at 1000d.

\begin{figure}
\includegraphics[width=1\linewidth]{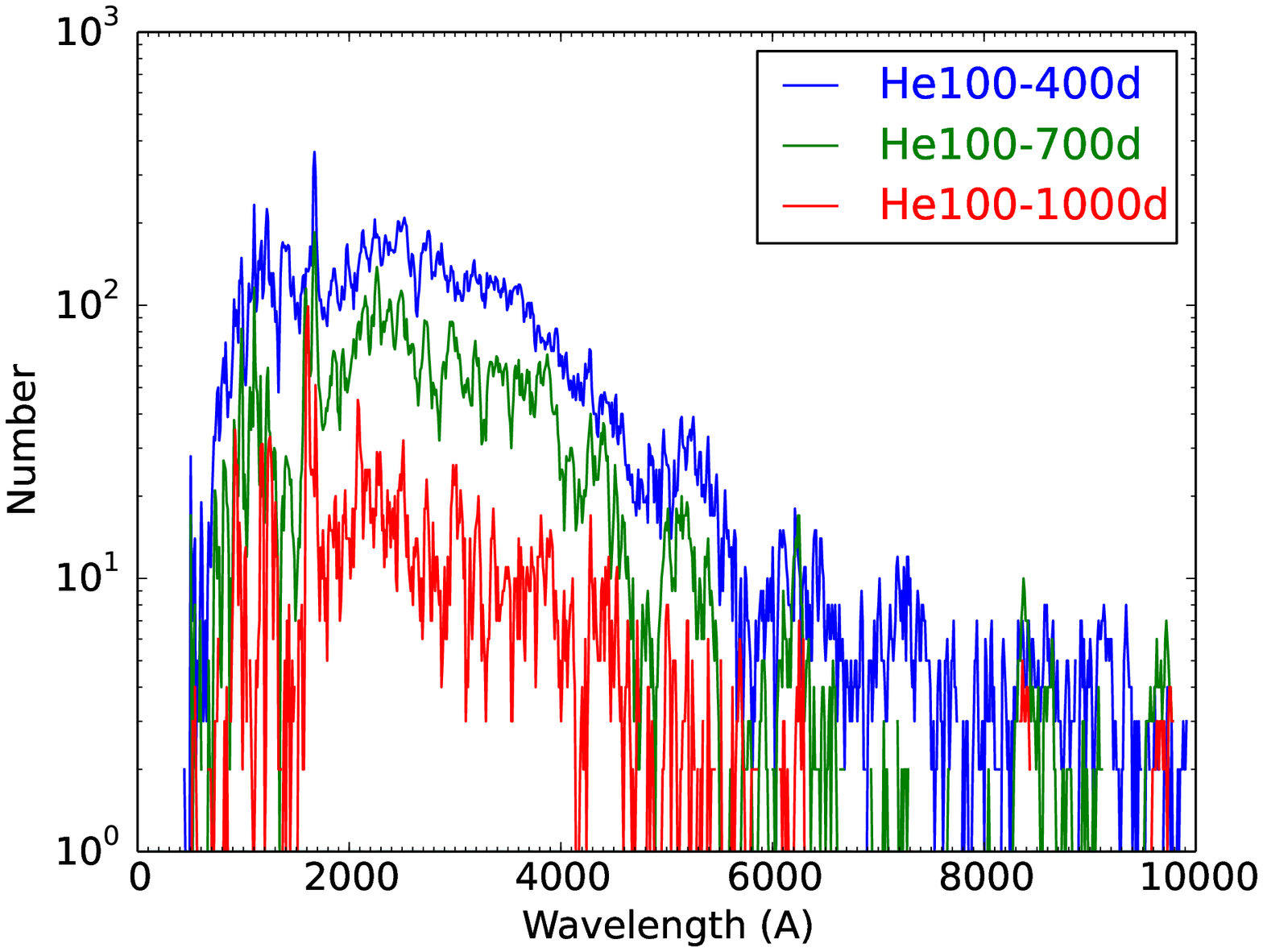} 
\caption{Number of optically thick lines at different epochs, Model He100. The value is the number of optically thick lines that a photon emitted at a given wavelength encounters between the centre of the ejecta and the edge.}
\label{fig:numberHe100}
\end{figure}
\begin{figure}
\includegraphics[width=1\linewidth]{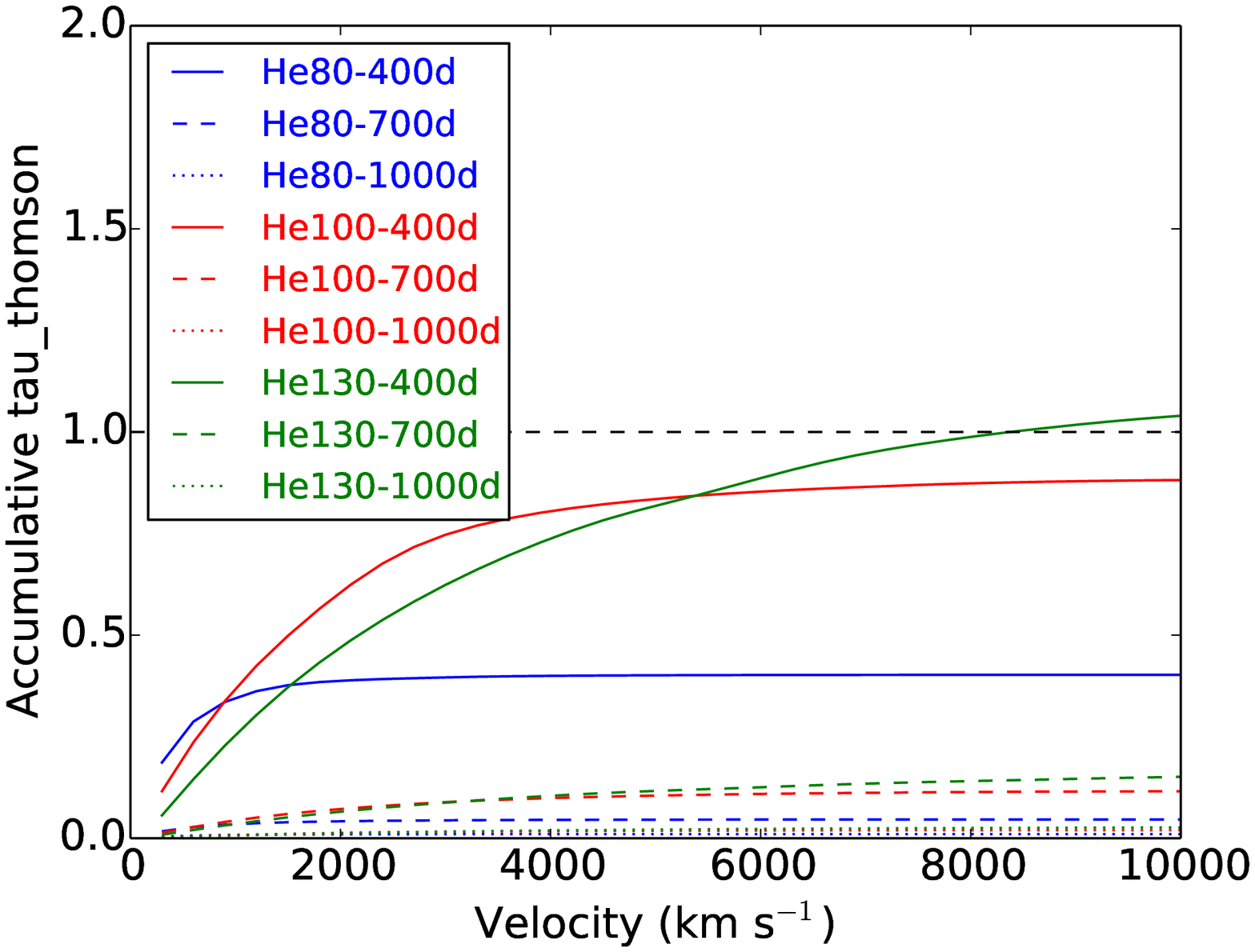} 
\caption{Accumulated electron scattering optical depth.}
\label{fig:thomson}
\end{figure}

The free-streaming radiation transport timescale is simply the time to traverse the ejecta, $\tau_{\rm rad,free} = V_{\rm ej}t/c$. Because $V_{\rm ej}/c \sim 0.02$ for PISN ejecta, $\tau_{\rm rad,free}/\tau_{\rm 56Co} \sim 0.02 t/\tau_{\rm 56Co}$, which is much smaller than unity up to 1000d. 
Scattering will lengthen the transport time scale by some factor. If there are $N_\lambda$ optically thick lines, of order $N_\lambda^2$ scatterings will occur assuming no fluorescense. Each scattering involves a spatial transport of $V_{ej}t/N_\lambda$, so the time scale is $\tau_\lambda = N_\lambda \left(V_{ej}/c\right) t$. Figure \ref{fig:numberHe100} shows the number of optically thick lines as function of wavelength for He100. In the UV, $N_\lambda \sim 10^2$ at 400d, and pure scattering thus gives a transport time-scale of $\tau_{\rm rad,UV}\sim 800$d (or $\tau_{\rm rad,UV}/\tau_{\rm 56Co} = 5-10$). At 1000 days the number of optically thick lines has decreased to $N_{\rm UV} \sim 10$, and the time-scale is $\tau_{\rm rad,UV} \sim 200$d (or $\tau_{\rm rad,UV}/\tau_{\rm 56Co} \sim 2$). If no fluorescence or collisional deexcitation processes were to occur, the UV transport would thus not strictly fulfill steady-state as it would take several hundred days for UV photons to get out of the ejecta. Fluorescence occurs, however, to significant probability following each absorption, and the actual transport time-scale will be shorter than this limiting value. It goes beyond the scope of this paper to analyze what the fluorescence efficiencies and exact transport times-scales are, but it is likely that a factor several reduction of the time-scale will occur, bringing the $\tau_{\rm rad,UV}/\tau_{\rm 56Co}$ ratio down to $< 1$. 
Finally, note that the electron scattering optical depth of the ejecta is smaller than unity at all times (Fig. \ref{fig:thomson}), and so contributes neglegibly to the total transport time scale.


\section{Light curves and spectra}
\label{sec:lcands}

\subsection{Bolometric and photometric evolution}

\begin{figure}
\includegraphics[width=1\linewidth]{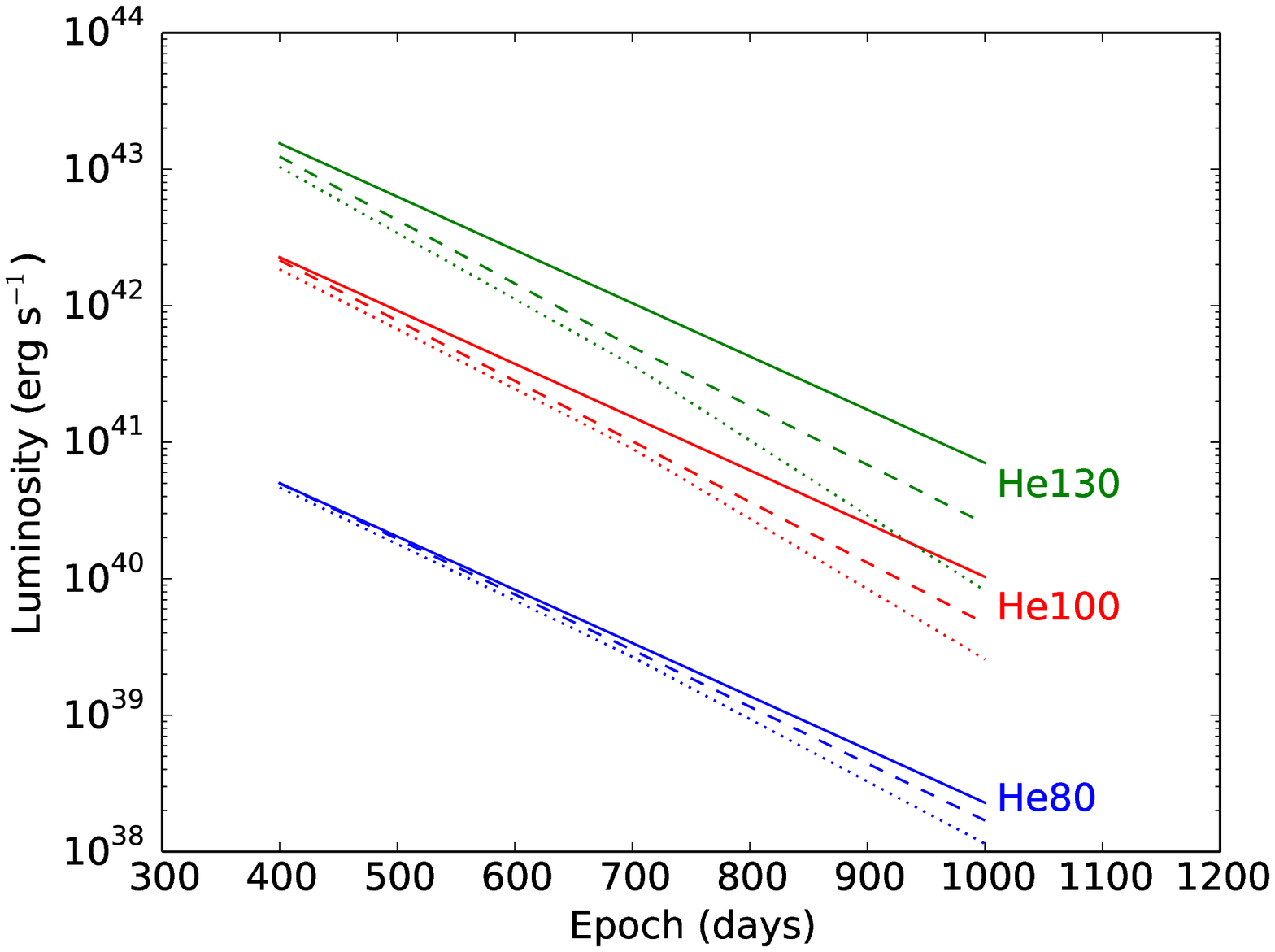} 
\caption{Radioactive decay luminosities (solid), bolometric luminosities (dashed) and quasi-bolometric $U-K$ luminosities (dotted). The last quantity is computed as the integrated flux between 3000 - 23,000 \AA.}
\label{fig:bol}
\end{figure}

\begin{figure}
\includegraphics[width=1\linewidth]{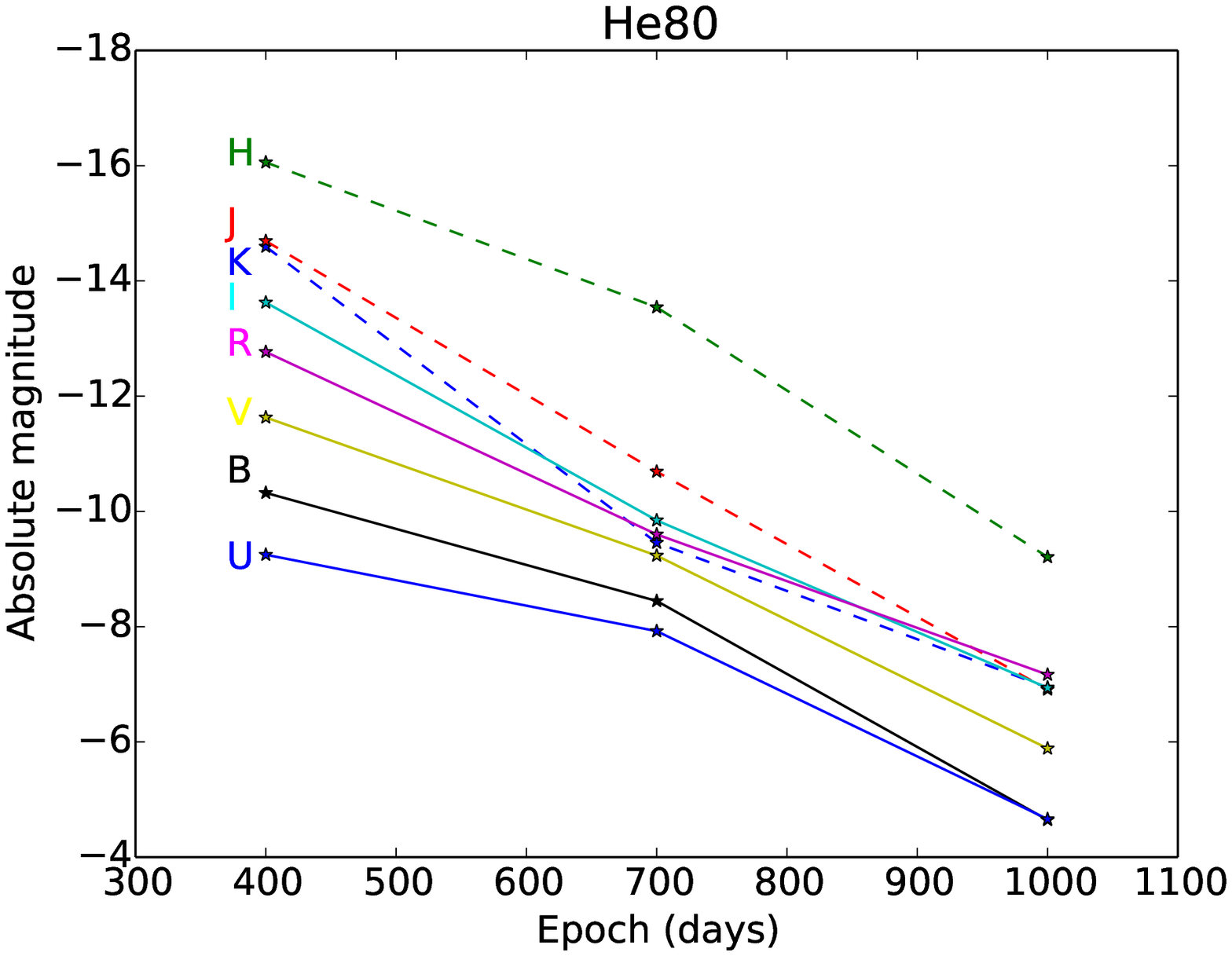} 
\includegraphics[width=1\linewidth]{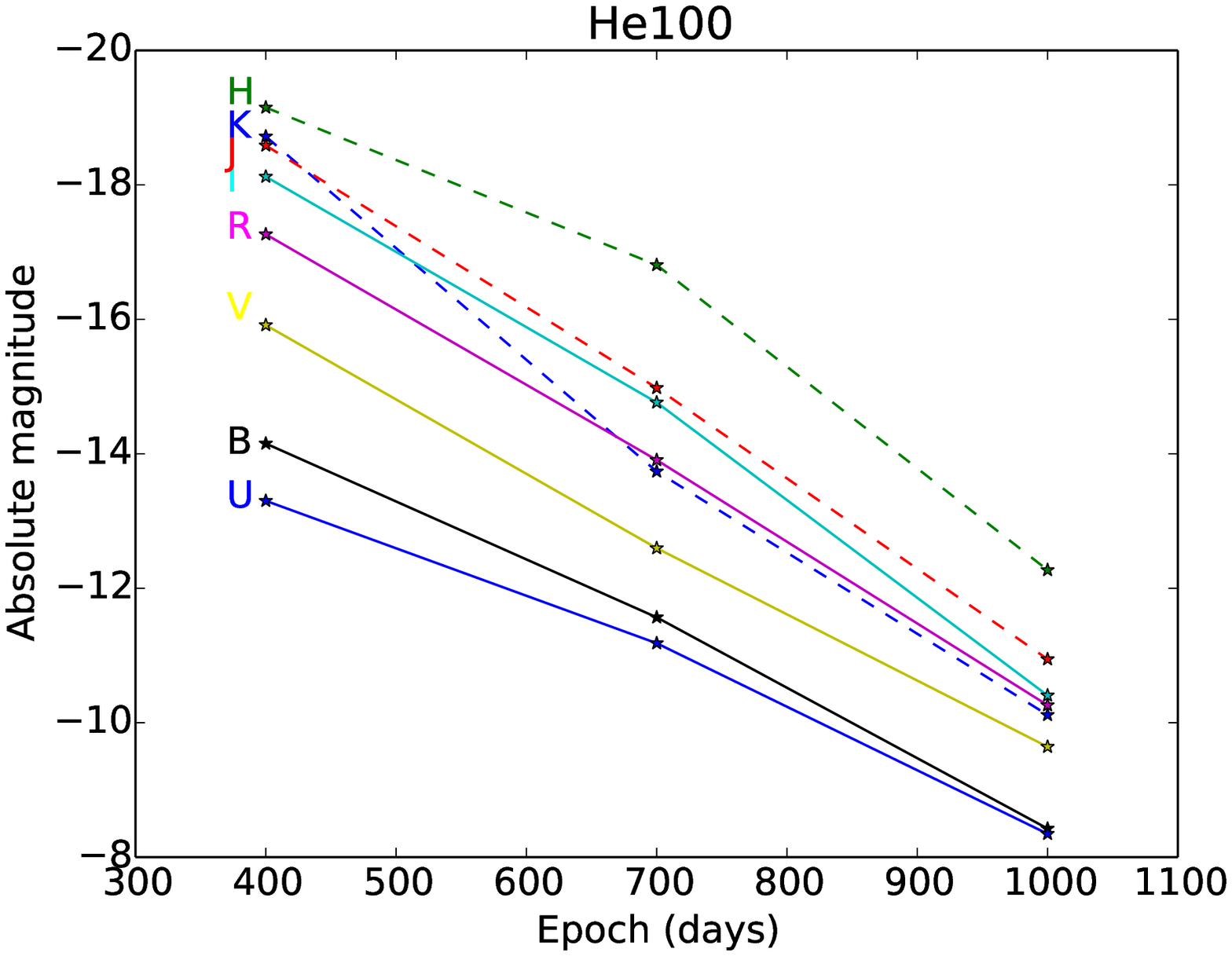} 
\includegraphics[width=1\linewidth]{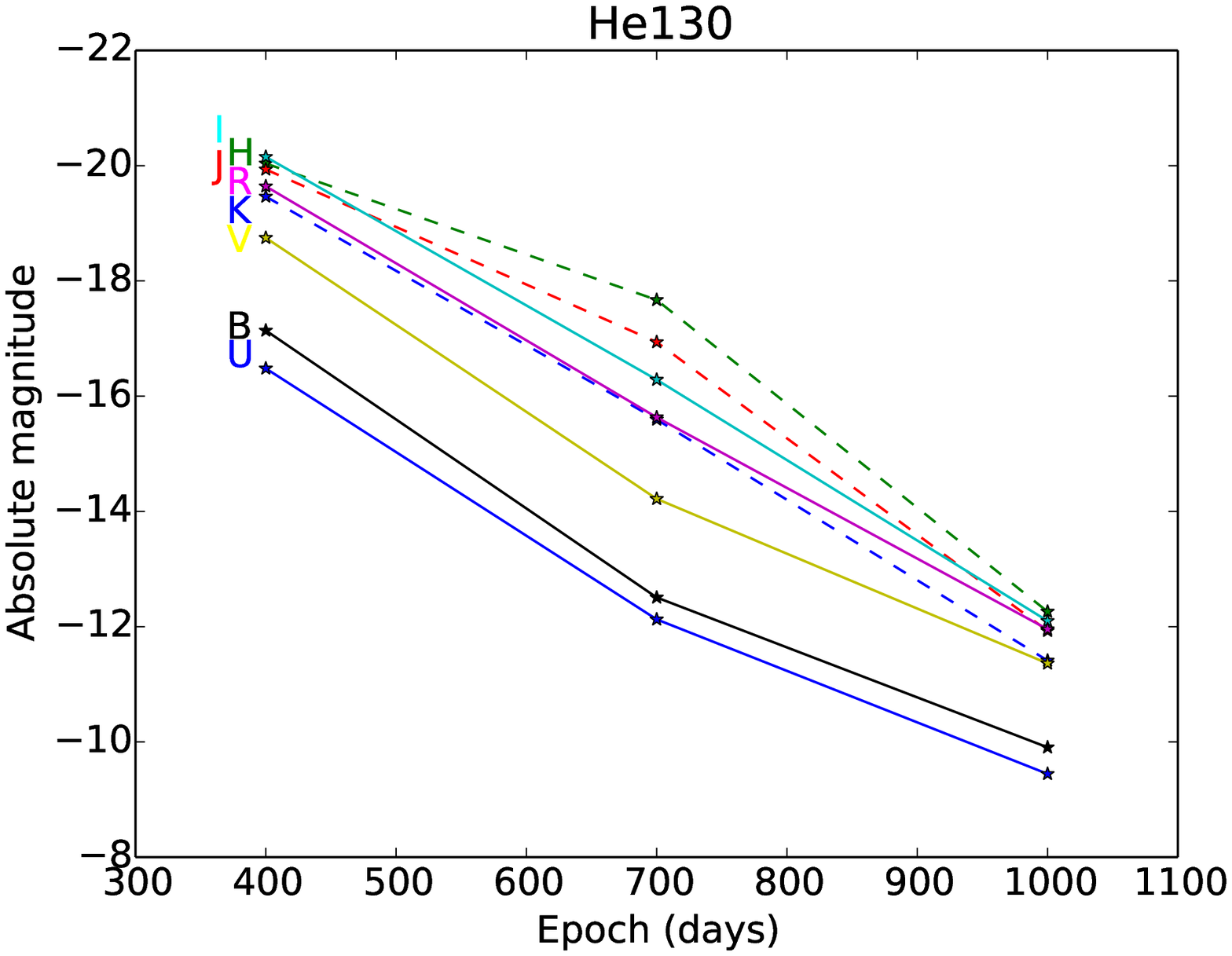} 
\caption{$UBVR_{\rm c}I_{\rm c}JHK$ (Johnson-Cousins and 2MASS) photometry of the models. The values, as well as SDSS $ugriz$ magitudes, are listed in table \ref{table:photometry}.}
\label{fig:photometry}
\end{figure}

Integrating the flux from our model spectra over standard broad band filters provides the photometric evolution. We calculate magnitudes in two systems - Vega-based magnitudes in the Johnson-Cousins $UBVR_{c}I_{c}$ and 2MASS $JHK$ system and AB magnitudes in the SDSS $ugriz$ system.
Figure \ref{fig:bol} shows the bolometric and quasi-bolometric $U-K$ luminosities. Figure \ref{fig:photometry} shows the photometric evolution. All ejecta are still optically thick to gamma-rays at 400 days (Sect. \ref{sec:gammadep}), so the bolometric luminosity is close to the radioactive decay luminosity of $^{56}$Co. For model He80 this holds true up to 1000d, whereas He100 and He130 experience some non-negligible gamma-ray escape after 400d. 
The He100 model loses half the decay luminosity of $^{56}$Co and  He130 loses about two thirds at 1000d. The quasi-bolometric $U-K$ curves are close to the bolometric curves up to 700 days in all models, implying a small fraction of the luminosity emerging in the far-UV and mid/far-infrared during the first 2 years. The fraction outside $U-K$ increases during the third year and by 1000d has reached 35\% (He80), 42\% (He100), and 65\% (He130).

As demonstrated by SN 1987A, $^{57}$Co takes over the powering of SNe after 3-4 years. The $^{57}$Ni/$^{56}$Ni ratio in these PISN models is between 0.2 and 0.8 times the solar $^{57}$Fe/$^{56}$Fe ratio of 0.023 \citep{Lodders2003}. At solar ratio, the $^{57}$Co decay luminosity at 1000d is 16\% of the $^{56}$Co decay luminosity including gamma-rays, and 65\% including only leptons and X-rays. Here, with $\sim$50\% gamma trapping at 1000d, the contribution by $^{57}$Co is $\lesssim$ 10\%, and can be ignored, but modelling into still later phases would see the light curves flatten due to its contribution and longer decay time (391d).

\begin{table*}
	\begin{center}
	\caption{Photometry (absolute magnitudes) of the models.}
	\label{table:photometry}
		\begin{tabular}{cccccccccc}
		
			\hline 
			Band & He80-400d & 700d & 1000d & He100-400d & 700d & 1000d & He130-400d & 700d & 1000d\\
						\hline
			$u$ & -8.13 & -6.97 & -3.72 & -12.20 & -10.20 & -7.51 & -15.60 & -11.20 & -8.57\\
			$g$ & -10.80 & -8.88 & -4.86 & -14.60 & -11.90 & -8.80 & -17.70 & -12.90 & -10.30\\
			$r$ & -12.40 & -9.51 & -7.15 & -16.80 & -13.30 & -10.20 & -19.30 & -15.10 & -11.90\\
			$i$ & -13.00 & -9.29 & -6.49 & -17.70 & -14.50 & -9.79 & -19.90 & -16.10 & -11.40\\
			$z$ & -14.30 & -10.40 & -7.50 & -18.80 & -14.90 & -10.90 & -20.50 & -16.50 & -12.50\\
			$U$ & -9.25 & -7.92 & -4.66 & -13.30 & -11.20 & -8.35 & -16.50 & -12.10 & -9.45\\
			$B$ & -10.30 & -8.45 & -4.65 & -14.20 & -11.60 & -8.42 & -17.10 & -12.50 & -9.91\\
			$V$ & -11.60 & -9.23 & -5.89 & -15.90 & -12.60 & -9.64 & -18.70 & -14.20 & -11.40\\
			$R_{\rm c}$ & -12.80 & -9.60 & -7.17 & -17.30 & -13.90 & -10.30 & -19.60 & -15.60 & -12.00\\
			$I_{\rm c}$ & -13.60 & -9.85 & -6.94 & -18.10 & -14.80 & -10.40 & -20.20 & -16.30 & -12.10\\
			$J$ & -14.70 & -10.70 & -6.91 & -18.60 & -15.00 & -10.90 & -19.90 & -16.90 & -11.90\\
			$H$ & -16.10 & -13.50 & -9.21 & -19.20 & -16.80 & -12.30 & -20.00 & -17.70 & -12.30\\
			$K$ & -14.60 & -9.46 & -6.94 & -18.70 & -13.70 & -10.10 & -19.50 & -15.60 & -11.40\\
				\end{tabular}
	\end{center}
\end{table*}

\subsection{Optical spectra at 400d}

\begin{figure*}
\includegraphics[width=1\linewidth]{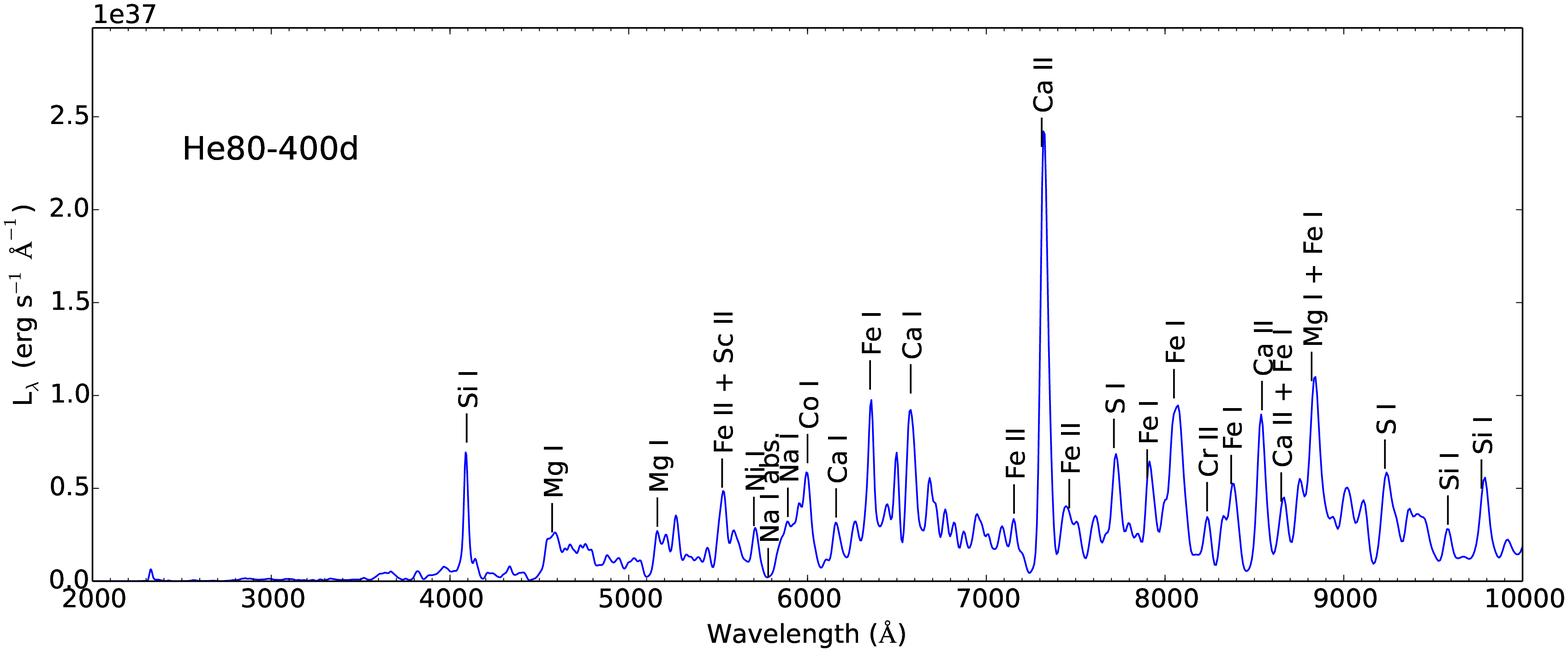} 
\includegraphics[width=1\linewidth]{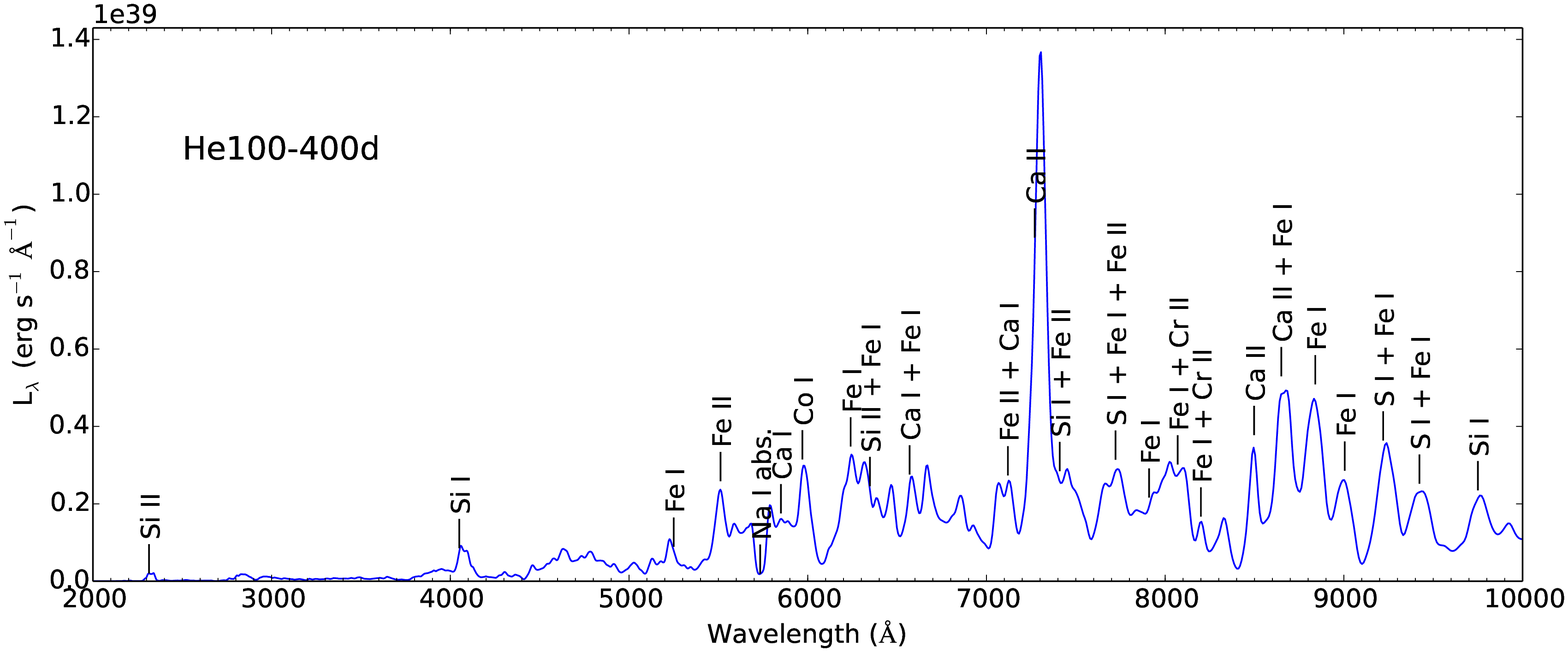} 
\includegraphics[width=1\linewidth]{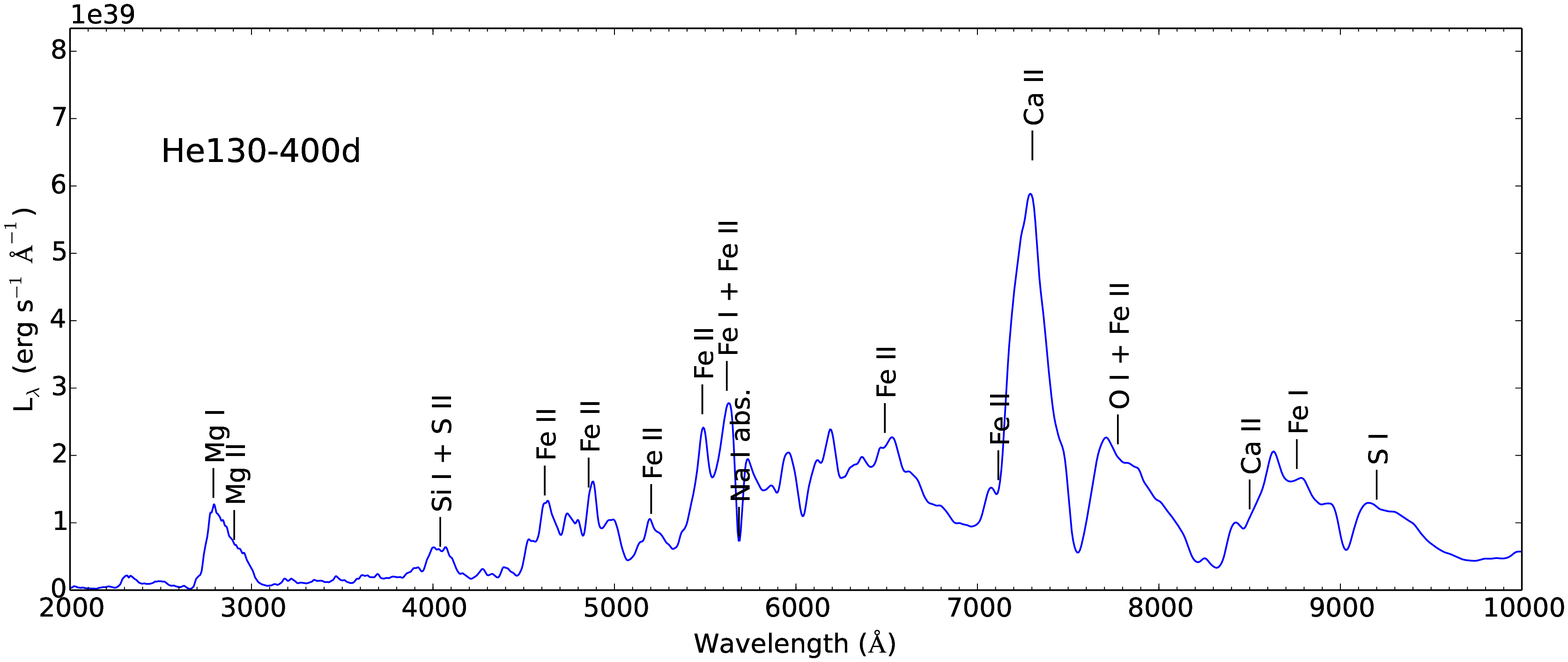} 
\caption{Model spectra at 400d.}
\label{fig:spectra400}
%
\end{figure*}

Figure \ref{fig:spectra400} (top) shows the spectrum of model He80 at
400 days. The spectrum is rich in distinct emission lines of Ca I, Ca
II, Si I, S I, Fe I, Fe II, and Co I with characteristic line widths of
2000 km s$^{-1}$ (the expansion velocity of the Fe/Si layers). 
Note the distinct Ca I \wl6572.
A few (somewhat broader) Mg I lines can also be seen, and a distinct Na I D
scattering line. There is no distinct emission from oxygen. Less than
2\% of the energy is deposited in the oxygen layers, and at the low
temperatures of $\sim$1500 K cooling is done by NIR and MIR lines
(mainly [Si I] 1.607, 1.646 $\mu$m and [O I] 63 $\mu$m). Among the strongest
lines are Si I 4103, Ca I \wl6572, [Ca II] \wll7291, 7323, S I]
\wl7725, and the Ca II NIR triplet. The three largest contributors to the optical
luminosity (taken as 2000-$10^4$ \AA) are Fe I (16\%), Ca II (13\%), and Si I (12\%). 
%

Figure \ref{fig:spectra400} (middle) shows the spectrum of model He100. Although the temperatures are somewhat higher than in He80, the spectrum maintains its red color as line blocking increases. A series of lines from Ca I, Ca II, Si I, S I, Fe I, Fe II, and Co I are seen, broader than in He80 (velocities of 2000-4000 km s$^{-1}$). The most distinct line is [Ca~II]~\wll7291, 7323. No distinct lines of C, O, or Mg can be seen, although much of the flux between 6100-6400 \AA~comes from a broad [O I] \wll6300, 6364 component, with narrow iron lines on top. The oxygen-zone temperature is high enough ($\sim 3500$ K) that [O I] \wll6300, 6364 is an important coolant. The Na I D absorption band centred at 5730 \AA~provides the only clear signature.
The three largest contributors are Fe I (21\%), Ca II (19\%), and Fe II (6\%). 
%

Fig. \ref{fig:spectra400} (bottom) shows the spectrum of model He130. Although the spectral colors remain similar, the higher expansion velocities of the Fe/Si layers (6000-8000 km s$^{-1}$) leads too a smoother spectrum with fewer distinct, unblended lines. The most distinct emission line is still [Ca~II] \wll7291,~7323. [O I] \wll6300, 6364 provides some flux ($T \sim 5000$ K so it is a coolant of the O-layers) but not enough to make a distinct emission line. However, oxygen is now revealed by a quite distinct P-Cygni-like O I \wl7774. Sodium gives a narrow absorption line centred at 5690 \AA. This absorption feature becomes centred at bluer wavelengths for more massive PISNe, as the Na-containing layers expand at higher velocities. Magnesium forms a Mg I + Mg II feature at 2800-3000 \AA, but no Mg I] \wl4571 line is produced. The neutral Mg abundance is less than $10^{-5}$ so Mg I] \wl4571 does not contribute to cooling, and its recombination emission is largely line blocked. 
A difference to the He80 and He100 models 
is that Fe II is now the largest contributor to the optical luminosity 
at 32\% (as the higher $^{56}$Ni mass leads to both higher iron mass and higher ionization state), with Fe I at 22\% and Ca II at 14\%.

\subsection{Optical spectra at 700d}

\begin{figure*}
\includegraphics[width=1\linewidth]{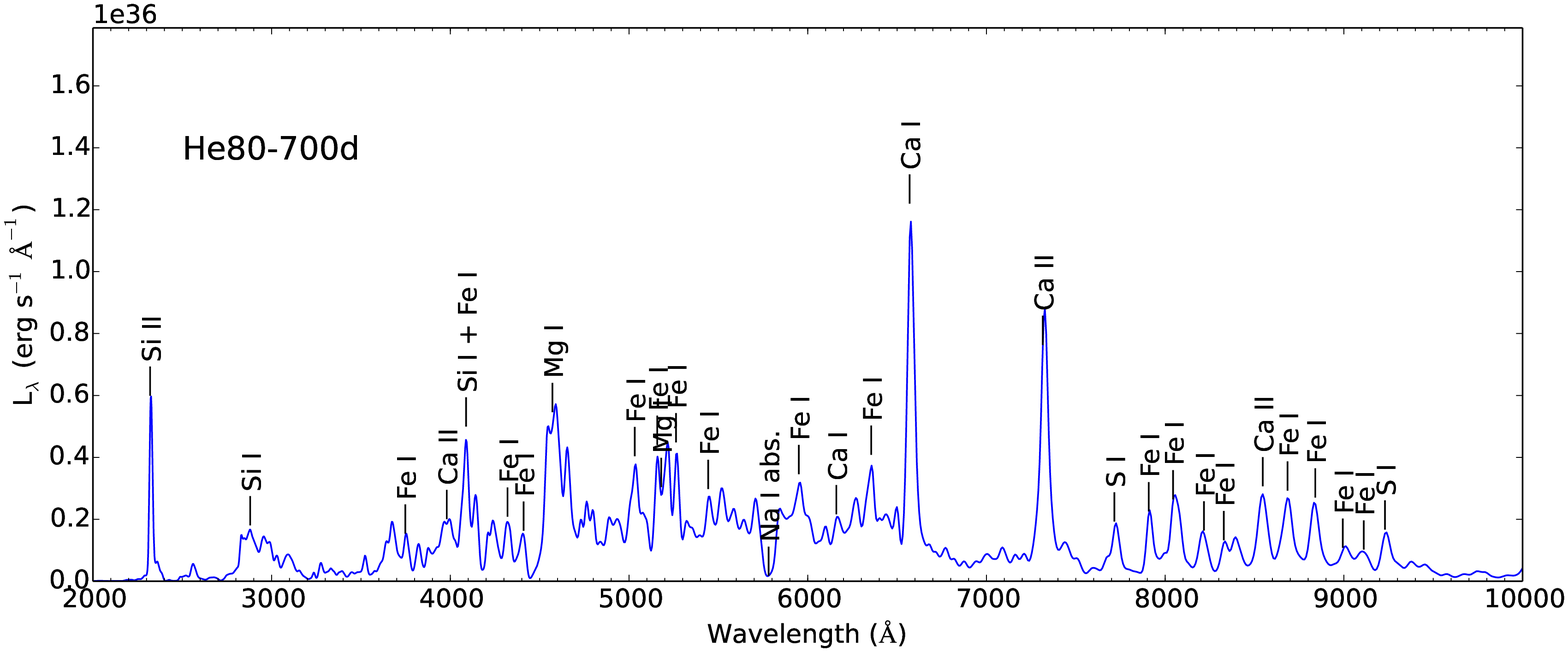} 
\includegraphics[width=1\linewidth]{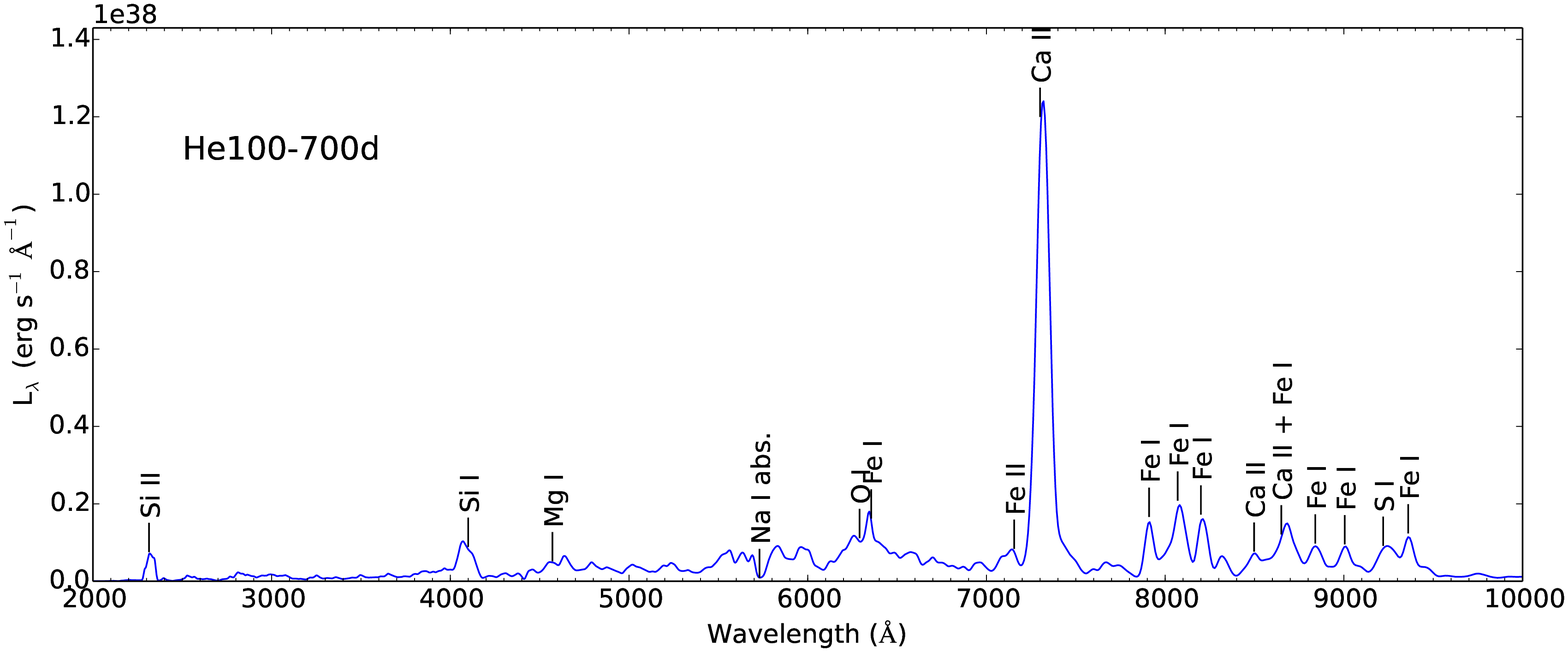} 
\includegraphics[width=1\linewidth]{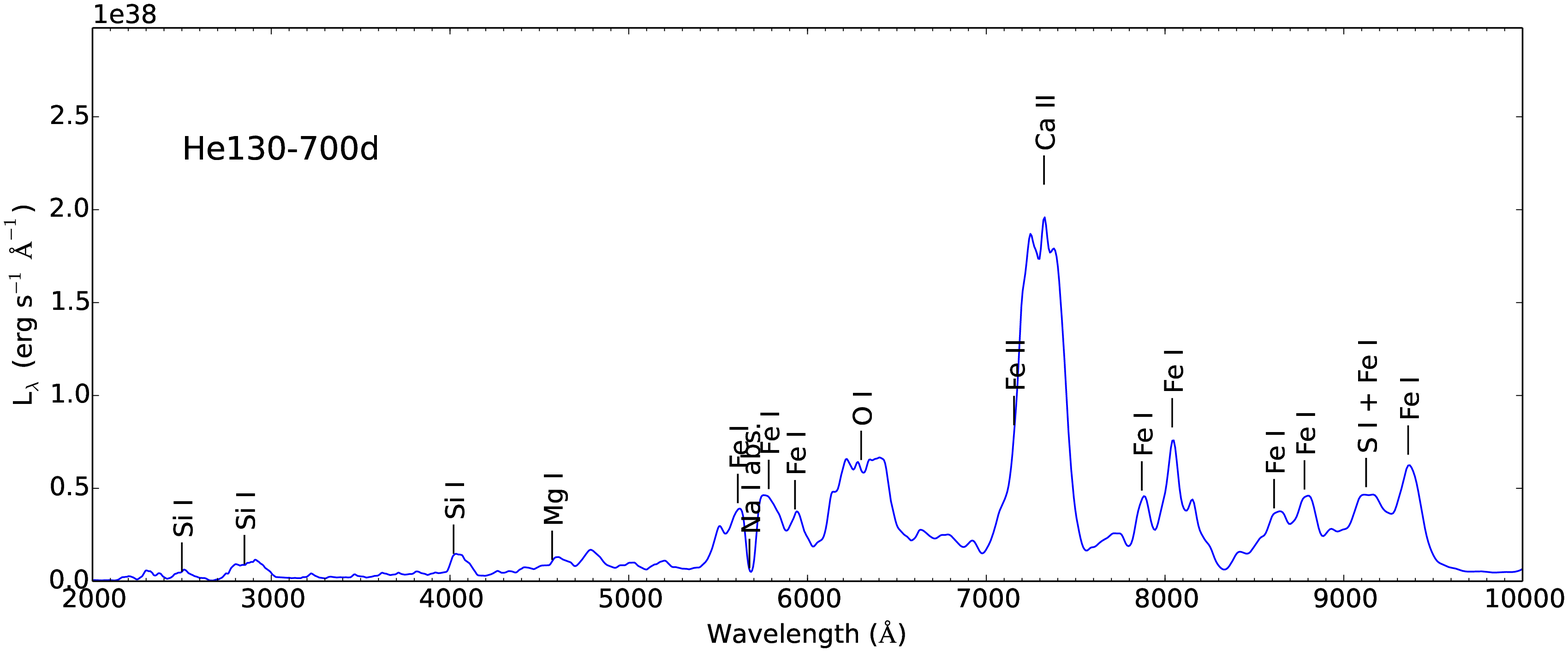} 
\caption{Model spectra at 700d.}
\label{fig:spectra700}
\end{figure*}

Figure \ref{fig:spectra700} (top) shows the spectrum of model He80 at 700 days. Compared to the 400d spectrum, Mg I] \wl4571 has strengthened, and the Ca II lines have weakened. Ca~I~\wl6572 is now the strongest line. 
The strongest contributors are Fe~I (31\%), Ca~I (8\%), and Ca~II (7\%).

Figure \ref{fig:spectra700} (middle) shows the spectrum of model He100. Compared to 400d, the forbidden calcium doublet has increased in prominance. The Ca II NIR triplet has decreased in strength to reveal a series of Fe~I lines between 8000-9500 \AA. A discernible [O~I] \wll6300,~6364 line is produced, although it is still not very distinct. Mg~I] \wl4571 is weak as well. As at 400 days, most of the spectrum is produced by Fe I (33\%), followed by Ca II (29\%) and Si I (5\%). Fe II makes neglegible contribution (3\%), the strongest line being [Fe II] \wl7155.

Figure \ref{fig:spectra700} (bottom) shows the spectrum of model He130. Compared to 400d, [O I] \wll6300, 6364 has become more distinct, but Mg~I] \wl4571 is still weak. The 5000-10000 \AA~range is dominated by Fe I lines, [O~I] \wll6300, 6364, and [Ca~II] \wll7291, 7323. Fe I has distinct emission clusters around 5700 \AA, 8000 \AA, and 8500-9500 \AA. Even the high $^{56}$Ni mass cannot maintain high ionization any longer at 700d, and the only significant Fe II emission is in [Fe~II] \wl7155, in the blue wing of [Ca~II] \wll7291, 7323. The strongest contributors to the spectrum are Fe I (43\%), Ca II (24\%), and O I (8\%).

\subsection{Optical spectra at 1000d}

\begin{figure*}
\includegraphics[width=1\linewidth]{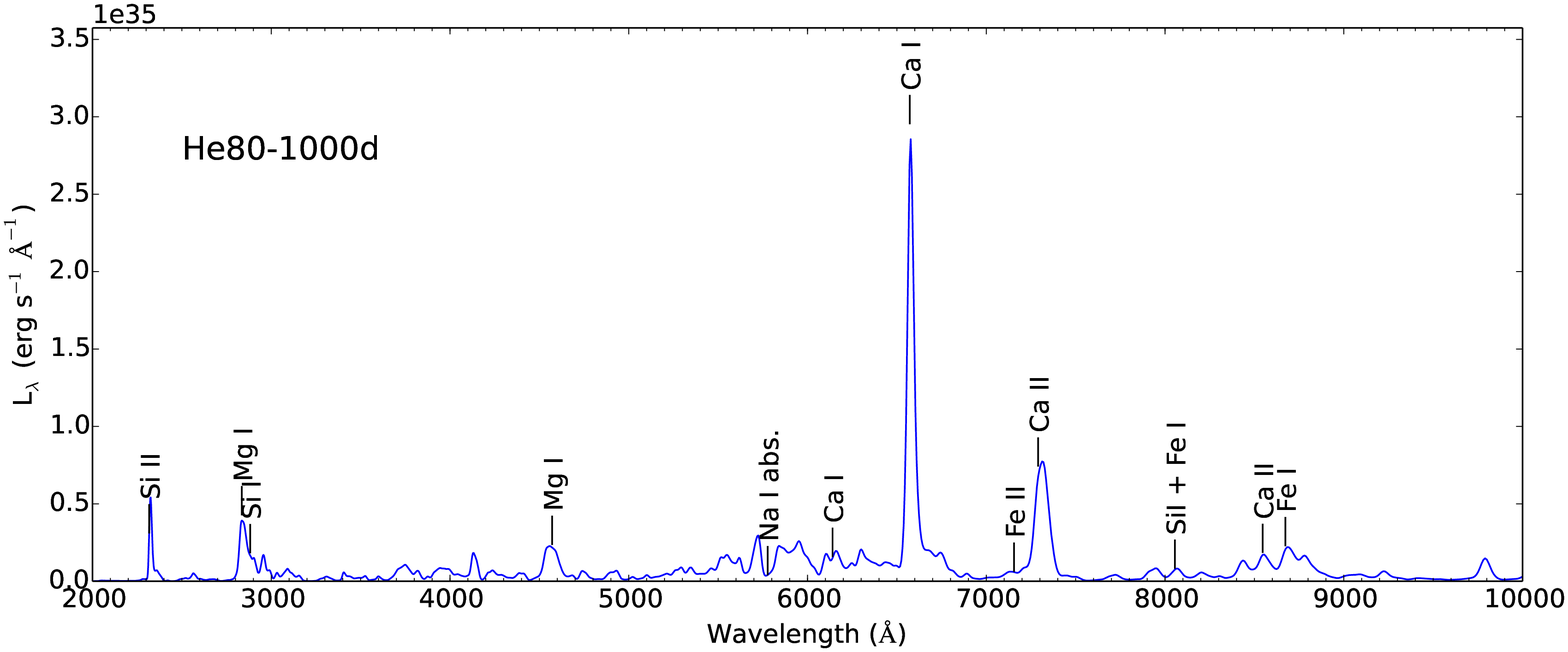} 
\includegraphics[width=1\linewidth]{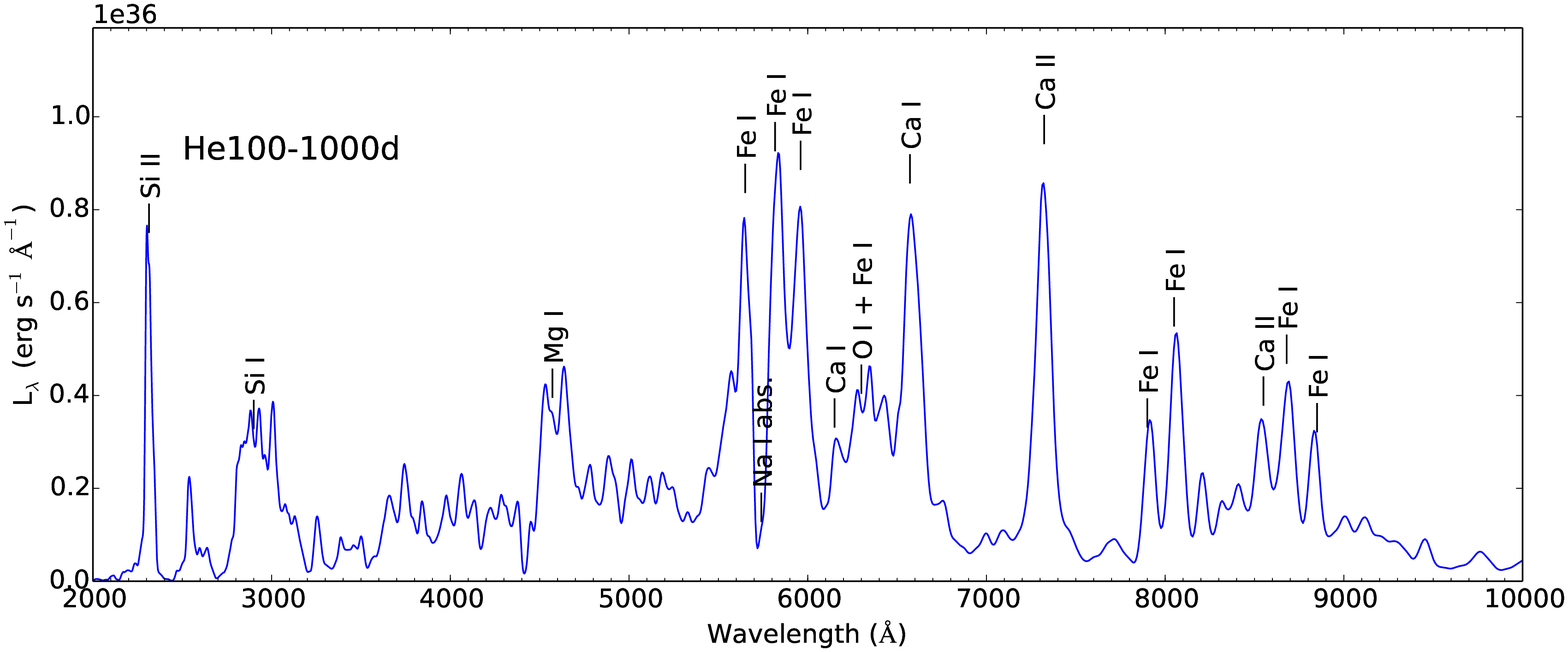} 
\includegraphics[width=1\linewidth]{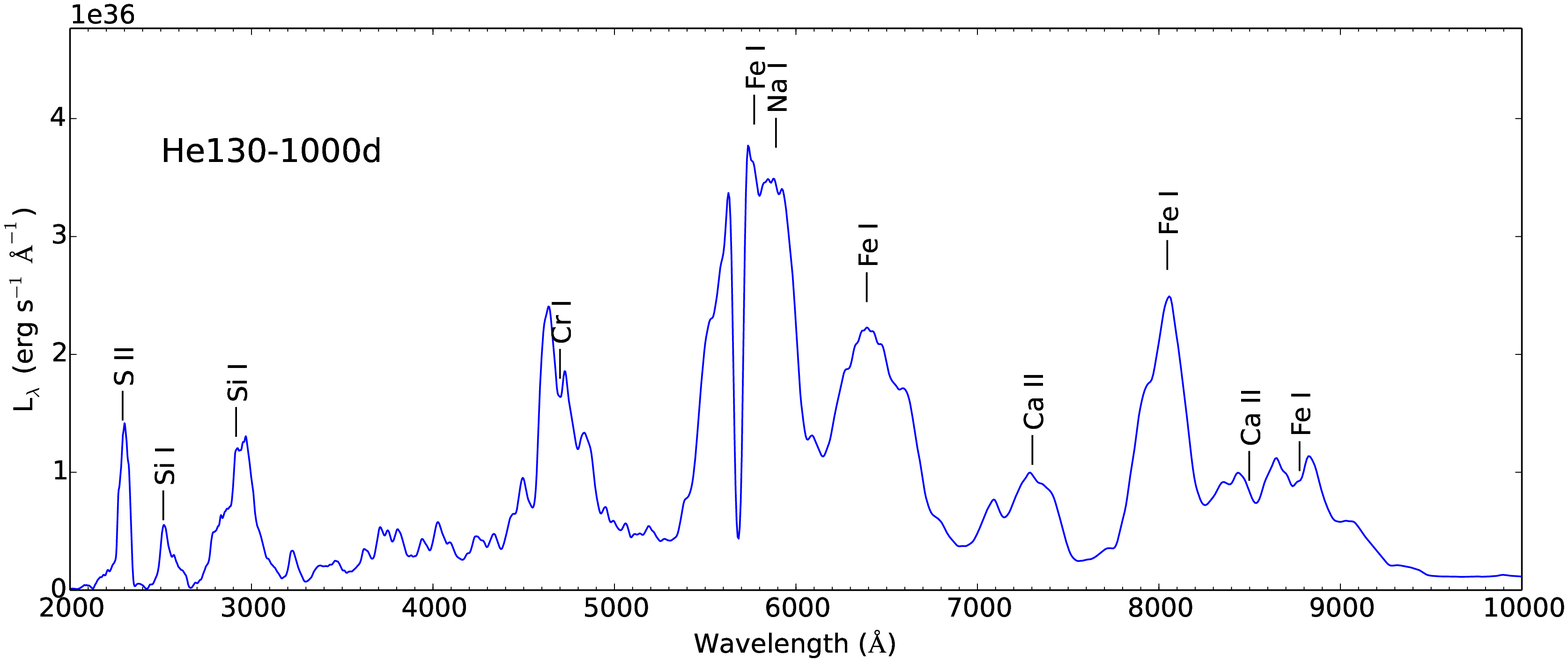} 
\caption{Model spectra at 1000d.}
\label{fig:spectra1000}
\end{figure*}

Figure \ref{fig:spectra1000} (top) shows the spectrum of model He80 at 1000 days. Ca I \wl6572 has increased its prominance and is now by far the strongest line. Also distinct are Mg~I] \wl4571, Ca~I \wl6160, and [Ca~II] \wll7291, 7323. The strongest contributors are Ca~I~(22\%), Ti~I~(18\%), and Ca~II~(10\%).

Figure \ref{fig:spectra1000} (middle) shows the spectrum of model He100. Compared to 700 days, the biggest difference is that [Ca II] \wll7291, 7323 has reduced its dominance as the ejecta are now too cool for it. Another difference is that Mg~I] \wl4571 has become prominent. [O I] \wll6300, 6364 is weak as the temperature has become too low to thermally drive it, with Fe I now dominating emission at that wavelength. Na~I~D still provides absorption and give a very distinct splitting of the Fe I 5700 \AA~feature, centred at 5720 \AA. The ejecta have become neutral enough that strong Ca I lines emerge at 6160 and 6572 \AA~(as at earlier epochs in He80). The strongest contributors are Fe I (40\%), Ca II (8\%), and Si I (7\%).

Figure \ref{fig:spectra1000} (bottom) shows the spectrum of model He130. The most prominent features are a Cr I + Fe I + Ti I line cluster at 4700 \AA~(likely a scattering/fluorescence effect), a broad Fe I feature at 5800 A, with a distinct Na I D absorption split centred at 5680 \AA, Fe I features at 6400 and 8000 \AA, as well as [Ca~II] \wll7291, 7323. The contributions by oxygen and magnesium are neglegible, as is contribution by Fe II. The strongest contributors are Fe I (56\%), Cr I (6\%), and Si I (6\%).

\subsection{Near-infrared spectra}
\begin{figure*}
\includegraphics[width=1\linewidth]{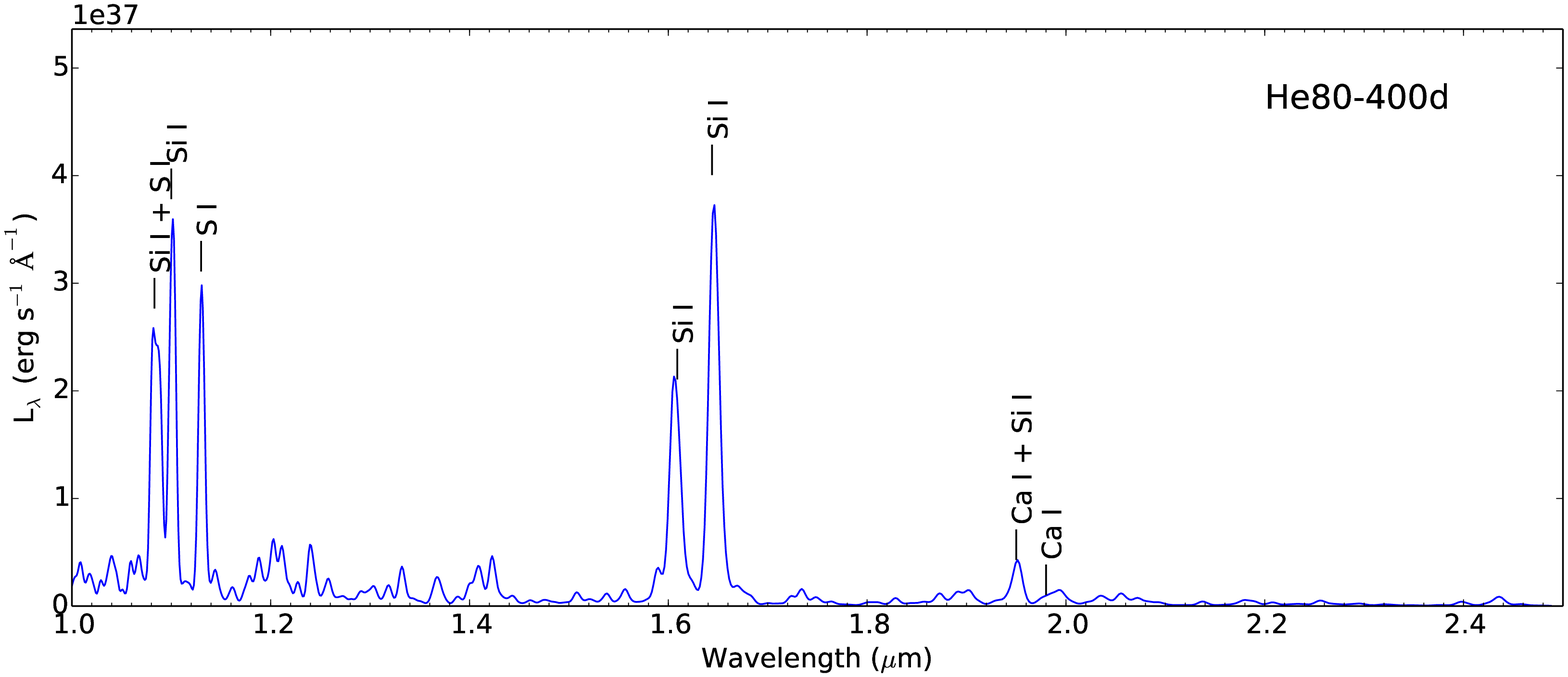} 
\includegraphics[width=1\linewidth]{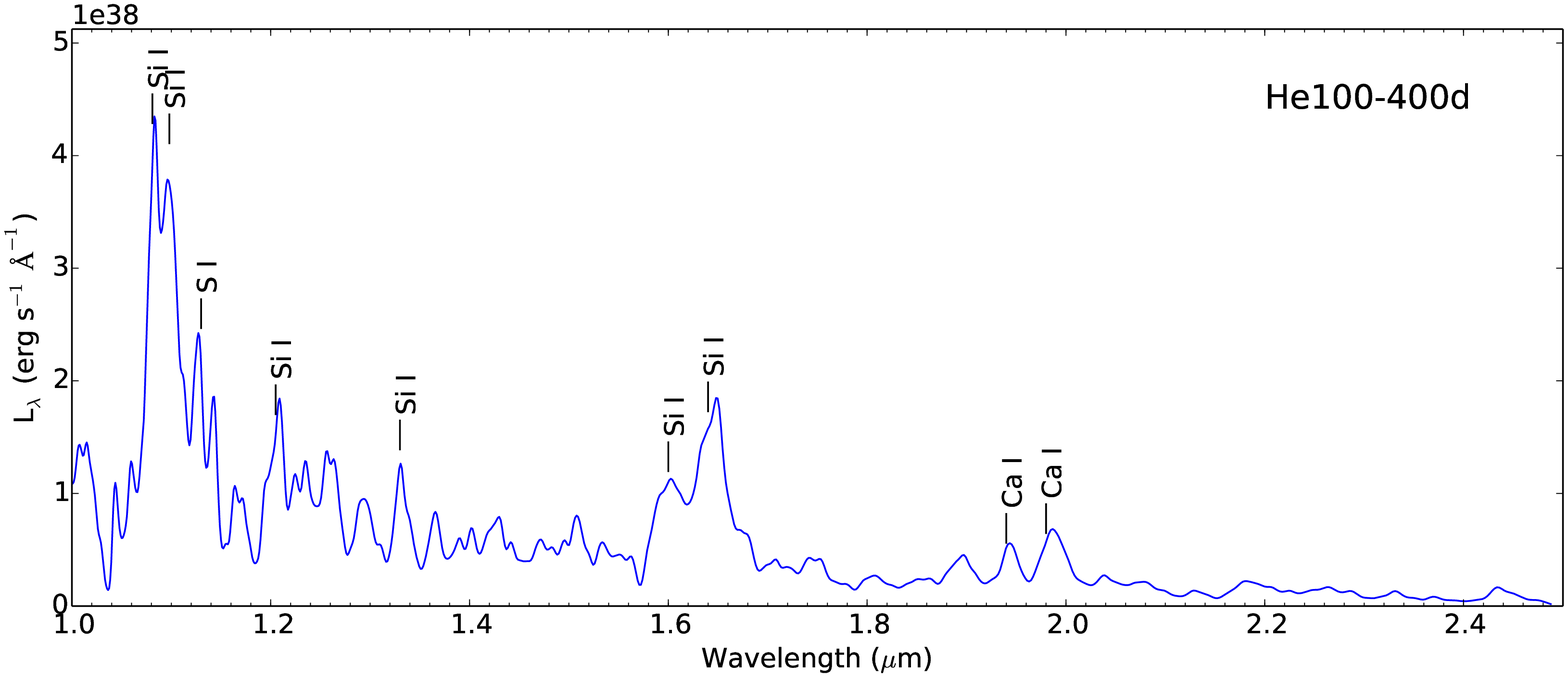} %
\includegraphics[width=1\linewidth]{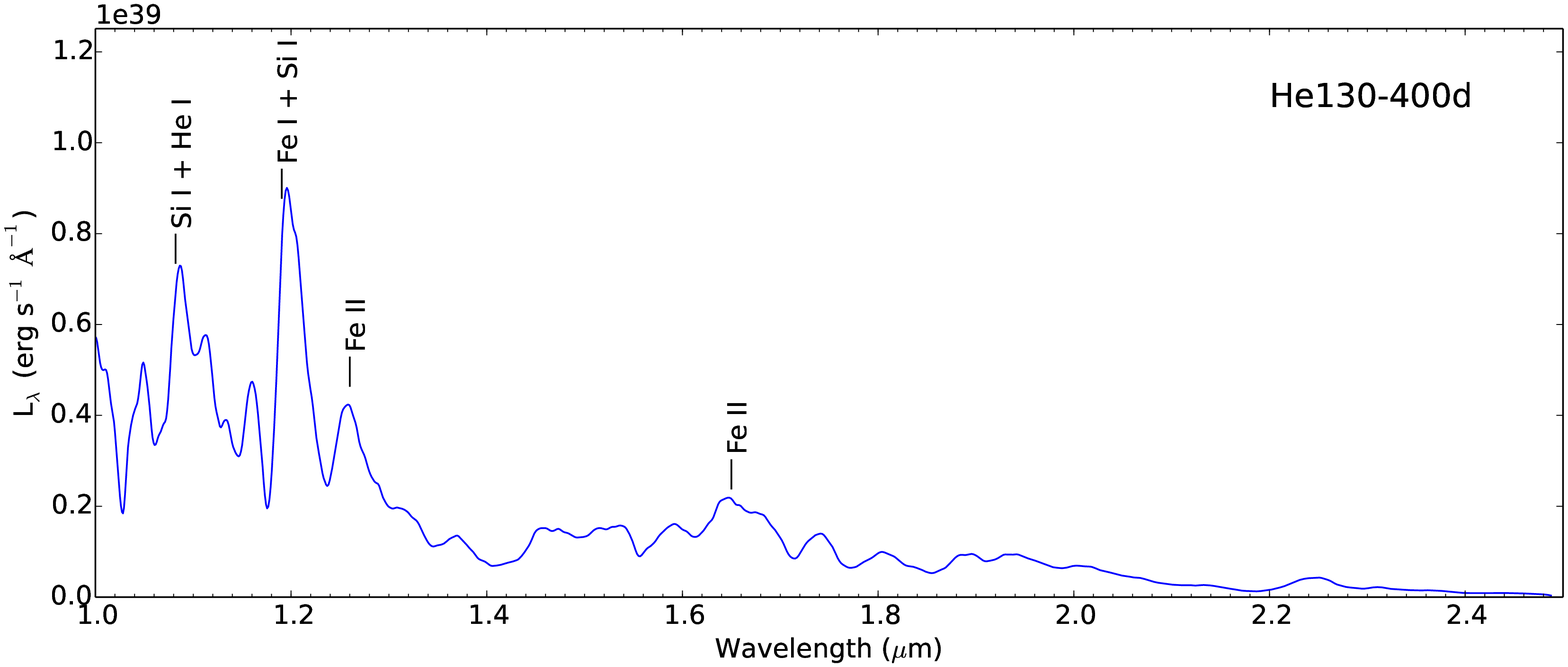} %
\caption{Model spectra in the near-infrared at 400d.}
\label{fig:nirspectra400}
\end{figure*}

The model spectra in the near-infrared at 400d are shown in Fig. \ref{fig:nirspectra400}. The He80 model is dominated by lines of neutral silicon and sulphur; [Si I] 1.088 $\mu$m, [Si I] 1.099 $\mu$m, [Si I] 1.607, 1.646 $\mu$m, and [S I] 1.082,~1.131 $\mu$m. There is relatively little iron-group emission as the $^{56}$Ni mass is low in this model. Ca~I lines from the 4s($^{3}$D) state are also produced  at 1.93-1.99 $\mu$m. 

In He100 the strongest  features are due to the same silicon and sulphur lines, now blended to a higher extent. Iron emission provides a stronger quasi-continuum as the $^{56}$Ni mass is 50 times higher than in He80. 

In He130 the strong Doppler blending leads to few distinct lines. The strongest feature is a Si I + Fe I blend at 1.2 $\mu$m. The iron quasi-continuum has strengthened further, with Fe II now making significant contribution.
Some of the emission around 1.08 $\mu$m is from He I in the core (produced in $\alpha-$rich freeze-out). Observationally it will be hard to distinguish He 1.083 $\mu$m emission from Si I and S I emission. He I 2.058 $\mu$m does not emit strongly enough to facilitate a helium identification.

The NIR spectra evolve with time, but without significant qualitative changes. The 700d and 1000d model spectra are displayed in Appendix \ref{NIRlate}. Throughout the NIR at all epochs, emission is dominated by Si~I, S~I, Ca~I, Fe~I, and Fe~II. There is no distinguishable emission from C, O, Na, or Mg.

\section{Comparison with PISN candidates}
\label{sec:compdata}

\subsection{SN 2007bi}

The superluminous SN 2007bi at redshift $z=0.1279$ had a bright and a
slowly declining light curve in good agreement with pair-instability
models of 100-130 \msun~He cores \citep{Gal-Yam2009}. We
compare nebular Very Large Telescope (VLT) and Keck spectra, taken at +367d and +471d rest-frame post-peak, respectively, of this SN with the models. There are several issues with the calibration process of these spectra. 
The VLT spectrum presented in \citet{Gal-Yam2009} is, contrary to an incorrect
statement in the supplementary information section, not corrected for
host galaxy contamination, which is significant. This in turn compromises the flux calibration, which was done against template-subtracted photometry (excluding the galaxy light).
As we now have access to high-quality host photometry \citep{Chen2015}, host
subtraction can be done by fitting galaxy models to the host photometry, and we have
recalibrated both the VLT and Keck spectra, which together with additional modelling is described in an upcoming paper (Jerkstrand et al., in prep.). Our new spectra are significantly different to the ones presented in \citet{Gal-Yam2009}, and we recommend using
these new calibrations for future analysis.
The spectrum has been corrected for Milky Way extinction (using the \citet{Cardelli1989} law with $R_V=3.1$ and $E(B-V)=0.03$ mag (estimated from the $A_R=0.07$ mag value reported by \citet{Gal-Yam2009}) and scaled to rest-frame spectral luminosity using $z = 0.1279$, $H_{\rm 0}=72$ km s$^{-1}$ Mpc$^{-1}$, and standard $\Lambda$CDM cosmology (giving a luminosity distance 583 Mpc).

\begin{figure*}
\includegraphics[width=0.75\linewidth]{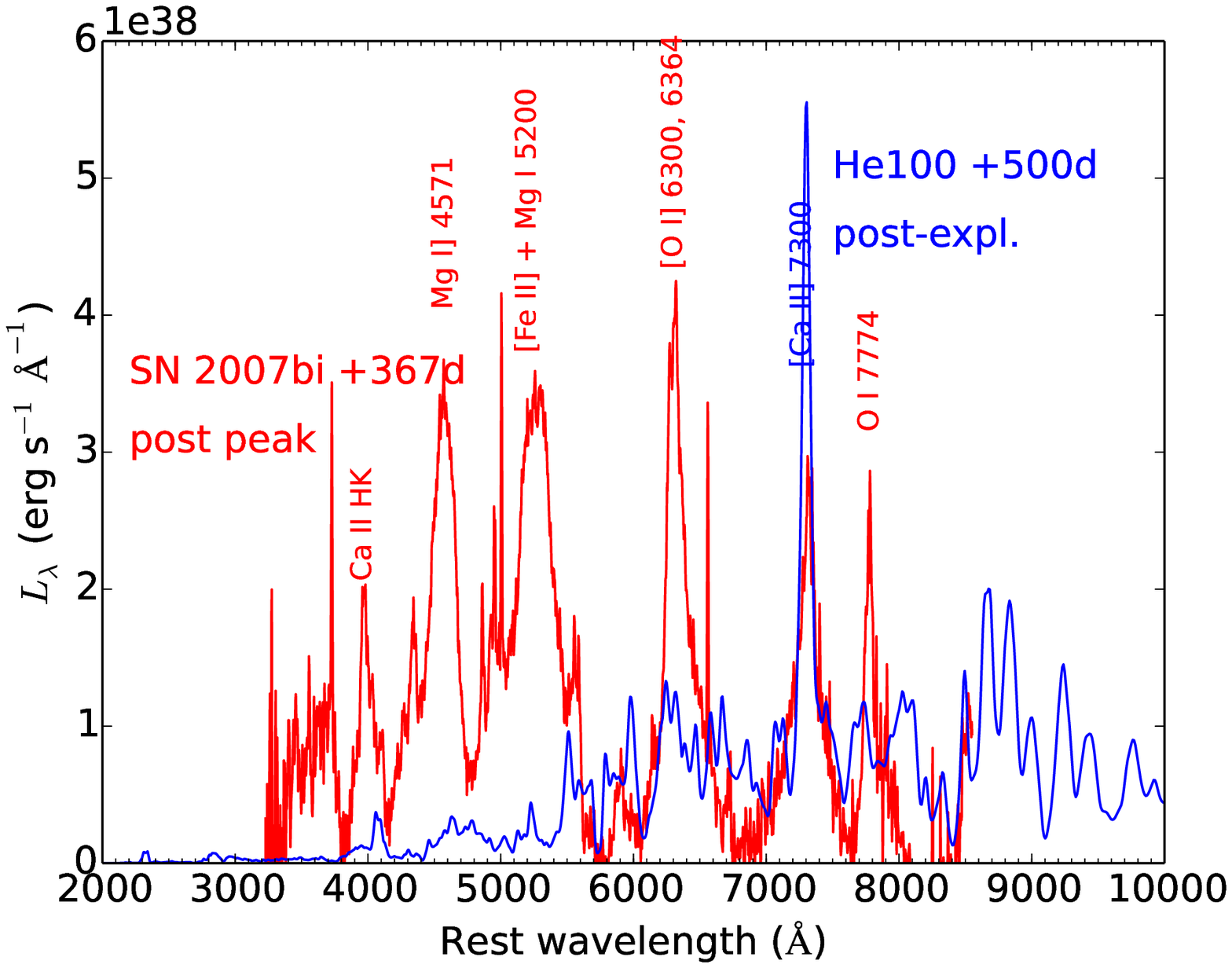} 
\caption{SN 2007bi VLT spectrum at +367d rest-frame post-peak (red), and He100 at +500d post-explosion (blue, 400d model scaled with $e^{-100/111}$).}
\label{fig:2007bi_1}
\end{figure*}
\begin{figure*}
\includegraphics[width=0.75\linewidth]{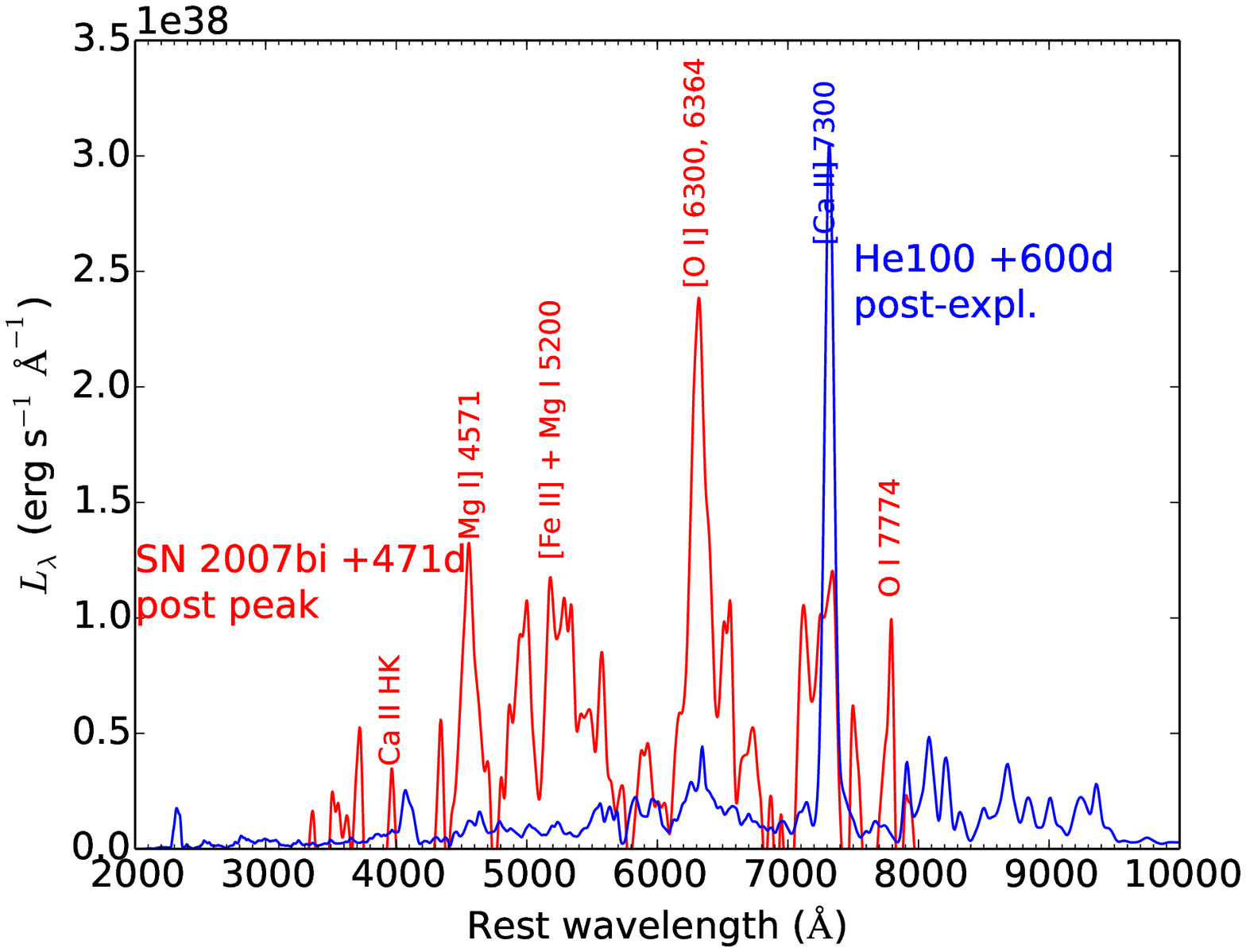} 
\caption{SN2007bi Keck spectrum at +471d rest-frame post-peak (red), and He100 at +600d post-explosion (blue, 700d model scaled with $e^{+100/111}$).}
\label{fig:2007bi_3}
\end{figure*}

 A complication for the model comparisons is that the explosion epoch of SN 2007bi is unknown, with detection already close to peak. SN 2007bi had its first photometric observation on JD 2454130, rising to R-band peak around JD 2454160. The VLT spectrum was obtained on JD 2454566, +436d after discovery in observer frame (+387d rest frame). The minimum rest-frame age of the SN at the time of the VLT spectrum is thus 387d. For the possibility of a PISN interpretation for the early light curve, helium cores of mass 100-130 \msun~exploding $\sim$100d (100 \msun~He core) to $\sim$250d (130 \msun~He core) in the rest-frame before the first data point are required\footnote{By inspection of Fig. 2 (bottom) in \citet{Gal-Yam2009}, to fit the post-maximum light curve, He100 needs to be shifted by about +40 days (giving an explosion epoch of $\sim$100d before the first data point), and He120 needs to be shifted by about -60 days (giving an explosion epoch of $\sim$200d before first data point). He130 with a 70\% higher $^{56}$Ni mass than He120 would need a further shift of about -50 days (giving an explosion epoch of $\sim$250 days before the first data point)}. Thus, for comparison with the model spectra, the relevant model epochs are $\sim$500d (387d+100d) for a He100 comparison and $\sim$650d (387d+250d) for a He130 comparison. We are limited to model calculations at the grid epochs at 400d, 700d and 1000d. However, physical conditions and spectral character do not change significantly on time-scales of 100-150d, and we can therefore use spectra at the closest available epoch, and scale the flux levels with the $^{56}$Co decay factor. We use here the He100 model at 400d, scaled with the exponential decay factor $\exp{\left(-100/111\right)}$ to 500d, and the He130 model at 700d, scaled with $\exp{\left(+50/111\right)}$ to 650d. For comparison with the 104d later Keck spectrum, we use the He100 model at 700d, scaled with $\exp{\left(+100/111\right)}$ to 600d, and the He130 model at 700d, scaled with $\exp{\left(-50/111\right)}$ to 750d. Of course, the He80 model is not relevent for comparison as it is too faint (by several magnitudes) to match the observed diffusion-phase luminosity. 

The model spectra are convolved with a Gaussian with $FWHM=600$ km s$^{-1}$, which is similar to the resolution of the observed spectra (500-1000 km s$^{-1}$). Note, however, that the Keck spectrum we display has been additionally smoothed with a 50 \AA~FWHM Gaussian due to its very high noise levels, corresponding to a further $\sim$3000 km s$^{-1}$ convolution.

\subsubsection{Comparison to He100}
Figure \ref{fig:2007bi_1} shows the SN 2007bi VLT spectrum compared to the He100 model at 400d, scaled  with $\exp{\left(-100/111\right)}$ to adjust the flux levels to 500d. The observed spectrum matches the total optical luminosity reasonably well. The model does not, however, reproduce the major emission lines of Ca II HK, Mg I] \wl4571, [Fe II] + Mg I \wl5200, and O~I~\wl7774 seen in the observed spectrum. Conversely, the multitude of narrow ($\sim 3000$ km s$^{-1}$) Fe I lines produced by the model are not seen in the observed spectrum. The Na I absorption trough also appears dissimilar, being narrower in the model. The model makes about twice as strong [Ca II] \wll7291, 7323 as observed. In general, the model spectrum is redder than the observed spectrum, although by less than the original spectrum which was not galaxy-contamination corrected. In general the lines appear too narrow in this model, although the observed [Ca~II]~\wll7291,~7323 and O~I~\wl7774 lines are narrower than the observed lines at shorter wavelength (Mg I] \wl4571 and [Fe II] + Mg I 5200). Comparison with the Keck spectrum (Fig. \ref{fig:2007bi_3}) shows similar discrepancies. 
In the data, [O I] \wll6300, 6364 continues to strengthen and [Ca II] \wll7291, 7323 weaken, which is opposite to the trend in the model.

\begin{figure*}
\includegraphics[width=0.75\linewidth]{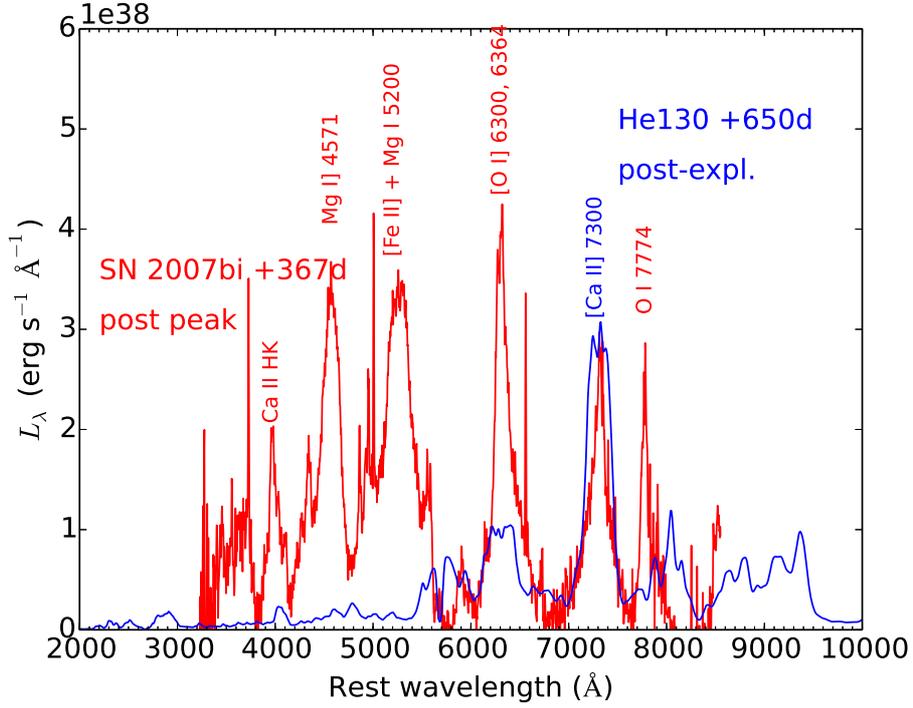} 
\caption{SN 2007bi VLT spectrum at +367d rest-frame post-peak (red), and He130 at +650d post-explosion (blue, 700d model scaled with $e^{+50/111}$).}
\label{fig:2007bi_2}
\end{figure*}
\begin{figure*}
\includegraphics[width=0.75\linewidth]{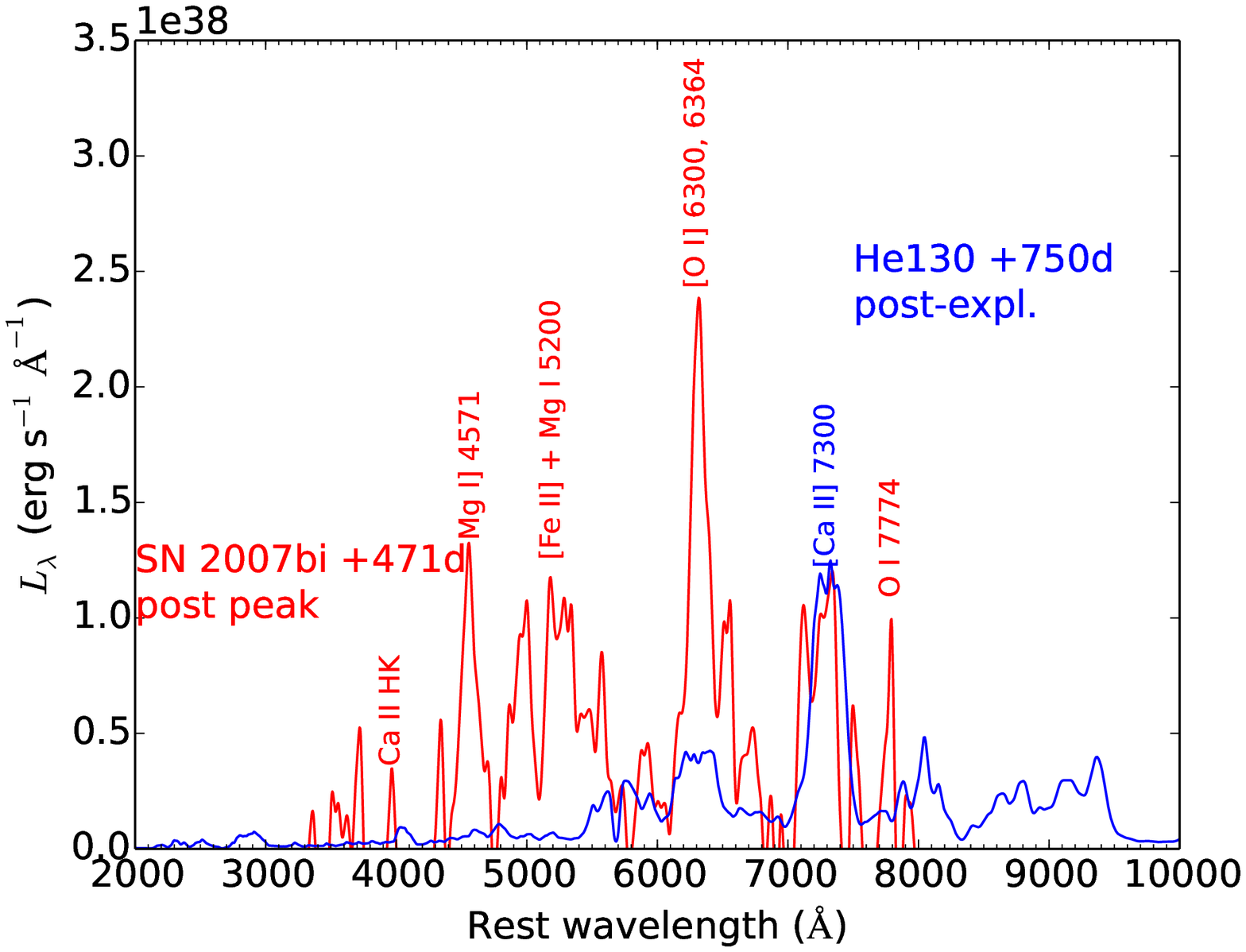} 
\caption{SN2007bi Keck spectrum at +471d rest-frame post-peak (red), and He130 at +750d post-explosion (blue, 700d model scaled with $e^{\left(-50/111\right)}$).}
\label{fig:2007bi_4}
\end{figure*}

\subsubsection{Comparison to He130}
Figure \ref{fig:2007bi_2} shows the SN 2007bi VLT spectrum compared with the He130 model at 700d, scaled with $\exp{\left(+50/111\right)}$ to adjust the flux level to 650 days. Again the bright observed lines below 6000 \AA, as well as O~I~\wl7774, are not reproduced at all by the model. Only the base of [O I] \wll6300, 6364 is reproduced, and [Ca II] \wll7291, 7323 has reasonable flux but is too flat-topped. The model also predicts narrow Na I D absorption and a strong Fe I feature at 8000 \AA, not seen in the data.
One should also note the differences between the line profiles. Compared to He100, the He130 model has significantly broader lines ($\sim 6000-10^4$ km s$^{-1}$), more similar to the widths of the observed lines in the blue. The observed narrow cores of [O~I]~\wll6300,~6364, [Ca~II] \wll7291,~7323, and O~I~\wl7774 are not reproduced in the He130 model, which has its oxygen and calcium at high velocities ($V \gtrsim 5000$ km s$^{-1}$ for Ca and $V \gtrsim 8000$ km s$^{-1}$ for O) giving broad, flat-topped line profiles. Thus, while the iron lines have broadened and make less of a discrepancy compared to He100, the discrepancy has increased for the intermediate-mass elements. Comparison with the Keck spectrum (Fig. \ref{fig:2007bi_4}) gives a similar picture. 

\subsection{PTF12dam}
PTF12dam is a well observed superluminous SN whose decline slope is similar to 
$^{56}$Co decay, and therefore we should consider comparing it to PISN models. \cite{Nicholl2013} argued that this could not be a PISN due to measured rise time (around 50 days) being faster than in quantitative models \citep[more than 150 days,][]{Kasen2011, Dessart2013}. Nevertheless, as apart from SN 2007bi this is the the only other well observed SNe that has a light-curve resembling PISN models, we compare the nebular spectrum of PTF12dam to our models.

The nebular spectrum of PTF12dam is taken from \cite{Chen2015}, and is of poorer quality than the SN 2007bi spectra, largely due to a higher amount of galaxy contamination. The galaxy is about two orders of magnitude brighter than the SN at the observation epoch, and the SN spectrum is therefore hard to extract in a reliable way. Throughout the evolution, the position of the SN is unresolved from the host in ground based imaging and spectra. Difference imaging allowed \cite{Chen2015} to recover the photometric lightcurve until +399d after the $r-$band peak (all epochs are quoted in the restframe). The nebular spectrum was taken at +509d after the $r-$band peak with the 10.4m Gran Telescopio CANARIAS (GTC). Because the SN contribution at this epoch is minor ($\lesssim$1\%), the spectrum was calibrated to SDSS pre-explosion photometry of the galaxy. By subtracting a Starburst99 model for the galaxy continuum, as well as individual narrow galaxy emission lines, \cite{Chen2015} were able to recover a noisy spectrum of the SN. The spectrum was finally scaled to an estimated SN $i$-band magnitude, extrapolated from the last measurement at +399d. This final scaling factor was large (3.5 magnitudes, T.-W. Chen, priv. comm.), implying difficulties in the galaxy subtraction processs and high uncertainty in the final flux levels.

For model comparison, we have chosen to plot the observed spectrum at its flux level before the final $i$-band calibration in \citet{Chen2014}, as the SN + host spectrum is already flux calibrated to the SDSS photometry and should be reliable since the galaxy is two orders of magnitude brighter. The final 3.5 mag scaling therefore likely stems from the galaxy model being too dim in the $i$-band. By avoiding the final $i-$band calibration, the quasi-continuum levels may be wrong (due to uncertainties in the galaxy subtraction), but the fluxes in the distinct SN emission lines (Mg I] \wl4571, [O I] \wll6300, 6364 and [Ca II] \wll7291, 7323) should be accurate. To avoid a misleading discrepancy with the uncertain quasi-continuum level, we shift the observed spectrum with $-10^{38}$ erg s$^{-1}$ \AA$^{-1}$ to bring it into reasonable agreement with the models. 

\begin{figure}
\centering
\includegraphics[width=1\linewidth]{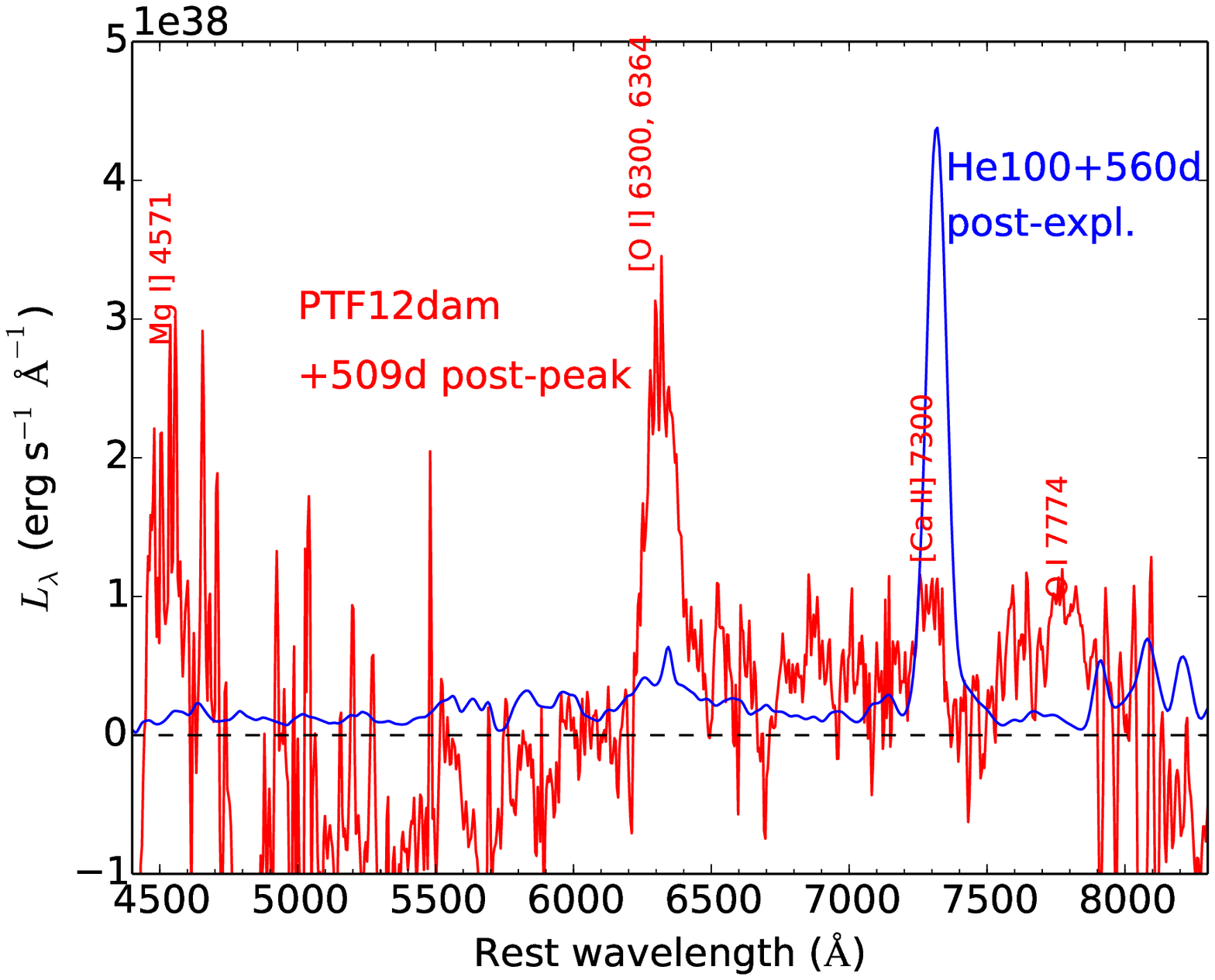} 
\includegraphics[width=1\linewidth]{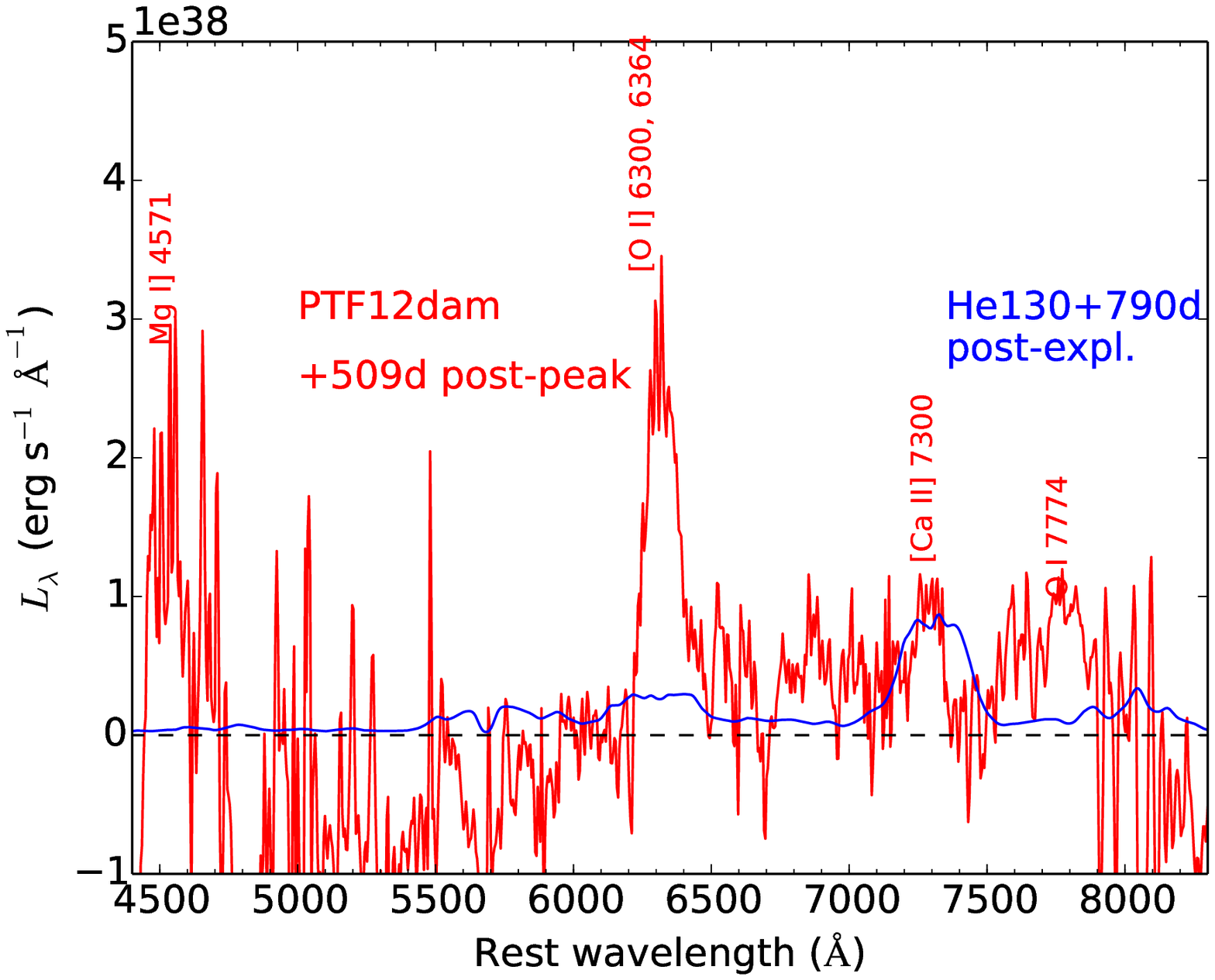} 
\caption{Comparison of the nebular spectrum of PTF12dam (red) with models (blue). Top: Comparison with He100 at +560d (700d model scaled with $e^{+140/111}$). Bottom: Comparison with He130 at +790d (700d model scaled with $e^{-90/111}$). See the text for explanations of the adapted scalings.}
\label{fig:12dam}
\end{figure}

The spectrum is compared to the models in Fig. \ref{fig:12dam}. The spectra were scaled to rest-frame spectral luminosity using $z = 0.107$, $H_{\rm 0}=72$ km s$^{-1}$ Mpc$^{-1}$, and standard $\Lambda$CDM cosmology (giving a luminosity distance of 481 Mpc). We have not attempted any correction due to dust as the Milky Way extinction is neglegible \citep[$A_V = 0.037$ mag,][]{Nicholl2013} and the estimated host extinction of $A_V=0.1$ mag \citep{Nicholl2013} would give a correction much smaller than the uncertainty in the final spectrum due to the host subtraction issues.

As for SN 2007bi, we assume an explosion epoch for each model that gives a good match of the post-peak light curve. From Fig. 4 in \citet{Nicholl2013}, Model He100 is 1-2 mags dimmer than PTF12dam at peak, and needs to be shifted by about +110d to match the later parts of the declining light curve. With a rise time of 160d, the epoch of the GTC spectrum becomes +560d (509d+160d-110d) post-explosion. For He130, Fig. 4 in \citet{Nicholl2013} shows that a -100 day shift is needed, which together with a 180d rise-time implies an epoch for the GTC spectrum of +790d (509d+180d+100d) post-explosion. Note that, as for SN 2007bi, this treatment leaves the peaks mismatched. 
In the figure, the He100 model is thus the 700d model scaled with $\exp{\left(+140/111\right)}$ to +560d, and the He130 model is the 700d model scaled with $\exp{\left(-90/111\right)}$ to +790d.

Given the issues with the data described above, the most relevant comparisons with the models are the luminosities in the distinct emission lines, Mg I] \wl4571, [O I] \wll6300, 6364 and [Ca II] \wll7291, 7323. As SN 2007bi, PTF12dam has significantly brighter Mg I] \wl4571 and [O I] \wll6300, 6364 emission than the models, whereas [Ca II] \wll7291, 7323 is dimmer. The strong Fe II + Mg I \wl5200 feature in SN 2007bi appears absent in PTF12dam. Irrespective of flux levels, PTF12dam shows a high [O I] \wll6300, 6364 / [Ca II] \wll7291, 7323 ratio of $\sim$3, compared to $0.1-0.3$ in the models.
The observed [O I] \wll6300, 6364 line has a width corresponding to an expansion velocity of $\sim$5000 km s$^{-1}$, somewhat too narrow for He100 (in which the O layer lies between 5000 and 9000 km s$^{-1}$), and much too narrow for He130 (in which the O layer lies between 8000 and 12000 km s$^{-1}$). 

Note that, as for SN 2007bi, there is no discernable redshift or distortion of [O I] \wll6300, 6364 or any other line profile. [O I] \wll 6300, 6364 is well fit with an optically thin Gaussian line profile (3:1 ratio between 6300 \AA~and 6364 \AA~components) with $FWHM=5600$ km s$^{-1}$. This suggests that there is little or no dust formation in the ejecta in the sense that any dust formed must have a neglegible opacity at optical wavelengths. This in turn removes an important uncertainty in the modelling regarding dust formation (see Sect. \ref{sec:discussion} for more discussion about this point).

The quasi-continuum appears to be redder than in SN 2007bi, but it is unclear how robust this property of the data is, as it relies sensitively to details in the galaxy subtraction process. We therefore ascribe little relevance to comparison of the spectral color with the models. From the large differences in line luminosities, line profiles, and line ratios described above, we conclude that the PTF12dam spectrum shows little resemblence to the PISN models.
 

\subsection{Non-superluminous supernovae}
\label{sec:normal-lum}
If PISNe exist, applying a normal stellar mass distribution to their progenitor range implies that most of them will produce small or moderate amounts of $^{56}$Ni and make normal or sub-luminous SNe. Thus, it is important to search for candidates among low-luminosity events not fitting into standard model schemes.

Synthesized $^{56}$Ni masses of less than 1 \msun~are in general obtained from He cores between 65-90 \msun.
These low $^{56}$Ni mass explosions still have large ejecta masses, $\gtrsim$ 65 \msun, and the diffusion time \citep{Arnett1982}
\begin{equation}
\tau_{\rm d} = 177\mbox{d}\left(\frac{M_{\rm ej}}{100~M_\odot}\right)^{3/4} \left(\frac{E_k}{10^{52}~\mbox{erg}}\right)^{-1/4} \left(\frac{\kappa}{0.1~\mbox{cm}^2~\mbox{g}^{-1}}\right)^{1/2}
\end{equation}
gives long diffusion light curves $\gtrsim 100$d as long as ionization (opacity) is maintained. The He80 model of \citet{Kasen2011} peaks in the optical after $\sim$100d at absolute magnitude -16, and the $M_{\rm ZAMS}=90$ \msun~model of \citet{Chatzopoulos2015} shows similar properties. The longest-duration (non-superluminous) Type Ib/c SN detected so far is SN 2011bm \citep{Valenti2012}, but even this had a rise time of 40 days and an inferred ejecta mass of 10 \msun, far below the PISN regime. Thus, there is currently no observed Type Ib/c SN that could be a PISN. 

Turning to Type II SNe, H-rich PISN models with small ($<$1 \msun) $^{56}$Ni production include those of \citet[][metal-free models R150 and R175]{Kasen2011} and \citet[][model 150M with $Z=0.001$]{Kozyreva2014b}. If the progenitor is a RSG, the optical light curves are complex, with an initial 20-30d cooling phase blending into a recombination/diffusion phase peaking at 200-300d (e.g. Fig. 7 in \citet{Kasen2011}). Model R175 shows a 200d plateau-like light curve, followed by settling on the $^{56}$Co tail. 
The \citet{Kozyreva2014b} model contains much less hydrogen (5 \msun~compared to 50 \msun, due to the higher metallicity) and shows a qualitatively different optical light curve, peaking after 70d and settling onto a $^{56}$Co tail at 150d. 

No known Type II SN shows either the $>200$d quasi-plateau obtained in the \citet{Kasen2011} models, or an evolution matching in detail the \citet{Kozyreva2014b} model. As \citet{Kozyreva2014b} discusses, it is not immediately obvious, however, that some Type II SNe could not be PISNe, in particular suggesting the very luminous Type IIP SN 1992am \citep{Hamuy2003}, SN 1992H \citep{Clocchiatti1996}, and SN 2009kf \citep{Botticella2010}, which are not easily understood in terms of normal core-collapse scenario due to the high explosion energies of $\gtrsim 10^{52}$ erg inferred \citep{Utrobin2010} \citep[see a discussion on energy limits for core-collapse in][]{Ugliano12}. 

No late-time data exists for SN 1992am or SN 2009kf, but for SN 1992H a 400d spectrum exists \citep{Clocchiatti1996}. As our models are H-free, and SN 1992H is dominated by H lines, we refrain from an explicit comparison, but several illuminating points can be made. The SN 1992H spectrum is dominated by the usual nebular core-collapse lines of [O I] \wll6300, 6364, H$\alpha$, [Fe II] \wl7155, and [Ca II] \wll7291, 7323, with no resemblence to our He80 model at 400d, which is characterized by a large variety of narrow silicon/sulphur and iron-group lines. H$\alpha$ in SN 1992H has a narrow peak, implying mixing down to small velocities, as obtained in core-collapse simulations \citep{Herant1991, Mueller1991, Kifonidis2006}, but in stark contrast to PISN simulations \citep{Joggerst2011,Chen2015,Whalen2014}. The spectrum is also blue, which cannot be achieved in PISN models due to their high ejecta masses and strong line blocking, with or without hydrogen (Sect. \ref{sec:opacity}). We conclude that the nebular properties of SN 1992H are inconsistent with a PISN interpretation. 


\section{Discussion}
\label{sec:discussion}

A limitation of the modelling is the lack of molecule and dust formation networks in the code, and the impact of these on the physical conditions in the ejecta and the radiative transport. 

Models for molecule and dust formation in PISN ejecta have been presented by \citet{Nozawa2003,Cherchneff2008,Cherchneff2009,Cherchneff2010}. Given the local nature of the gamma-ray deposition (Sect. \ref{sec:gammadep}) and consequential dominance of the inner iron-silicon-sulphur layers for the nebular spectral formation process, it is chiefly molecule and dust formation in the inner layers that is of immediate concern. In the $M_{\rm ZAMS}=170$ \msun~(80 \msun~He core\footnote{The mapping between $M_{\rm ZAMS}$ and He core mass depends sensitively on the assumed metallicity, rotation, and prescribed mass loss recipe. In the \citet{Whalen2014} models a 170 \msun~star makes a 120 \msun~He core, whereas in the \citet{Cherchneff2009} model a 170 \msun~star makes a 80 \msun~He core.}) unmixed PISN model of \citet{Cherchneff2009}, the innermost zone contains 20 \msun~of material, which is all the $^{56}$Ni (3.6 \msun\footnote{Note also difference in $^{56}$Ni mass compared to He80 model used here.}) and most of the silicon and sulphur. In this zone, molecule formation occurs around 400d, with an efficient production of SiS. Some S$_2$ is also formed, but is efficiently depleted by 600-700d. Other molecules (SiO, CO, SO, CS) are only formed in trace amounts ($< 10^{-5}$ by number). 

Dust pre-cursor formation in this zone begins around 400d, with Si$_4$ reaching 1\% number abundance \citep{Cherchneff2010}. Other congregates (e.g. FeS$_{2-4}$ and Fe$_{2-4}$) form with abundances $\lesssim 10^{-4}$. Overall only a small fraction of the iron and silicon atoms are depleted onto molecules or dust precursors, but sulphur is almost completely depleted by 500d. Thus, our predicted sulphur line emission might be overestimated. Inspection of the cooling channels in our Si/S zone shows that the cooling is roughly equally shared between Si, S, Ca, and Fe. Thus, removal of sulphur will only have a limited impact on the atomic cooling capability, and likely the total cooling efficiency will increase with the formation of SiS. This would lead to a lower temperature than in our models, which in turn would lead to lower ionization (because recombination coefficients increase with decreasing temperature). Since our models are already predominantly neutral in this zone, this should have no strong impact on the spectral formation. 

The strongest effect is possibly the opacity effect of dust formation. Inspection of the line profiles of e.g. [O I] \wll6300, 6364 in SN 2007bi and PTF12dam shows no discernable line shifts or distortions, which suggests that  dust has a minor impact. In the models, the total dust precursor formation at 1000d is 6 \msun, mainly of silicates, with little carbon dust formed \citep{Cherchneff2010}. Inclusion of dust opacity would lead to dimming and reddening of the optical model spectra, which would further increase discrepancy with PISN candidates (SN 2007bi and PTF12dam).



\begin{figure}
\includegraphics[width=1\linewidth]{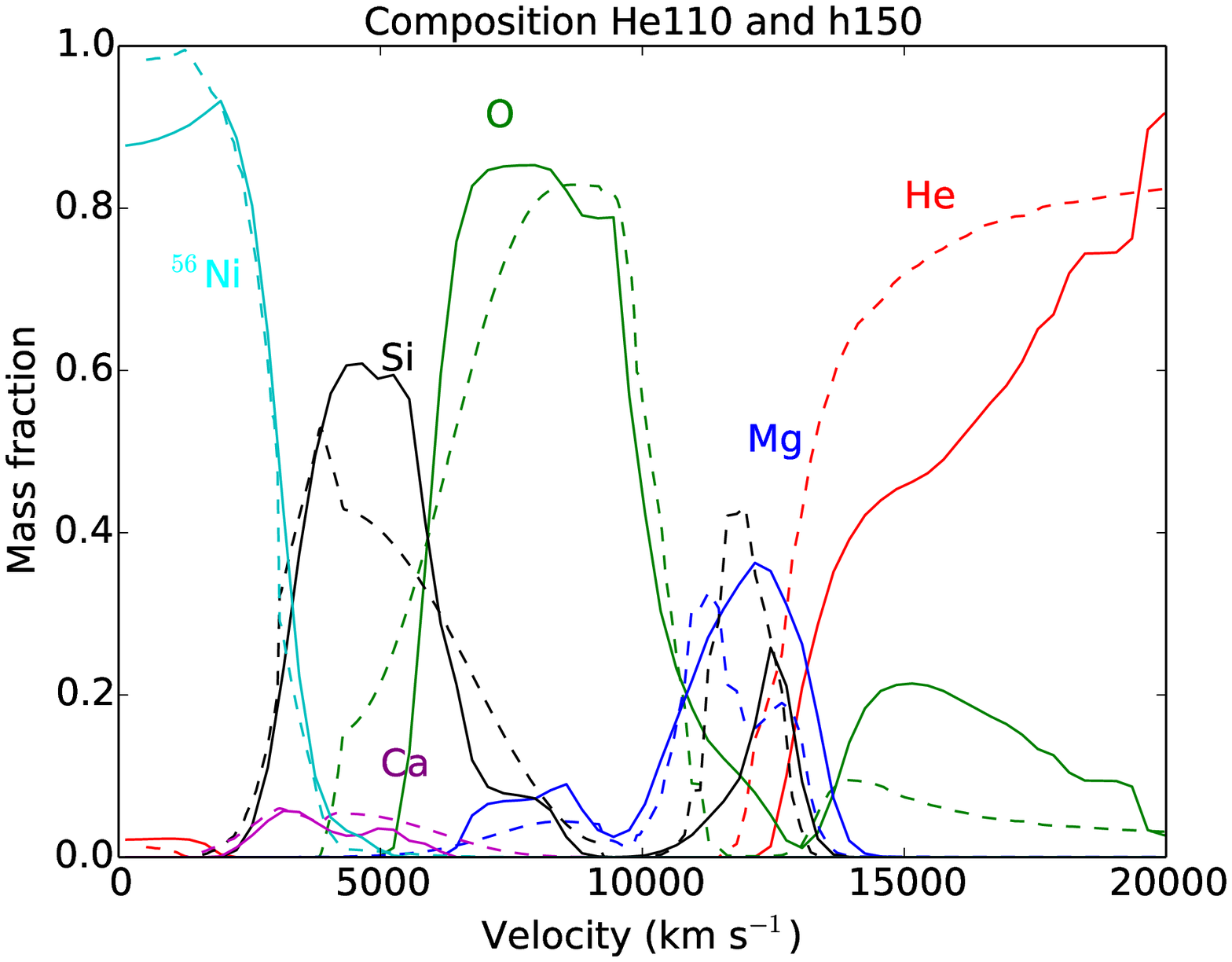} 
\includegraphics[width=1\linewidth]{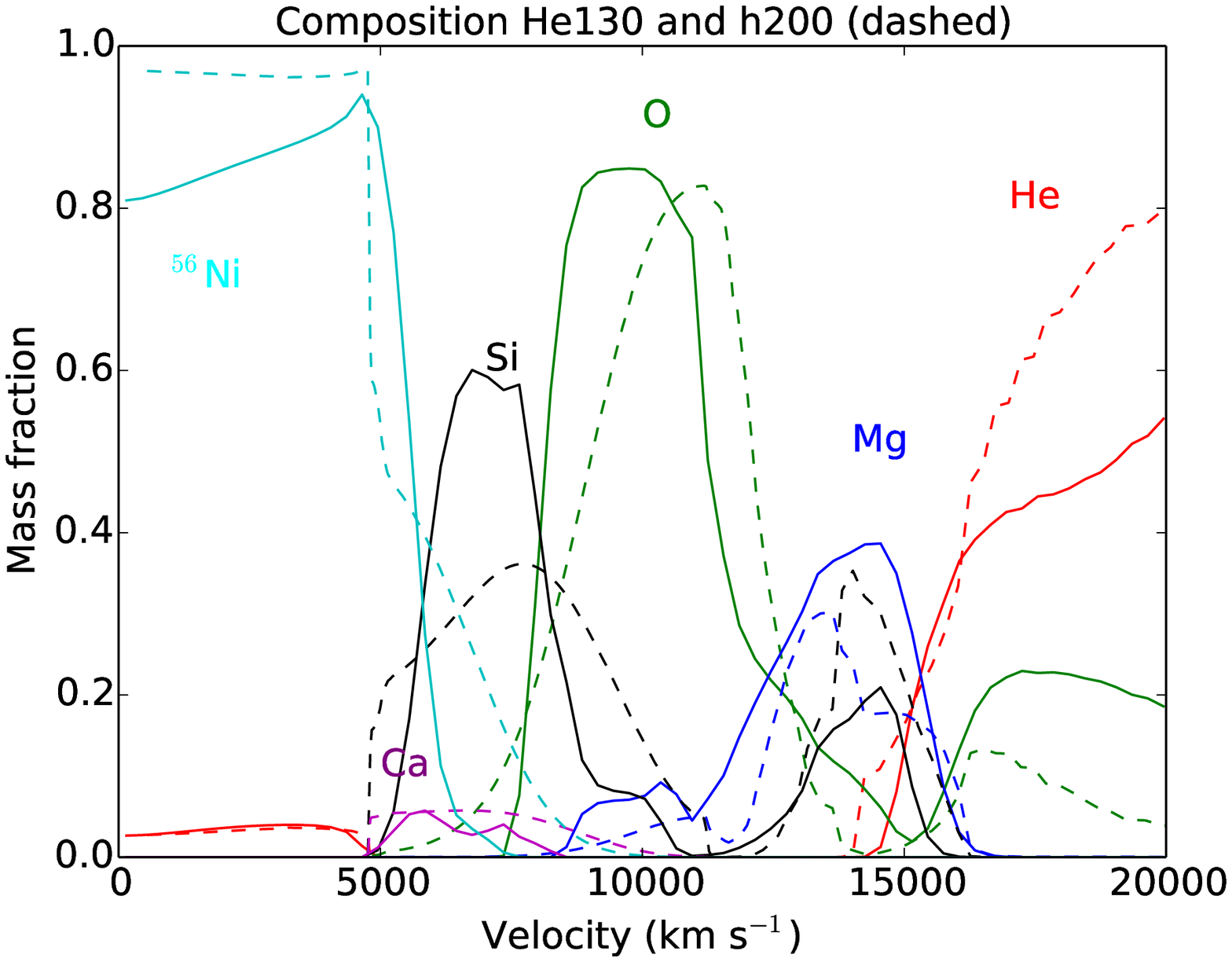} %
\caption{Comparison of models He110 and He130 of HW02 (solid lines) to models h150 and and h200 of \citet{Whalen2014} (dashed lines), which have corresponding He core masses.}
\label{fig:compwhalen}
\end{figure}

When it comes to the robustness of rejecting a pair-instability hypothesis for observed candidates (SN 2007bi and PTF12dam), the robustness of the predicted nucleosynthesis and explosion dynamics is of course a critical issue. Figure \ref{fig:compwhalen} compares the ejecta structures of the 110 and 130 \msun~He core models of WH02 with the h150 and h200 models of \citet{Whalen2014}, which explode with corresponding He core masses. The \citet{Whalen2014} models differ not only in metallicity, but also have rapid rotation (0.4 times break-up rotation). Despite these two differences, it is clear from Fig. \ref{fig:compwhalen} that the nucleosynthesis and dynamic structures are only moderately different. He110/h150 have almost identical distributions of $^{56}$Ni, Si, and O. The Si and O regions are somewhat more mixed in h150 than in He110. He130/h200 show somewhat larger, but still moderate, differences. In h200 a small fraction of the $^{56}$Ni is mixed out to the 7500 -10,000 km s$^{-1}$ range, whereas in He130 the $^{56}$Ni stays below 7500 km s$^{-1}$. A small amount of high-velocity $^{56}$Ni may have significant effects on the early light curve. In the nebular phase, however, line formation roughly follows the $^{56}$Ni distribution and this high-velocity tail would have a small influence.

It therefore appears safe to compare local PISN candidates with the HW02 metal-free models. Because the \citet{Whalen2014} models are rapidly rotating, we can also rule out any strong sensitivity of the ejecta structure due to rotation (for the same final He core mass). See \citet{Chatzopoulos2013} for an in-depth discussion of rotation effects. 


\begin{figure}
\includegraphics[width=1\linewidth]{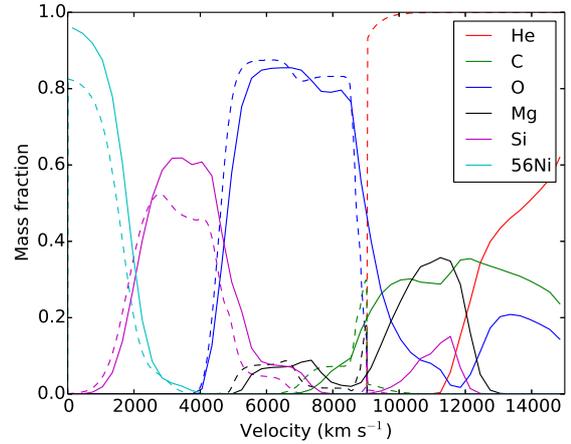} 
\caption{The composition (most abundant elements) of the He100 model of HW02 used here (solid lines) compared to the He100 model computed by \citet{Dessart2013} (dashed lines).}
\label{fig:compwaldman}
\end{figure}

It is also important to compare the output of different codes for the same stellar model. In Fig. \ref{fig:compwaldman} we compare the He100 model of HW02 with the He100 model of \citet{Dessart2013}. Apart from the low-density He atmosphere at the highest velocities ($>$ 9000 km s$^{-1}$), the two codes give similar ejecta. In the outermost layers, HW02 appears to obtain some explosive He burning, which the \citet{Dessart2013} simulation does not. This will have some impact on photospheric phase spectra, which are formed out in these layers. The impact in the nebular phase would however be minor, as spectral formation occurs deep in the core (Sect. \ref{sec:physical}). The same reasoning applies to sensitivity to the metallicity of the atmosphere, which depends on the original metallicity of the star. Nevertheless, it is desirable to confirm these arguments by computing further model suites, using explosion models with different metallicities, rotation rates, and physical approximations for the stellar evolution and explosion. Models that would be of interest include the h150 and h200 models of \citet{Whalen2014}, and the 250M model of \citet{Kozyreva2014b}. 

Previous theoretical work on late-time PISN emission has been done by \citet{Dessart2013}, who presented models for four PISN explosions covering both photospheric and nebular phases (up to 1000d). Of these models three contain hydrogen, whereas the fourth one is a 100 \msun~He core, directly comparable to the He100 model investigated here. 
Apart from some small differences for the dynamic structure and composition, \citet{Dessart2013} also applies some mixing/smearing (mainly to soften composition edges), which we do not for our models.

The \citet{Dessart2013} model spectra (Figs. 14 and 19 in their paper) are qualitatively similar to the model spectra computed here, with little flux emerging below 5000 \AA, and the 5000-10,000 \AA~range dominated by a strong [Ca~II]~\wll7291,~7323, with a forest of relatively narrow lines making up the rest of the spectrum. Some differences are also apparent, for instance does our model produce stronger Ca~II~NIR lines at 400d, whereas \citet{Dessart2013} has a more pronounced Mg I] \wl4571. The computer codes are similar in many respects, both computing full NLTE solutions including non-thermal and radiative rates. Our zoning here is 300 km s$^{-1}$ compared to a variable 100-1000 km s$^{-1}$ in \citet{Dessart2013}. 
SUMO relies on the Sobolev approximation which CMFGEN does not, but in the nebular phase this approximation should be accurate because the break-down region at $\lesssim 10$ km s$^{-1}$ is smaller than the vast majority of line separations. 
SUMO also lacks time-dependent terms which CMFGEN includes for both radiation field and statistical equilibrium. There are also differences in the atomic data libraries. For the particular application of nebular-phase PISNe, the spectra are dominated by Fe~I, which is computed with 496 levels and 23,000 transitions in SUMO, compared to 412 (super)-levels and 72,000 transitions in the \citet{Dessart2013} model, so Fe I should be well modelled by both simulations. SUMO also includes Co I, Ni I, Sc I, Sc II, Ti I, Ti II, V I, V II, Cr I, Cr II, Mn I, and Mn II which are absent in CMFGEN. These elements can play significant roles in the line transfer as well as in charge transfer reactions between the metals. 

Despite these code differences, it is encouraging for the robustness of predicted nebular PISN spectra that the He100 model computed here is relatively similar to the one of \citet{Dessart2013}. 
In particular, the same conclusion follows when comparing a He100 model with the best candidate so far, SN 2007bi; the modelled and observed spectra differ significantly in several key aspects. Some specific points in the comparison are, however, different because we compare our model with a recalibrated spectrum in which the host galaxy has been subtracted. This removes some of the color discrepancy, but several other discrepancies remain. Spectral dissimilarities are also present in models for the earlier phases \citep{Dessart2012}.We have here also extended the comparison of the candidates to other PISN models. 


What kind of ejecta structure would a PISN need to appear more similar to SN 2007bi and PTF12dam? With the strong observed emission lines of Mg I] \wl4571, [O I] \wll 6300, 6364, and O I \wl7774, a higher energy deposition in the O/Mg zone appears necessary. This would require more mixing between the $^{56}$Ni and O/Mg layers than seen in the multi-D simulations performed so far (which show almost none). The [O I] \wll6300, 6364 and [O I] \wl7774 lines have also quite narrow peaks which in spherically symmetric models at least requires strong mixing of oxygen to the centre of the nebula. It is still unclear, however, whether even such mixed ejecta would be able to produce blue enough colors as there is no obvious reason why the line blocking should decrease with mixing. The nebular model presented by \citet{Gal-Yam2009} does not include line blocking, and would probably also become too red if this was added. This model also assumes complete mixing both macroscopically (on large scales) and microscopically (at atomic level). Unless any serious shortcomings in the hydrodynamic mixing simulations performed so far can be identified, there is no known physical mechanism to produce the macroscopic mixing.

\section{Conclusions and summary}
\label{sec:conclusions}
We have computed spectral synthesis models for the late-time (400d-1000d) appearance of PISNe. 
The models show several unique signatures of these explosions. The high column densities, combined with weak mixing of $^{56}$Ni evidenced by multi-D simulations, lead to gamma-ray trapping in the deep-lying Fe/Si/S layers for several years post explosion. The energy deposition in the overlying O/Mg/C/He layers is never more than 10-20\%. As we show steady state to be valid, and $^{57}$Co gives neglegible contribution before 1000d, the consequence is that the bolometric luminosity closely follows the $^{56}$Co decay power until at least two years post-explosion.

The temperature and ionization obtained in the nebular phase varies
strongly with the amount of $^{56}$Ni produced in the explosion.  The
lowest mass PISNe (65 \msun~He cores) produce no $^{56}$Ni at all, whereas the 
the most massive ones (133 \msun~He cores) produce almost almost 60 \msun.
Our lowest mass model here, He80, makes 0.13 \msun~and  is cool
($T<$ 3000 K) and neutral ($x_e < 0.1$). The higher mass models He100 and
He130 are hotter and more ionized, but still stay within $T <
5000$ K and $x_e < 1$. Combined with the localized gamma-ray
deposition, this leads to emission of neutral heavy element species
(Fe I, Si I, S I) at characteristic core velocities, from 2000
km s$^{-1}$ at the low-mass end to 8000 km s$^{-1}$ at the
high-mass end. At the low-mass end, this core velocity is low enough to
produce a forest of distinct emission lines, dominated by Fe~I in the
optical and by Si~I and S~I in the NIR. At the upper end of the mass
range, the increased Doppler broadening leads to a smoother spectrum
with fewer clean emission lines. In all cases are the ejecta optically
thick from line blocking throughout the UV and much of the optical
regime for several years, and radiative transfer must be included to
obtain accurate model results.

The He80, He100, and He130 models presented here give clear and testable predictions for what PISNe are expected to look like in the nebular phase. The main characteristics of the spectra are that they are dim in the UV/blue (due to the line blocking operating for years or decades) and are dominated by a multitude of Fe~I, Si I and S I lines that have characteristic velocities from 2000 km s$^{-1}$ for low-mass PISNe up to 8000 km s$^{-1}$ for high-mass ones. Other distinct properties are strong [Ca II] \wll7291, 7323, narrow Na I D absorption troughs, and for some models a strong Ca~I~6572 \AA~line (which can be confused with H$\alpha$).

Comparison of these models with the most promising
PISN candidate so far, SN 2007bi,
shows several key discrepancies. None of the distinct characteristics of our model spectra  are seen in the observed spectra, which instead display other strong emission lines not produced by the models. A second, well observed PISN candidate PTF12dam also does not show any quantitative resemblance to our models, in particular having a much too high [O I] \wll6300, 6364 / [Ca II] \wll7291, 7323 ratio. 
With recent simulations showing no
strong impact of rotation or magnetic fields on the PISN structure
\citep{Chatzopoulos2015}, and constraints on the possible influence of
hydrodynamic mixing \citep{Joggerst2011,Chen2014,Whalen2014}, it
seems unlikely that models varying in physical parameters compared
to the ones investigated here would be able to significantly improve
the fits to these two candidates.

Whether PISNe occur or not in Nature will perhaps not be settled by observations of super-luminous supernovae. If they occur at the upper end of the theoretical mass range, $\mbox{M(He)} = 100-130$ \msun, they presumably also occur at the lower end, $\mbox{M(He)} = 65-100$ \msun, in similar or larger numbers. These low-mass PISNe have more typical SN luminosities, and nebular spectra of candidate events can be searched for the characteristic signatures of narrow Si, S, Ca, and Fe lines predicted by the He80 model. A distinct prediction of this model is an increasingly prominent Ca I \wl6572, eventually dominating the nebular spectrum.

\appendix

\section{NIR evolution}
\label{NIRlate}

\begin{figure*}
\includegraphics[width=1\linewidth]{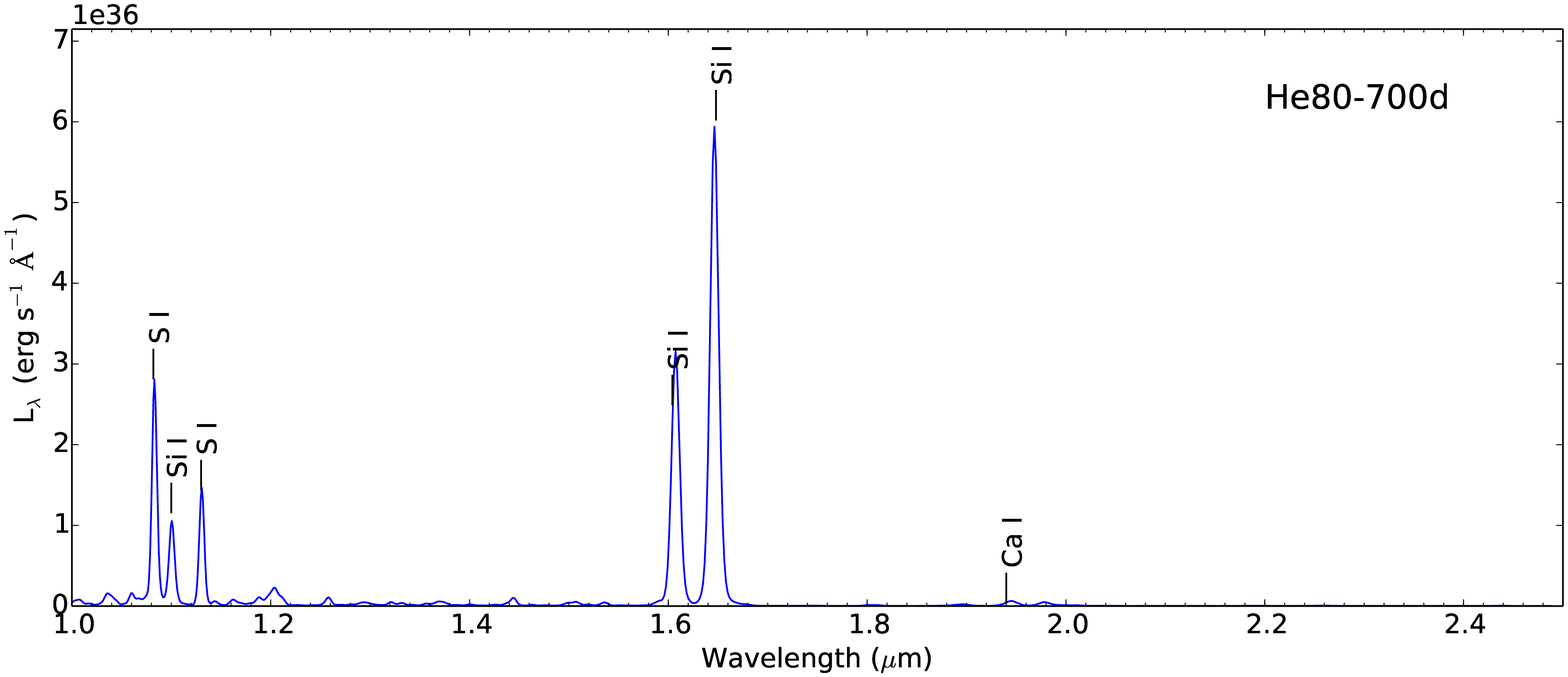} 
\includegraphics[width=1\linewidth]{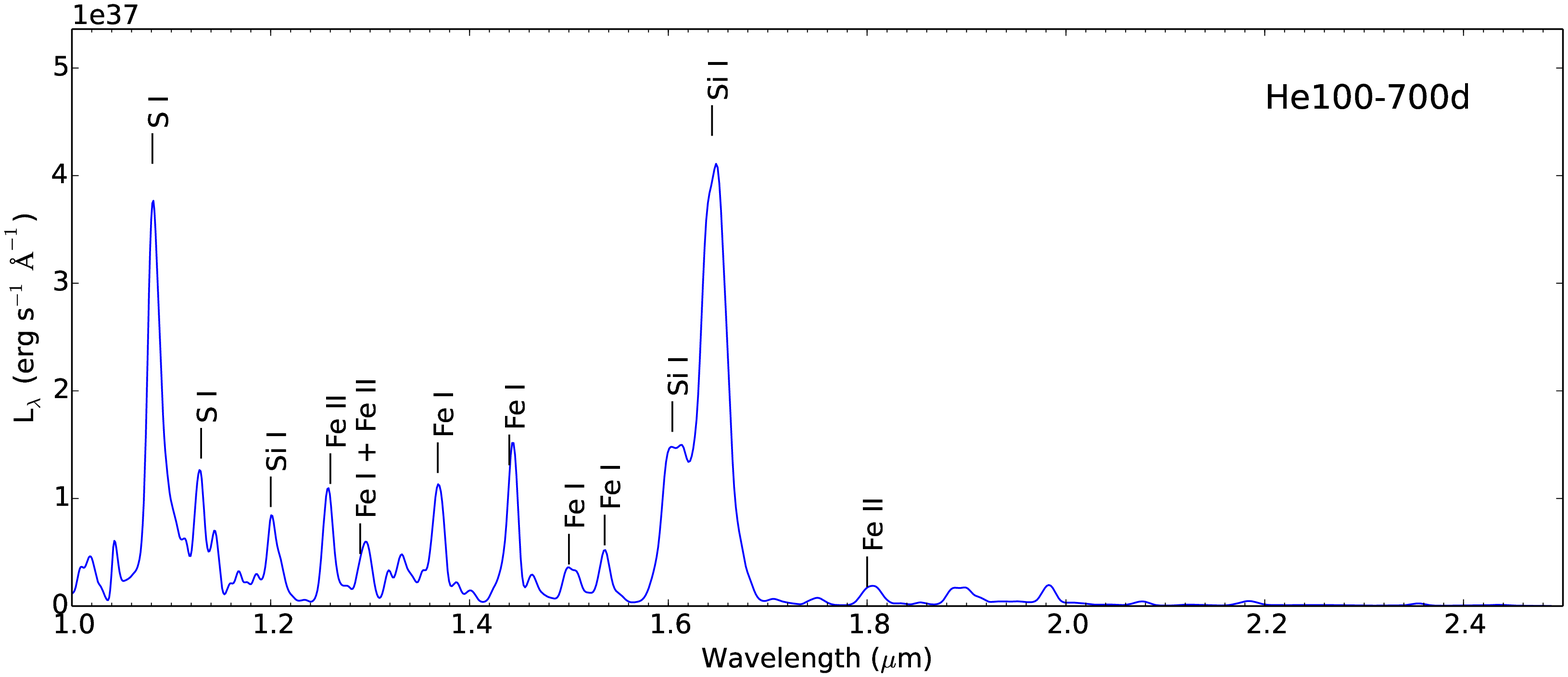} 
\includegraphics[width=1\linewidth]{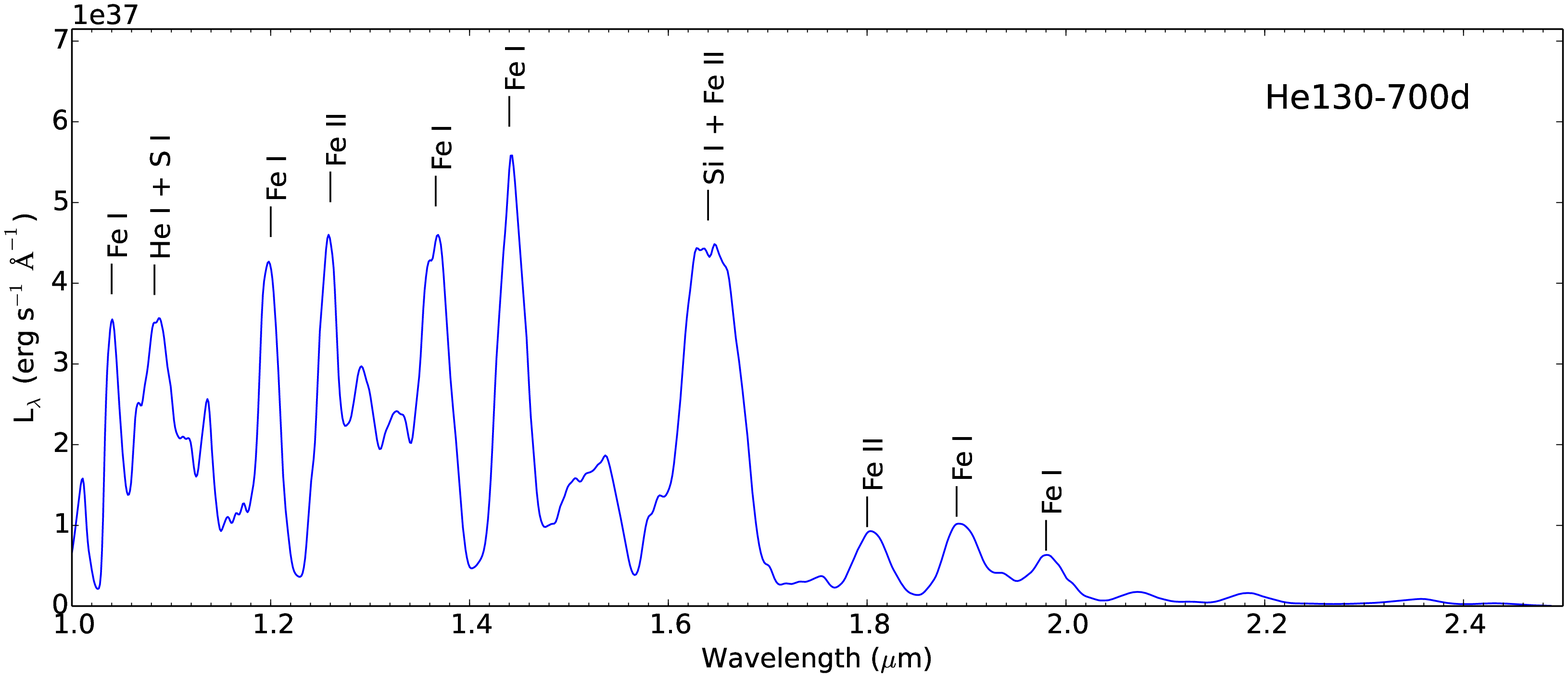} 
\caption{Model spectra in the near-infrared at 700d.}
\label{fig:nirspectra700}
\end{figure*}

\begin{figure*}
\includegraphics[width=1\linewidth]{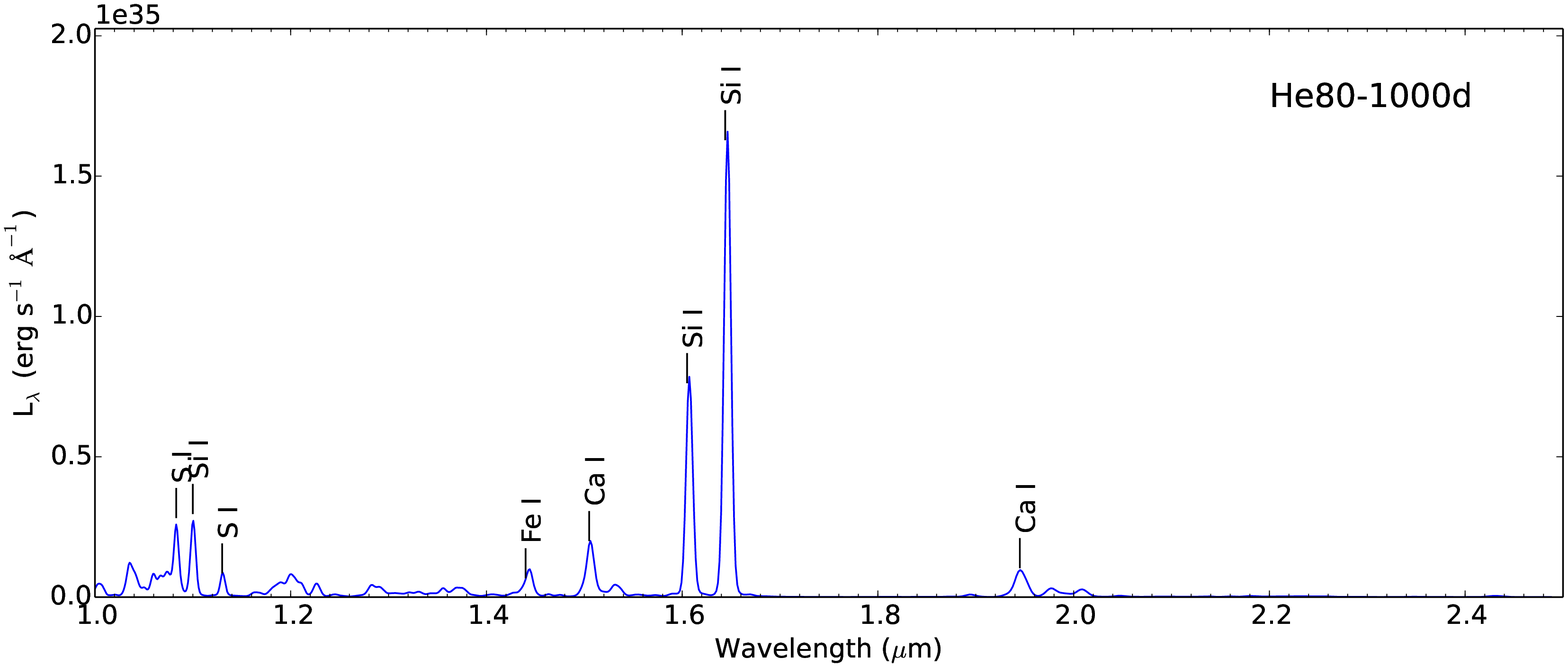} 
\includegraphics[width=1\linewidth]{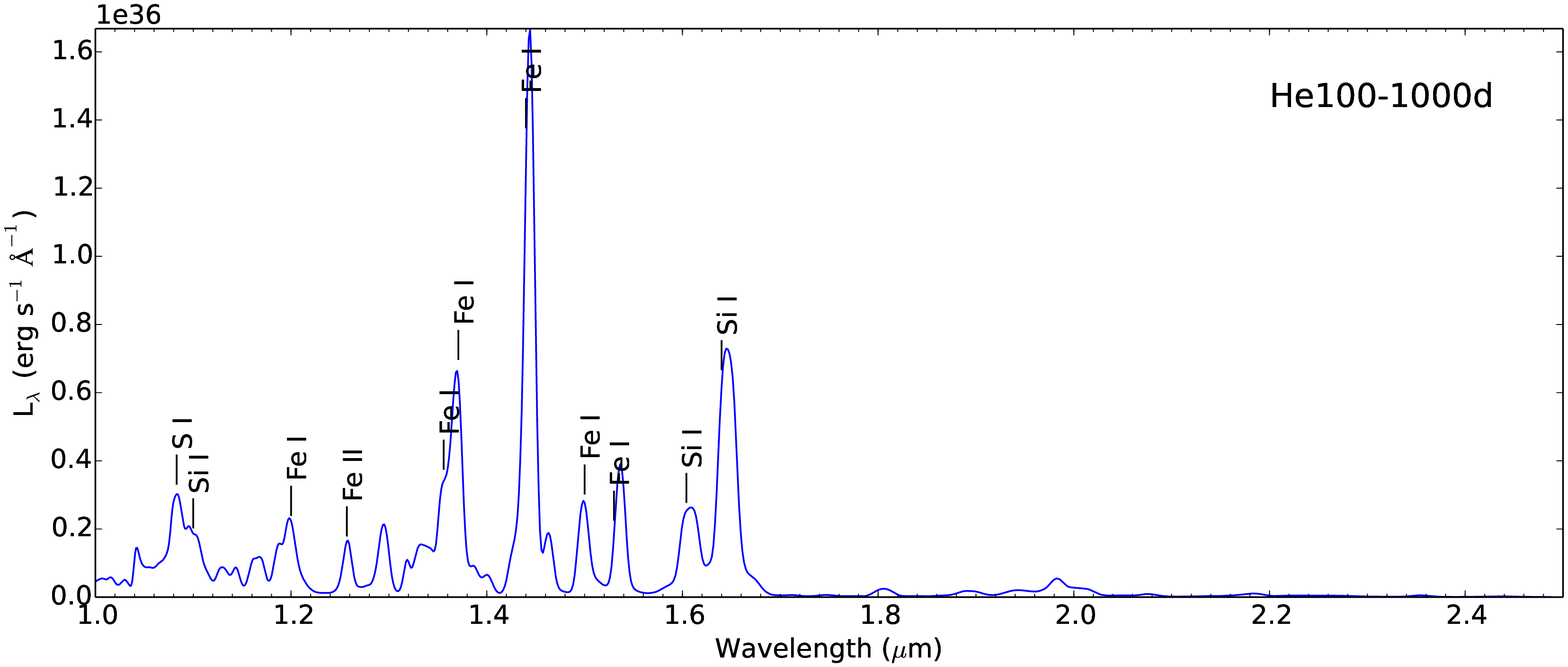} 
\includegraphics[width=1\linewidth]{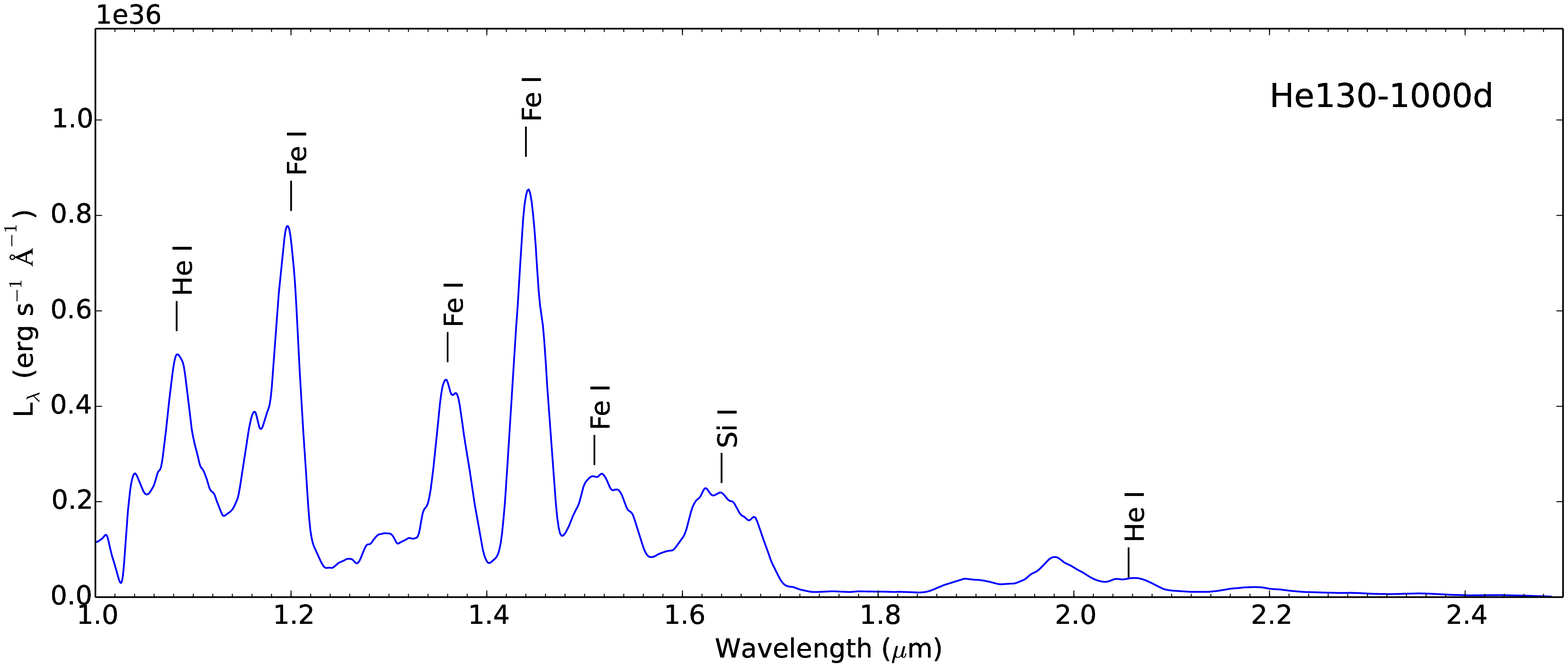} 
\caption{Model spectra in the near-infrared at 1000d.}
\label{fig:nirspectra1000}
\end{figure*}

Figures \ref{fig:nirspectra700} and \ref{fig:nirspectra1000} show the models in the NIR at 400d and 1000d, respectively.

\section*{Acknowledgments}
We acknowledge useful discussions with R. Waldman, R. Hirschi, A. Kozyreva, S. Valenti, A. Gal-Yam, P. Mazzali, M. Sullivan, L. Dessart, T.-W. Chen, and J. Vink. We also thank the referee for very useful comments that improved the manuscript. The research leading to these results has received funding from the European Research Council under the European Union's Seventh Framework Programme (FP7/2007-2013)/ERC Grant agreement n$^{\rm o}$ [291222]. AJ acknowledges STFC DIRAC supercomputing grant ACSP74 ``Spectral modelling of superluminous Type Ic supernovae''. AH was supported by an Australian Research Council (ARC) Future Fellowship (FT120100363) and NSF grant PHY-1430152 (JINA-CEE).

\bibliographystyle{mn2e3}
\bibliography{bibl}

\end{document}